\documentclass[a4paper,fleqn,usenatbib]{mnras}

\usepackage{amsmath}	
\usepackage{amssymb}	
\usepackage{newtxtt,newtxmath}

\usepackage[T1]{fontenc}
\usepackage{ae,aecompl}

\usepackage{graphicx}	
\usepackage{epstopdf}
\usepackage{subcaption}
\usepackage{multicol}
\usepackage{multirow}
\usepackage{float}
\usepackage[usenames,dvipsnames,svgnames,table]{xcolor}
\usepackage{enumitem}
\usepackage{bm}
\usepackage{tablefootnote}

\usepackage[noabbrev]{cleveref}

\crefname{equation}{}{}
\crefname{figure}{}{}

\newcommand{\wo}{$W_0$}
\newcommand{\so}{$\sigma_0$}
\newcommand{\lya}{Ly$\alpha$}
\newcommand{\ha}{H$\alpha$}

\newcommand{\oiii}{[O{\sc iii}]}
\newcommand{\oii}{[O{\sc ii}]}

\newcommand{\msol}{M$_\odot$}

\newcommand{\cgs}{erg s$^{-1}$ \AA$^{-1}$}

\def\arraystretch{1.0}%

\title[EW Distributions of $z = 0.47$ \ha~Emitters]{Correlations between \ha~Equivalent Width and Galaxy Properties at $z = 0.47$: Physical or Selection-driven?}
\author[Khostovan et al.]{A.~A.~Khostovan$^{1}$\thanks{NASA Postdoctoral Program Fellow}\thanks{E-mail:
		akhostov@gmail.com}, S.~Malhotra$^{1,2}$, J.~E.~Rhoads$^{1,2}$, S.~Harish$^{2}$, C.~Jiang$^{3}$,  J.~Wang$^{4}$, I.~Wold$^{1}$,  \newauthor Z.-Y.~Zheng$^{3}$, L.~F.~Barrientos$^{6}$, A.~Coughlin$^{2,5}$, W.~Hu$^{4}$, L.~Infante$^{6,7,8}$, L.~A.~Perez$^{2}$, J.~Pharo$^{2}$, \newauthor F.~Valdes$^{9}$, A.~R.~Walker$^{10}$\\ 
	$^{1}$Astrophysics Division, NASA Goddard Space Flight Center, Greenbelt, MD 20771, United States of America\\
	$^{2}$School of Earth and Space Exploration, Arizona State University, Tempe, AZ 85287, United States of America\\
	$^{3}$CAS Key Laboratory for Research in Galaxies and Cosmology, Shanghai Astronomical Observatory, Shanghai 200030, China\\
	$^{4}$CAS Key Laboratory for Research in Galaxies and Cosmology, University of Science and Technology of China,
	Hefei, Anhui 230026\\
	$^{5}$Chandler-Gilbert Community College, 2626 East Pecos Road, Chandler, AZ 85225-2499\\
	$^{6}$Instituto de Astrof\'isica, Facultad de F\'isica, Pontificia Universidad Cat\'olica de Chile, Santiago, Chile\\
	$^{7}$Las Campanas Observatory, Observatories of the Carnegie Institution of Washington, La Serena, Chile\\
	$^{8}$N\'ucleo de Astronom\'ia, Universidad Diego Portales, Santiago, Chile\\
	$^{9}$National Optical Astronomy Observatory, 950 N. Cherry Avenue, Tucson, AZ 85719, United States of America\\
	$^{10}$Cerro Tololo Inter-American Observatory, National Optical Astronomy Observatory, Casilla 603, La Serena, Chile}

\date{}

\pubyear{2021}
\begin{document}

\label{firstpage}
\pagerange{\pageref{firstpage}--\pageref{lastpage}}
\maketitle

\begin{abstract} 
The \ha~equivalent width (EW) is an observational proxy for specific star formation rate (sSFR) and a tracer of episodic, bursty star-formation activity. Previous assessments show that the \ha~EW strongly anti-correlates with stellar mass as $M^{-0.25}$ similar to the sSFR -- stellar mass relation. However, such a correlation could be driven or even formed by selection effects. In this study, we investigate how \ha~EW distributions correlate with physical properties of galaxies and how selection biases could alter such correlations using a $z = 0.47$ narrowband-selected sample of 1572 \ha~emitters from the Ly$\alpha$ Galaxies in the Epoch of Reionization (LAGER) survey as our observational case study. The sample covers a 3 deg$^2$ area of COSMOS with a survey comoving volume of $1.1\times10^5$ Mpc$^3$. We assume an intrinsic EW distribution to form mock samples of \ha~emitters and propagate the selection criteria to match observations, giving us control on how selection biases can affect the underlying results. We find that \ha~EW intrinsically correlates with stellar mass as $W_0 \propto M^{-0.16\pm0.03}$ and decreases by a factor of $\sim 3$ from $10^{7}$ \msol~to $10^{10}$ \msol, while not correcting for selection effects steepens the correlation as $M^{-0.25\pm0.04}$. We find low-mass \ha~emitters to be $\sim 320$ times more likely to have rest-frame EW$ > 200$\AA~compared to high-mass \ha~emitters. Combining the intrinsic \wo -- stellar mass correlation with an observed stellar mass function correctly reproduces the observed \ha~luminosity function, while not correcting for selection effects underestimates the number of bright emitters. This suggests that the \wo -- stellar mass correlation when corrected for selection-effects is physically significant and reproduces three statistical distributions of galaxy populations (line luminosity function, stellar mass function, EW distribution). At lower stellar masses, we find there are more high-EW outliers compared to high stellar masses, even after we take into account selection effects. Our results suggest that high sSFR outliers indicative of bursty star formation activity are intrinsically more prevalent in low-mass \ha~emitters and not a byproduct of selection effects.

\end{abstract}


\section{Introduction}

Understanding the star formation history of galaxies is of great importance in order to investigate how galaxies form and evolve. Currently, observations show that galaxies generally follow a correlation between their star formation rate (SFR) and stellar mass, commonly referred to as the `main sequence' (e.g., \citealt{Daddi2007,Noeske2007,Whitaker2012,Whitaker2014,Speagle2014}). Low-mass galaxies are observed to have higher specific star formation rates ($\textrm{sSFR} = \textrm{SFR}/\textrm{M}$; comparison between current-to-past star-formation activity) in comparison to high-mass galaxies suggesting they are dominated by young stellar populations and undergoing recent star-formation activity, while massive galaxies created the majority of their stellar population at earlier times (e.g., \citealt{Juneau2005,Zheng2007,Damen2009,Fumagalli2012}). The sSFR is found to increase as $(1+z)^{2.4-3.5}$ up to $z \sim 2$ (e.g., \citealt{Karim2011,Fumagalli2012,Faisst2016}) suggesting that galaxies at early cosmic times were more actively producing stars and dominated by young stellar populations. This is steeper than the $(1+z)^{2.25}$ cold gas accretion-dominated growth predicted by analytical models and cosmological simulations (e.g., \citealt{Neistein2008,Dekel2009}). Some studies at $z>2$ find higher sSFRs (e.g., \citealt{Stark2013,deBarros2014,Salmon2015,Faisst2016,Jiang2016}), while other studies find a flat evolution in the sSFRs at $z > 2$ (e.g., \citealt{Stark2009,Gonzalez2010,McLinden2011,Gonzalez2014,Heinis2014,Tasca2015,Marmol2016}). However, the effects of nebular emission lines in broadband photometry is found to cause an overestimation in stellar mass measurements (e.g., \citealt{Gonzalez2010,Schenker2013,deBarros2014}). Correcting for this effect, \citealt{Faisst2016} found an increase of $(1+z)^{1.5}$ at $z > 2$ significantly shallower than what is predicted by simulations (e.g., \citealt{Weinmann2011}) and empirical models (e.g., \citealt{Khochfar2011,Speagle2014}). At a given redshift, the sSFR exhibits a U-shaped scatter where it increases towards both low and high stellar masses (e.g., \citealt{Willett2015,Davies2019}) where simulations attribute such an effect to short-timescale variations in the star formation activity of galaxies (e.g., \citealt{Hopkins2014,Sparre2017,Katsianis2019,Matthee2019}).

The extensive work on measuring the sSFR at varying cosmic epochs has provided useful information regarding the intensity of star-formation activity at different stellar masses. However, several caveats arise when using sSFR to investigate SF histories. First, the choice of SF indicator is important as calibrations that trace long timescales (e.g., UV continuum $\sim 100$ Myr) could have signatures of episodic SF washed out while indicators sensitive to instantaneous activities (e.g., emission lines such as \ha; $\sim 5 - 10$ Myr) would be able to observe bursty SF activity. Second, stellar mass measurements could be overestimated by nebular emission line contributions, hence, affecting the measured sSFR (e.g., \citealt{deBarros2014}) and are model-dependent based on the best-fit SEDs. Another method of capturing episodic SF activity involves using \ha/UV~ratios (e.g., \citealt{Glazebrook1999,Iglesias2004,Lee2011,Weisz2012,Dominguez2015,Guo2016,Emami2019}), however these measurements are also susceptible to systematics arising from dust correction assumptions (e.g., \citealt{Kewley2002,Lee2009,Shivaei2015,Broussard2019, Faisst2019}; Pharo et al. {\it in prep}) and initial mass function (IMF) variations (e.g., \citealt{Meurer2009,Mehta2017}).

An alternative approach is using the \ha~equivalent width (EW), which is defined as the ratio of \ha~flux (tracing instantaneous SF activity) and continuum flux density (tracer of stellar mass), therefore, making it a model-independent, observational proxy of sSFR. It is also independent of dust corrections in the case where nebular and stellar $E(B - V)$ are equal (e.g., \citealt{Kashino2013,Reddy2015,Puglisi2016}). Furthermore, the \ha~EW is even more sensitive to bursty star formation histories in comparison to \ha/UV line ratios given the continuum at $6563$\AA~is redwards of the $4000$\AA~break and traces the old, low-mass stellar population. 

Measurements of \ha~EW distributions have extended up to $z \sim 2$ using narrowband (e.g., \citealt{Sobral2014}) and spectroscopic surveys (e.g., \citealt{Erb2006,Fumagalli2012,Reddy2018}). Recently, studies have used {\it Spitzer} IRAC color excess associated with strong \ha~emission to study EW at $z > 2$ (e.g., \citealt{Shim2011,Labbe2013,Stark2013,Smit2014,Faisst2016,Marmol2016,Rasappu2016,Smit2016,Caputi2017,Faisst2019}). \citet{Fumagalli2012} investigated \ha~EW properties up to $z = 2.6$ using samples drawn from SDSS, VVDS, 3D-{\it HST}, and the spectroscopic sample of \citet{Erb2006} and found a correlation between the typical \ha~EW and stellar mass at all redshifts. They also found the typical \ha~EW at a given stellar mass increases as $(1+z)^{1.8}$ and the redshift evolution shows little mass dependence. Using the narrowband HiZELS survey, \citet{Sobral2014} also found EW$\sim M^{-0.25}$ and an increasing EW with redshift scaled as $(1+z)^{1.72\pm0.06}$ up to $z \sim 2$, in agreement with \citet{Fumagalli2012}. Recently, \citet{Reddy2018} used the spectroscopic MOSDEF survey and found an EW -- stellar mass correlation of $M^{-0.378\pm0.004}$ and $M^{-0.286\pm0.003}$ at $z \sim 1.5$ and $2.3$, respectively. The redshift evolution at $z > 2$ is less clear where IRAC color excess measurements find no evolution \citep{Marmol2016} and $(1+z)^{1.5 - 1.8}$ \citep{Faisst2016,Smit2016}.

The correlation between \ha~EW -- stellar mass is somewhat expected given that continuum luminosity and stellar mass should be directly correlated. Indeed the EW -- stellar mass correlation should mimic how sSFR and stellar mass trend. The increase in rest-frame EW with redshift is also similar to the cosmic sSFR evolution (e.g., \citealt{Fumagalli2012}). However, a major question that needs to be addressed is to what degree are these trends shaped by selection effects? For example, narrowband surveys are line flux-limited with a rest-frame EW cutoff, where the former is set by the narrowband detection limit and the latter ensures that sources have a narrowband color excess consistent with emission lines rather than stellar continuum features in the filter profile. The rest-frame EW limit removes low EW sources at bright continuum (high mass). The line flux selection limit selects sources at high EW towards faint continuum while transitioning towards lower EW with increasing continuum luminosity. This essentially makes narrowband-selected samples increasingly incomplete towards lower stellar masses and lower EWs systems and can easily form an EW -- stellar mass correlation. Grism surveys also have the added complexity of a variable line flux limit depending on the position of the emission line in respect to the grism throughput and typically higher EW selection limits (e.g., \citealt{Momcheva2016}). This too can result in an EW -- stellar mass correlation formed/shaped by selection effects. Therefore, to investigate the EW -- stellar mass correlation requires that we also understand how selection limits affects the underlying measurement and whether an `intrinsic' correlation is still in place after correcting for said biases.
	
Past studies of \lya~emitters and Lyman Break Galaxies (LBGs) report a correlation between \lya~EW and rest-frame UV continuum and a lack of high EW, bright UV continuum sources (e.g., \citealt{Ouchi2003,Shapley2003,Stanway2007,Stark2010,Kashikawa2011,Zheng2014,Hashimoto2017,Santos2020}), known as the `Ando effect' \citep{Ando2006}. Similar results have been reported in previous \ha~studies as mention above, as well as \oiii~and \oii~studies (e.g., \citealt{Fumagalli2012,Sobral2014,Khostovan2016,Reddy2018}).  \citet{Ando2006} suggests a physical origin for the deficiency arguing that massive LBGs have older stellar populations and experience the majority of their star-formation activity at earlier times. This is also reinforced by clustering studies showing bright UV \lya~emitters residing in massive dark matter halos ($>10^{12}$ \msol; e.g., \citealt{Khostovan2019}) similar to \ha, \oiii, and \oii~emitters (e.g., \citealt{Sobral2010,Cochrane2018,Khostovan2018}). However, several studies that modeled \lya~EW distributions report selection functions and low survey volumes (sample variance) can result in a EW -- stellar mass correlation with a lack of bright continuum, high EW systems \citep{Nilsson2009,Zheng2014,Hashimoto2017}. Therefore, to assess if the EW -- stellar mass correlation is an intrinsic properties of star-forming galaxies requires that we take into account selection biases and use samples covering large comoving volumes to mitigate sample/cosmic variance effects.

Furthermore, observations of starburst galaxies (high SFR outliers in the main sequence) are also seen as extreme emission-line galaxies (EELGs) with \ha~and \oiii~EWs $>200$\AA~at different cosmic epochs (e.g., \citealt{Atek2011,Atek2014,Maseda2014,Calabro2017}). \citet{Atek2014} found that such sources are in a state of bursty star formation that can double their stellar mass within $100$ Myr and can contribute as much as $\sim 30$ percent to the total SFR at $z \sim 1 - 2$ for emission-line-selected samples. They also conclude that the contribution of starbursts increases towards lower stellar masses. However, to accurately quantify the high EW outlier/starburst fraction requires an estimation of the intrinsic population of star-forming galaxies at a given stellar mass where selection biases and sample variance issues are taken into account.

In this paper, we present a new methodology to constrain \ha~EW distributions by simulating the intrinsic \ha~distributions, propagating selection effects, and comparing them to observations. We use the $z = 0.47$ \ha~narrowband-selected sample from the Ly$\alpha$ Galaxies at the Epoch of Reionization (LAGER; \citealt{Zheng2019}) selected in \citet{Khostovan2020} as our case study. The sample consists of 1572 \ha~emitters within a survey comoving volume of $1.1\times10^5$ Mpc$^3$ which allows for us to robustly probe wide ranges of EWs, including rare emitters, and decreasing cosmic/sample variance effects. With the observationally-constrained simulations, we investigate the intrinsic correlations between \ha~equivalent width and galaxy properties, analyze how selection effects can shape the correlations, quantify the high equivalent width outlier (bursty SF) fractions, and discuss the implications of our results for future surveys as well as SFR -- stellar mass correlation measurements.

The organization of the paper is as follows: We present the LAGER \ha~sample in \S\ref{sec:sample} as the observation for which we will constrain the EW distribution using our simulations, which we describe in \S\ref{sec:modeling}. We then show our main results in \S\ref{sec:results} where we investigate the correlations between EW, \ha~luminosity, and $R$-band luminosity, followed by an analysis of the high EW fraction for different $R$-band luminosities. We then discuss in \S\ref{sec:discussion} the implications of our results in terms of the lack of massive, high EW \ha~emitters, the effects of selection on the main sequence, and what our results imply regarding future surveys. Lastly, we present our main conclusions in \S\ref{sec:conclusions}.

We assume a flat $\Lambda$CDM cosmology where $H_0 = 70$ km s$^{-1}$ Mpc$^{-1}$, $\Omega_m = 0.3$, and $\Omega_\Lambda = 0.7$.  Unless otherwise explicitly stated, all magnitudes follow the AB magnitude system and stellar masses assume a \citet{Chabrier2003} initial mass function (IMF).

\section{Sample}
\label{sec:sample}

We use our samples of \ha, \oiii, and \oii~emission-line galaxies from \citet{Khostovan2020} that were selected as part of the Ly$\alpha$ Galaxies in the Epoch of Reionization (LAGER) survey \citep{Zheng2019}. Observations were done with a custom designed narrowband NB964 filter ($\lambda = 9640$\AA; FWHM = $92$\AA; \citealt{Zheng2019}) in a single 3 deg$^2$ pointing of the COSMOS field using the DECam imager on the 4-m Blanco CTIO telescope. Corresponding archival broadband DECam $z$ ($\lambda = 9138.16$\AA; FWHM = $1478.68$\AA; \citealt{Abbott2018}) imaging was obtained through the NOAO Science Archive. We note that, upon completion, the LAGER survey will comprise a total of 8 independent fields for a combined survey area of 24 deg$^2$. The survey reaches down to a $5\sigma$ limiting flux of $8.2 \times 10^{-18}$ erg s$^{-1}$ cm$^{-2}$, corresponding to $10^{39.83}$, $10^{40.55}$, and $10^{41.13}$ erg s$^{-1}$ for \ha, \oiii, and \oii~emitters, respectively. We refer the reader to \citet{Hu2019} for details regarding the data reduction and source extraction. 

Sample selection is explained in great detail in \citet{Khostovan2020}. In brief, narrowband excess sources that exhibit potential emission line features are selected based on three selection criteria. First, a $5\sigma$ narrowband cut of 25.45 mag is applied to remove any false detections/potential artifacts. Second, a rest-frame equivalent width cut of $35$\AA~is applied to remove sources that could mimic an emission-line feature, such as a strong continuum break (e.g., 4000\AA~break). The last criteria is a color significance cut (`Bunker' parameter; \citealt{Bunker1995}) $\Sigma > 3$ which ensures that the narrowband excess of emission line galaxy candidates is not dominated by photometric scatter $>3\sigma$. Each of these criteria are crucial when we model \ha~emitters in the LAGER survey (see \S\ref{sec:modeling}).

The emission line identification is done by using archival spectroscopic redshifts, photometric redshifts from the COSMOS2015 catalog \citep{Laigle2016}, and an empirically-calibrated $BVzJ$ color-color selection that is designed based on the locations of spectroscopically confirmed sources. This last selection method is crucial in selecting high equivalent width sources for which their continuum is faint or the number of broadband detections were too few to robustly constrain their photometric redshifts. These will also be sources of interest when we investigate high EW outliers as potential bursty systems. Spectroscopic redshifts are taken from the wealth of ancillary spectroscopic observations done in the COSMOS field (\citealt{Lilly2009,Brammer2012,Cool2013,Balogh2014,Comparat2015,Kriek2015,Silverman2015,Momcheva2016,Masters2017,Hasinger2018,Straatman2018}).

In brief, any source with spectroscopic redshifts corresponding to the expected \ha, \oiii, and \oii~redshifts within the NB964 filter are automatically selected. If no spectroscopic redshift exists or the quality of the spectra is poor ($Q_f \le 2$), then candidates are identified based on their photometric redshifts from the COSMOS2015 catalog. The remaining candidates that were not selected by the above two methods are then identified based on their observed $BVzJ$ colors.

The total sample comprises of 1572 $z = 0.47$ \ha, 3933 $z = 0.93$ \oiii, and 5367 $z = 1.59$ \oii~emission line galaxies. Currently, there are 222 (14\%), 126 (3\%), and 104 (2\%) spectroscopically confirmed emitters for the \ha, \oiii, and \oii~samples, respectively. Although the LAGER DECam images cover a 3 deg$^2$ area, the final sample only covers a $2.4$ deg$^2$ survey area given the limited photometric areal coverage of the COSMOS2015 data. Given this limited coverage, the final comoving survey volumes are 1.1, 3.4, and $6.7 \times 10^5$ Mpc$^3$ for the \ha, \oiii, and \oii~samples, respectively. 

As the main objective of this study is to investigate the \ha~equivalent width distributions, we place our focus on only the $z = 0.47$ \ha~sample although we plan to investigate the other emission line samples in the near future using a similar analysis introduced in this paper. Furthermore, the large volume and depth of the samples provides us with a sample covering a wide range of \ha~luminosities, equivalent widths, and continuum luminosities (stellar masses) that allow us to robustly investigate the physical correlations of the equivalent width distributions with galaxy properties.

\section{Forward Modeling Equivalent Width Distributions}
\label{sec:modeling}

Here, we outline our methodology of forward modeling equivalent width distributions by simulating samples of \ha~emitters with intrinsic EW, line luminosity, and stellar mass properties and propagating observational and selection effects. The aim of our methodology is to investigate if correlations exist between EW and \ha~luminosity and/or continuum luminosity (proxy for stellar mass) and, if so, to what extent are the correlations selection-driven or intrinsic.

\subsection{Choice of Intrinsic Equivalent Width Distribution Model}
\label{sec:EW_model}

Given the lack of \ha~EW studies that model the underlying distribution, we look to the extensive \lya~EW studies done over the past decade at $z < 2$ (e.g., \citealt{Cowie2010,Wold2014,Wold2017}) and $z > 2$, (e.g., \citealt{Gronwall2007,Nilsson2009,Kashikawa2011,Stark2011,Ciardullo2012,Zheng2014,Oyarzun2016,Hashimoto2017,Santos2020}), which suggest an exponential distribution best represents observations. However, some studies also have explored a normal distribution as an alternative to represent EW distributions (e.g., \citealt{Ouchi2008,Guaita2010,Kashikawa2011,Zheng2014,Oyarzun2017}).

We define the exponential distribution as: 
\begin{eqnarray}
p({\rm EW}|W_0) = \frac{1}{W_0} e^{-{\rm EW}/W_0}
\label{eqn:exp_model}
\end{eqnarray}
where $p({\rm EW}|W_0)$ is the probability of having a source with a rest-frame equivalent width given a distribution with an $e$-scaling, \wo, which we refer to as the `characteristic equivalent width'. The normal distribution is defined as:
\begin{eqnarray}
p(\textrm{EW}|\sigma_0) = \frac{1}{\sqrt{2 \pi \sigma_0^2}} e^{-\textrm{EW}^2/2\sigma_0^2}
\label{eqn:normal}
\end{eqnarray}
where the shape of the distribution is set by the `characteristic width', $\sigma_0$. This assumes half-sided Gaussian with the center set to $0$\AA.

For a given simulation, we assume either an exponential or a normal distribution and assign EW per source by randomly sampling from the intrinsic distribution profiles. We later will investigate whether the exponential or the normal distribution best represents observations. After randomly assigning an EW to each source, we independently sample the second property which will be either line luminosity drawn from an intrinsic \ha~luminosity function or stellar mass from a stellar mass function, as discussed in detail below. Our simulations depend on two inputs: the assumed \wo~or \so~and the Schechter parameters that define the luminosity function (LF) and stellar mass function (SMF) assumed. Since the latter are already constrained based on observations and are fixed parameters, the only free parameter in making our mock samples is \wo~or \so~that defines the shape of the intrinsic EW distribution. As we will show below, these mock simulations will undergo a selection method to represent a $z = 0.47$ \ha~LAGER sample and are then fitted to the LAGER EW distributions. The advantage of our approach is that we have control over the intrinsic EW distribution via \wo~or \so~and can trace how selection biases affect the underlying shape of the EW distribution.

In the following sections below, we describe in detail the methodology and assumptions of our two different approaches where we keep the EW distributions independent of either \ha~luminosity or rest-frame $R$-band continuum luminosity (observational proxy for stellar mass).

\begin{figure}
	\centering
	\includegraphics[width=\columnwidth]{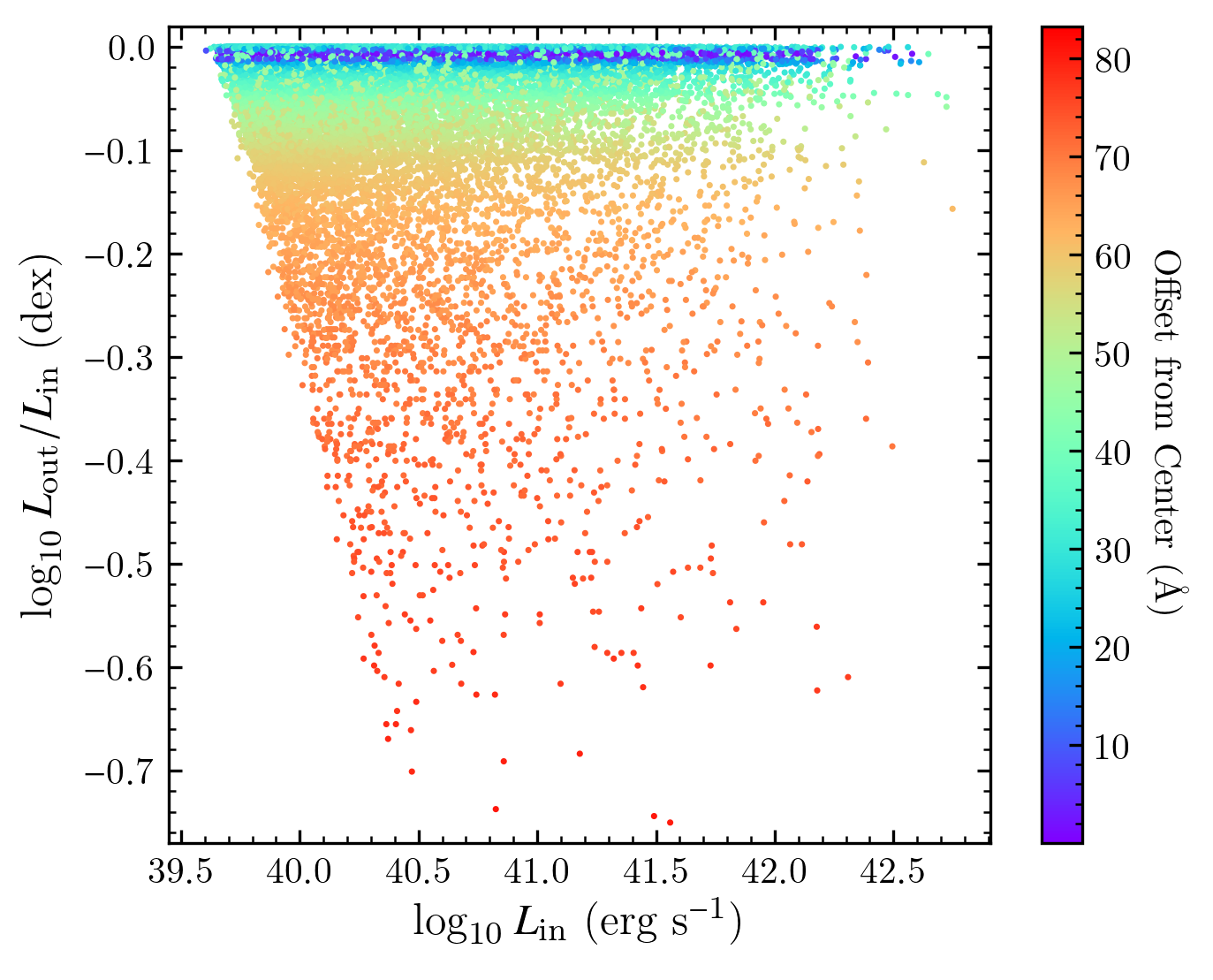}
	\caption{Demonstration of the filter profile effect on \ha~luminosity for mock sources that passed the narrowband selection criteria. An offset up to $\pm45$\AA~is consistent with the NB964 FWHM where the \ha~luminosity is reduced up to $-0.1$ dex by the filter profile. Sources intrinsically brighter than $>10^{40.5}$ erg s$^{-1}$ cover the full NB profile while  intrinsically fainter sources are more clustered around the central wavelength. This is because the faint emitters fail the NB magnitude selection limit towards the filter wings. This highlights the need of taking into account the filter profiles to mimic the LAGER observations, especially for sources spread further away from the central wavelength.}
	\label{fig:filter_profile_corr}
\end{figure}

\subsection{Modeling Associated Photometry \& Filter Profile Effects}
\label{sec:photometry}

Narrowband filter profiles are not perfect top-hat filters and, in some cases, resemble more of a Gaussian profile than a top-hat. This can affect the underlying \ha~luminosity and EW measurements depending on the position of the emitter in respect to the NB profile. An intrinsically bright \ha~emitter would be observable (above the survey NB limit) throughout most of the filter profile, although sources populating the wings of the filter would have observationally faint \ha~flux. However, intrinsically faint \ha~emitters would primarily be detected closer to the filter transmission peak while the same sources would be observationally fainter than the NB magnitude limit towards the filter wings. Typically these effects are taken into account when making statistical property measurements such as luminosity functions (e.g., \citealt{Sobral2013,Khostovan2020}). However, to properly generate a mock observation sample of \ha~emitters would require that we also mimic the filter profile effect on the modeled \ha~luminosity and equivalent width. 

We start by randomly assigning redshifts over the full wavelength coverage of the NB filter to each mock \ha~emitter by drawing from an uniform redshift distribution. For each source, we assume that the continuum flux density is flat in $f_\lambda$ over the wavelength range of both the NB and BB filter profile and that the emission line is a delta function centered on the assigned random redshift. We convolve the spectra with the NB and BB transmission curves to determine the modeled magnitudes. Since we require that a galaxy with no emission line has a nebular color excess, (BB -- NB$ = 0$) and that NB964 has an effective wavelength $\sim 500$\AA~redder than the DECam $z$ filter, we apply a correction to the NB magnitude of $5\log_{10}(\lambda_{\rm{BB}}/\lambda_{\rm{NB}})$. This would be the correction needed for the case of a galaxy with a flat continuum and no emission line flux. Errors on the modeled photometry are assigned based on the typical LAGER photometry errors as a function of NB/BB magnitude, where the typical error is $0.012\pm0.001$ and $0.089\pm0.008$ around 22 and 24.3 mag, respectively, in NB magnitudes.

An example of the filter profile effect is shown in Figure \ref{fig:filter_profile_corr} for mock sources that passed the LAGER selection criteria. Although we initially assumed a uniform distribution to assign redshifts, once the narrowband filter is convolved with the mock spectra and the selection criteria is applied our mock samples revert to redshift distributions tracing the narrowband filter profile as expected. Mock sources with redshifts (observed wavelengths) within the FWHM ($92$\AA; offsets up to $\pm45$\AA~in Figure \ref{fig:filter_profile_corr}) show at most a $0.1$ dex decrease in the observed \ha~luminosity. However, sources that are further towards the wings of the filter (high offsets from the central wavelength) are affected up to $0.75$ dex. The diagonal cut towards fainter intrinsic luminosity, $L_{\rm in}$, is based on the NB magnitude and $\Sigma$ cut such that sources with $L_{\rm in} < 10^{40.5}$ erg s$^{-1}$ are only detectable towards the central parts of the filter profile given the selection limits while brighter sources can be observed over the full wavelength coverage of the filter, although with significantly reduced luminosities. 

One assumption made in typical narrowband surveys, include the LAGER \ha~sample, is for top-hat narrowband and broadband filters with widths based on their respective FWHMs. This is due to the low resolution of the filter profiles ($R \sim 100$ for LAGER) such that the exact position of the emission line within the narrowband filter is not known. We therefore follow this assumption to keep consistency with a narrowband observation by using our modeled magnitudes which take into account the filter profiles (as in any observation) and measure the line fluxes and equivalent widths as:
\begin{align}
\centering
F_L & = \Delta\textrm{NB} \frac{f_{\textrm{NB}} - f_{\textrm{BB}}}{1 - (\Delta\textrm{NB}/\Delta\textrm{BB})} \nonumber \\
\textrm{EW}_\textrm{obs} &= \frac{F_L}{f_C} = \Delta\textrm{NB}\Bigg(\frac{f_\textrm{NB} - f_\textrm{BB}}{f_\textrm{BB} - f_\textrm{NB}(\Delta\textrm{NB}/\Delta\textrm{BB})} \Bigg)
\label{eqn:flux_EW}
\end{align}
with $F_L$ and $f_C$ being the emission line flux and continuum flux density, respectively, $f_\textrm{NB}$ and $f_\textrm{BB}$ are the observed narrowband and broadband flux densities, and $\Delta\textrm{NB}$ and $\Delta\textrm{BB}$ are the associated FWHMs of the two filters. We use the measurements from Equation \ref{eqn:flux_EW} to compare with the respective measurements from the LAGER survey, however we keep track of the intrinsic (modeled) properties such that measurements of \wo~published in this work are direct inputs in the model and are independent from observational effects.

\subsection{Generating Mock Galaxies}
In this section we describe how we populate our mock samples with modeled \ha~emitters. Both approaches presented below rely on an intrinsic distribution to randomly assign EW as discussed in \S\ref{sec:EW_model} and a secondary intrinsic distribution is used to assign either \ha~luminosity or rest-frame $R$-band continuum luminosity in approach 1 and 2, respectively. We compare the mock samples with the \ha~LAGER sample in bins of \ha~or continuum luminosity depending on the approach used to generate the mocks. Although we generate EW and the secondary property to be initially independent from each other, varying \wo~and $\sigma_0$ measurements between bins of \ha~and continuum luminosity would signify a correlation between EW and the secondary property is present. Therefore, each approach is designed to investigate if an intrinsic correlation between EW and \ha~luminosity (\S\ref{sec:approach1}) or continuum luminosity/stellar mass (\S\ref{sec:approach2}) exists.

\subsubsection{Approach 1: Line Luminosity}
\label{sec:approach1}
In this approach, we buildup our mock samples by assigning \ha~luminosity and EW with the aim to investigate if a correlation exists between the two observational properties. We assume the \citet{Khostovan2020} $z = 0.47$ \ha~luminosity function as the intrinsic line luminosity distribution of our mock sample.

We start by randomly assigning an \ha~luminosity from the \citet{Khostovan2020} LF and a rest-frame equivalent width from either an exponential or a normal distribution assuming a given \wo~or \so, respectively. \ha~luminosities are assigned within the range of $10^{39.6}$ and $10^{44}$ erg s$^{-1}$, where we ensure that the lower limit of our selection is slightly below the LAGER 5$\sigma$ \ha~luminosity limit of $10^{39.83}$ erg s$^{-1}$. Lowering the limit too much will result in fewer retained sources after selection criteria are applied as discussed below which would affect the underlying number statistics when matching to the LAGER EW distributions. Equivalent widths are randomly selected with a minimum limit of $0$\AA~to take into account all emission line possibilities. The rest-frame $R$-band continuum luminosity, $L_R$, centered at $6563$\AA~is then measured by the combination of the \ha~luminosity and EW, where EW$ = L_\textrm{\ha}/L_R$.

The narrowband and broadband photometry and associated photometric errors are modeled following the method we outlined in \S\ref{sec:photometry}. We then use the observed magnitudes to measure the filter-affected \ha~luminosity and EW which we use to compare with the observed associated properties in the LAGER \ha~sample; however, we keep track of the intrinsic properties for each source and the final \wo~and \so~measurements are the intrinsic characteristic EW widths. The LAGER selection criteria is then applied to our mock samples such that for a source to be considered an emission-line galaxy it must satisfy the following conditions: (1) NB magnitude must be brighter than the 5$\sigma$ NB cut of 25.45 mag, (2) the rest-frame EW (filter-affected) must be larger than 35\AA, and (3) the color significance must be $\Sigma > 3$. This last criteria tests the significance of the nebular color excess and makes use of the modeled photometric errors.

In total, we model $10^6$ sources per mock sample that are statistically designed to represent typical LAGER \ha-selected galaxies. These samples will be used to test for correlations between EW and line luminosity as will be shown in the sections below.

\subsubsection{Approach 2:  Continuum Luminosity (Stellar Mass)}
\label{sec:approach2}
Here we buildup our mock samples by assigning rest-frame $R$-band continuum luminosity centered at $6563$\AA~and EWs. As a proxy for the continuum luminosity distribution, we assume a stellar mass function and a mass-to-light ratio constrained using LAGER $R$-band luminosity and COSMOS2015 stellar mass measurements \citep{Laigle2016} to populate our samples.

We assume the HiZELS $z = 0.40$ \ha~stellar mass function of \citet{Sobral2014} as our intrinsic stellar mass distribution. The choice of this stellar mass function was based on two main factors. First, HiZELS is a narrowband survey similar to LAGER in that it covers a thin redshift slice ($\Delta z = 0.01$; same as LAGER) that best represents LAGER \ha~sources given that typical continuum-selected stellar mass functions cover a wider redshift window per measurement. Second, HiZELS is \ha-selected, such that the star-forming population used in constraining the stellar mass function is similar to LAGER. Continuum-selected stellar mass functions typically subdivide samples into star-forming and passive populations by using a color-color diagnostic (e.g., $UVJ$; \citealt{Ilbert2013,Muzzin2013,Tomczak2014,Davidzon2017}). Although this technique has been used to extensively study the stellar mass function at different cosmic epochs, it comes with the caveat that photometric scatter and poorly constrained photometric redshifts can introduce contaminants into the sample. However, the HiZELS SMF directly selects galaxies based on the \ha~line, which is a known tracer for star-formation activity and, therefore, negates the need for a color-color diagnostic to select star-forming galaxies.

\begin{figure}
	\centering
	\includegraphics[width=\columnwidth]{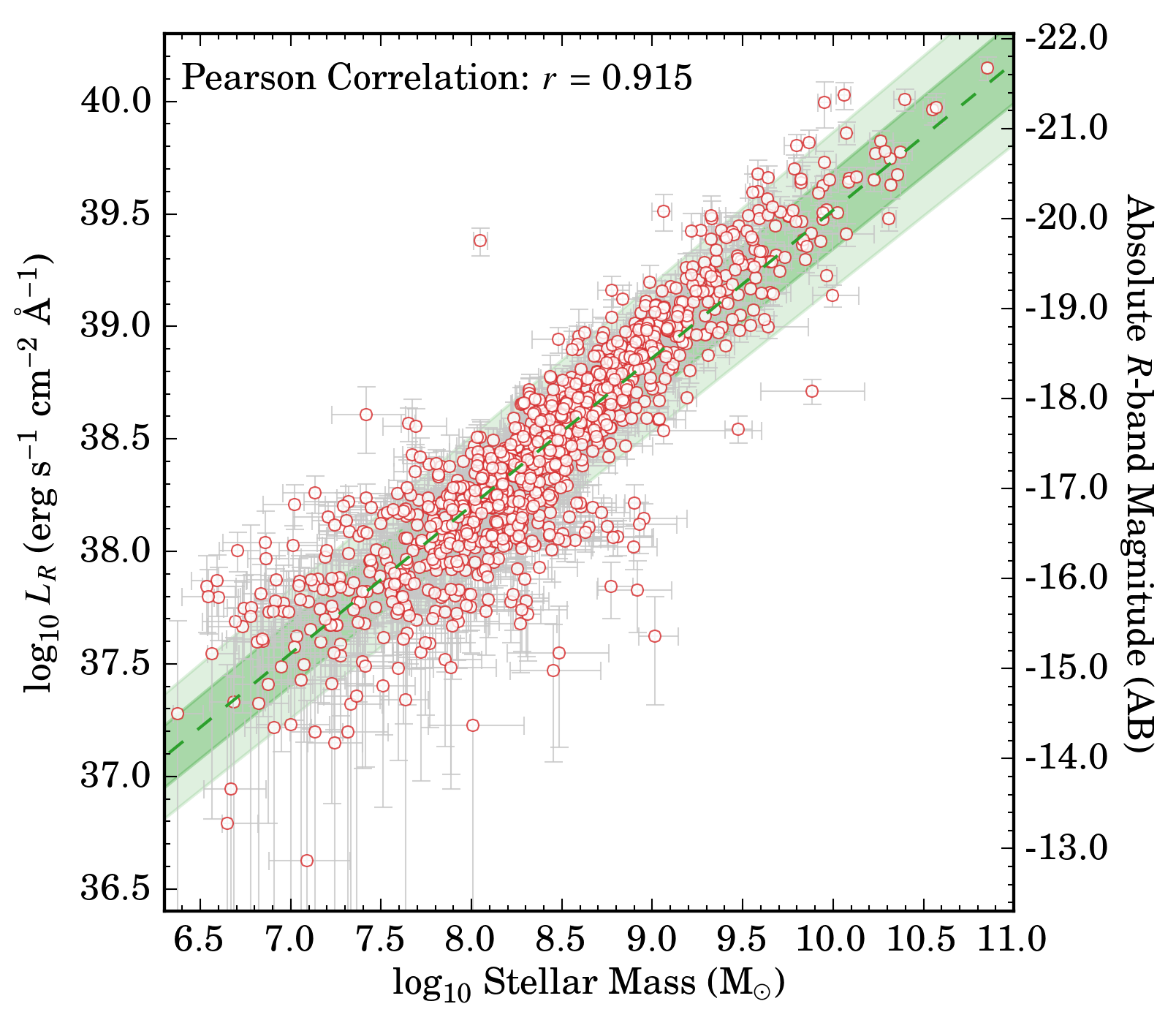}
	\caption{Comparison between stellar mass and rest-frame $R$-band luminosity. The stellar mass measurements are from \citet{Laigle2016} and the $R$-band luminosity is measured using the narrowband and broadband photometry from LAGER and de-redshifting to rest-frame. The \ha~sample follows a linear correlation between the continuum luminosity and stellar mass which we use in our Approach 2 (\S\ref{sec:approach2}). We note the scatter in stellar mass increases for $L_R < 10^{38.3}$ erg s$^{-1}$ \AA$^{-1}$ which corresponds to stellar masses $< 10^9$ \msol. These would be the dwarf-like population of \ha~emitters where their stellar masses are sensitive to the faint continuum used in SED fitting. Overlaid as a {\it green dashed} line is our empirical fit with the $1\sigma$ and $2\sigma$ levels shown as {\it dark} and {\it light green} shaded regions, respectively.} 
	\label{fig:ML_ratio}
\end{figure}

For a given mock source, we randomly select from the assumed stellar mass functions in a range between $10^{6.0}$ to $10^{12}$ \msol. Our lower limit is set slightly lower than the minimum stellar mass of the LAGER \ha~sample. Given that \ha~is redder than the 4000\AA~break, the continuum luminosity centered at 6563\AA~is a reliable proxy for stellar mass as it would trace the majority of the stellar population (e.g., low-mass, old stars). We convert our randomly selected stellar masses as described above to the corresponding $R$-band luminosities for our mock sample modeling. Figure \ref{fig:ML_ratio} shows the mass-to-light ratio of the observed LAGER \ha~emitters with stellar masses from the COSMOS2015 catalog. We find a tight correlation between the two properties with a Pearson correlation coefficient $r = 0.915$ and a best fit of:
\begin{eqnarray}
L_R = \Bigg(\frac{10^{32.96\pm0.08}}{\textrm{erg s}^{-1} \textrm{\AA}^{-1}}\Bigg) \Bigg(\frac{M}{M_\odot}\Bigg)^{0.656\pm0.009}
\label{eqn:ML_ratio}
\end{eqnarray}
where $L_R$ is $R$-band continuum luminosity centered at 6563\AA~and $M$ is stellar mass. For each mock source, we assign their continuum luminosity by using their randomly selected stellar mass and the correlation shown in Equation \ref{eqn:ML_ratio}. We also augment the correlation for each source based on the $1\sigma$ errors in order to take the scatter into account such that the mass-to-light ratio of our mock samples best represents the observed trend shown in Figure \ref{fig:ML_ratio}. Taking into account the scatter also incorporates variations in the mass-to-light ratio introduced from varying star-formation histories and IMFs, especially towards lower stellar masses.

The EW of each mock source is randomly selected as discussed in \S\ref{sec:EW_model}. The \ha~luminosity per source is then measured by using the combination of continuum luminosity and equivalent width ($L_\textrm{\ha} = $ EW$ \times L_R$). We then follow the same methodology as in \S \ref{sec:approach1} in modeling the narrowband and broadband photometry, propagating filter profile effects, and applying the LAGER selection criteria.

Our mock samples from this approach comprise a total of $10^6$ sources that are designed to represent LAGER \ha~emitters which will be used to test for an EW -- stellar mass correlation, as suggested by various studies (e.g., \citealt{Fumagalli2012,Sobral2014,Khostovan2016,Reddy2018}).

\subsection{Fitting the Mocks to Observations}

Each mock sample that we create is statistically designed to represent our observed \ha~sample where we have control over the intrinsic distributions set by \wo~or \so~and also how selection changes the underlying EW distribution. The mock samples are compared to the observations by finding the best-fit intrinsic \wo~or \so~that matches the LAGER \ha~equivalent distributions after selection criteria are applied to the mocks.

We fit the mocks to our observations following a maximum likelihood estimation (MLE) approach. We start by binning both the observations and mocks in equally sized EW bins and assume Poisson errors for both, where the mock samples have their EW distributions set by the intrinsic \wo~or \so. Both histograms are normalized to unity for the fitting procedure. We measure the likelihood of each mock sample as:
\begin{eqnarray}
\mathcal{L}(\textrm{\wo}|obs) \propto e^{-\chi^2/2}\\
\chi^2 = \sum_{i=1}^n \Bigg(\frac{N^{mock}_i - N^{obs}_i}{\sigma_i}\Bigg)^2
\end{eqnarray}
with $\mathcal{L}$ being the likelihood of our mock sample representing the observations with an EW distribution defined by \wo~or \so. The $\chi^2$ is measured by comparing the normalized number of mock, $N^{mock}_i$, and observed, $N^{obs}_i$, sources in each $i^{th}$ EW bin with $\sigma_i$ being the Poisson error associated with the observed number of sources in each bin.

To test whether a correlation exists between EW and galaxy properties, we subdivide the observations and the mocks in subsamples based on the respective galaxy properties (\ha~luminosity/stellar mass). For each subsample, we measure the underlying histograms and follow the same fitting procedure. As described in \S\ref{sec:approach1} and \S\ref{sec:approach2}, we initially randomly assign EWs and galaxy properties independent from one another. By subdividing them in bins of \ha~luminosity/stellar mass and measuring their respective \wo~or \so~using the fitting methodology described above, we can test whether the assumption that \ha~luminosity/stellar mass is independent of EW holds true or not. We can also test whether an exponential or normal distribution set by \wo~and \so, respectively, best represents the observed EW distributions.

\section{Results}
\label{sec:results}

\begin{figure}
	\includegraphics[width=\columnwidth]{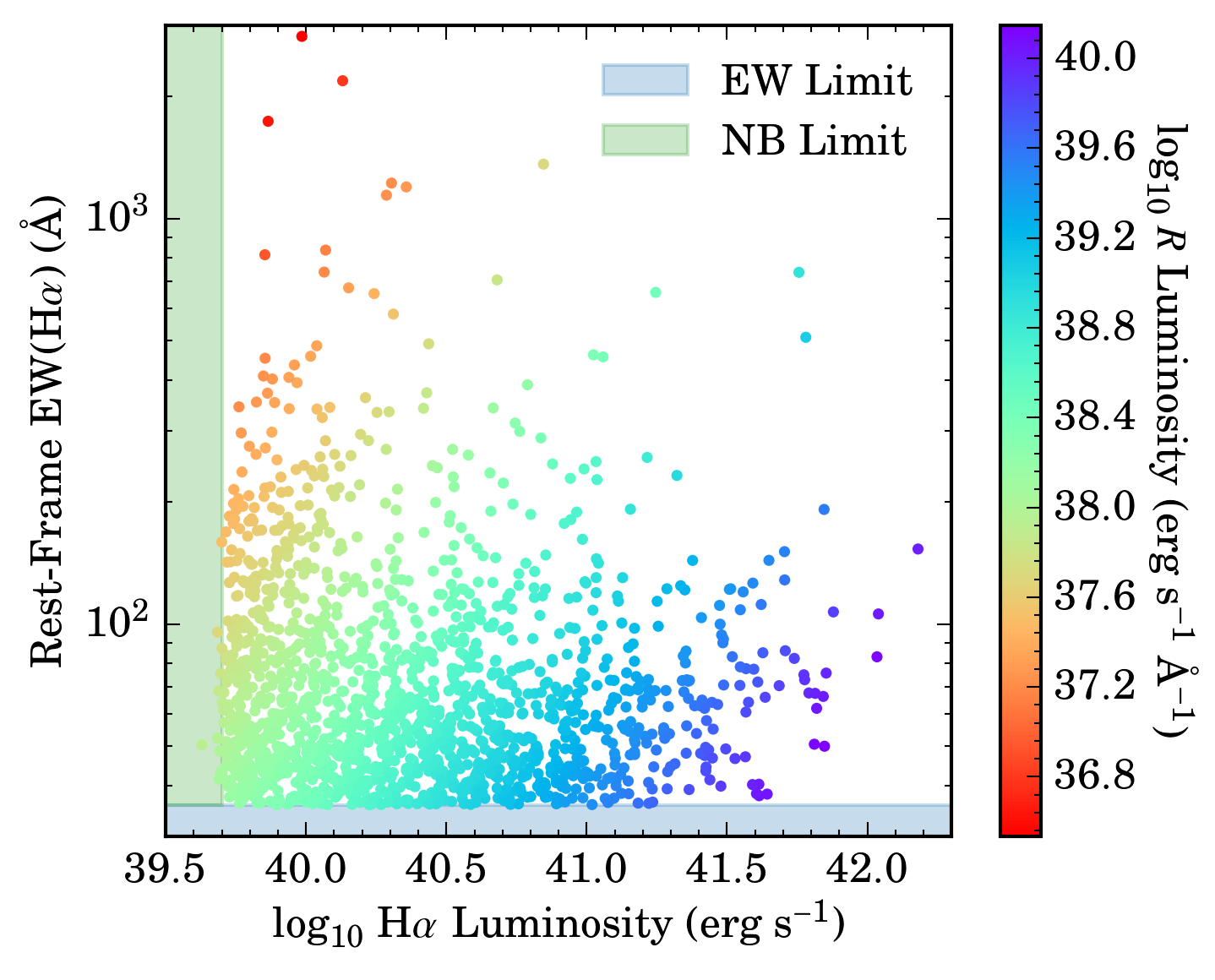}
	\caption{Comparison between rest-frame EW and \ha~luminosity of the $z = 0.47$ LAGER \ha~sample with the rest-frame $R$-band luminosity color coded per source. The sample has the \ha~luminosity bounded by the $5\sigma$ narrowband magnitude limit, shown as the {\it green} highlighted region, and the EW bounded by the $>35$\AA~limit, shown as the {\it blue} highlighted region. No clear, distinguishable trend is seen. We find several EW$> 10^3$\AA~\ha~emitters mostly at faint \ha~luminosities with significantly faint continuum. Such sources are interesting on their own as potential cases of bursty star-formation activity.}
	\label{fig:EW_Line}
\end{figure}

\subsection{Equivalent Width and Line Luminosity}
\label{sec:ew_line}

\begin{figure*}
	\centering
	\includegraphics[width=\textwidth]{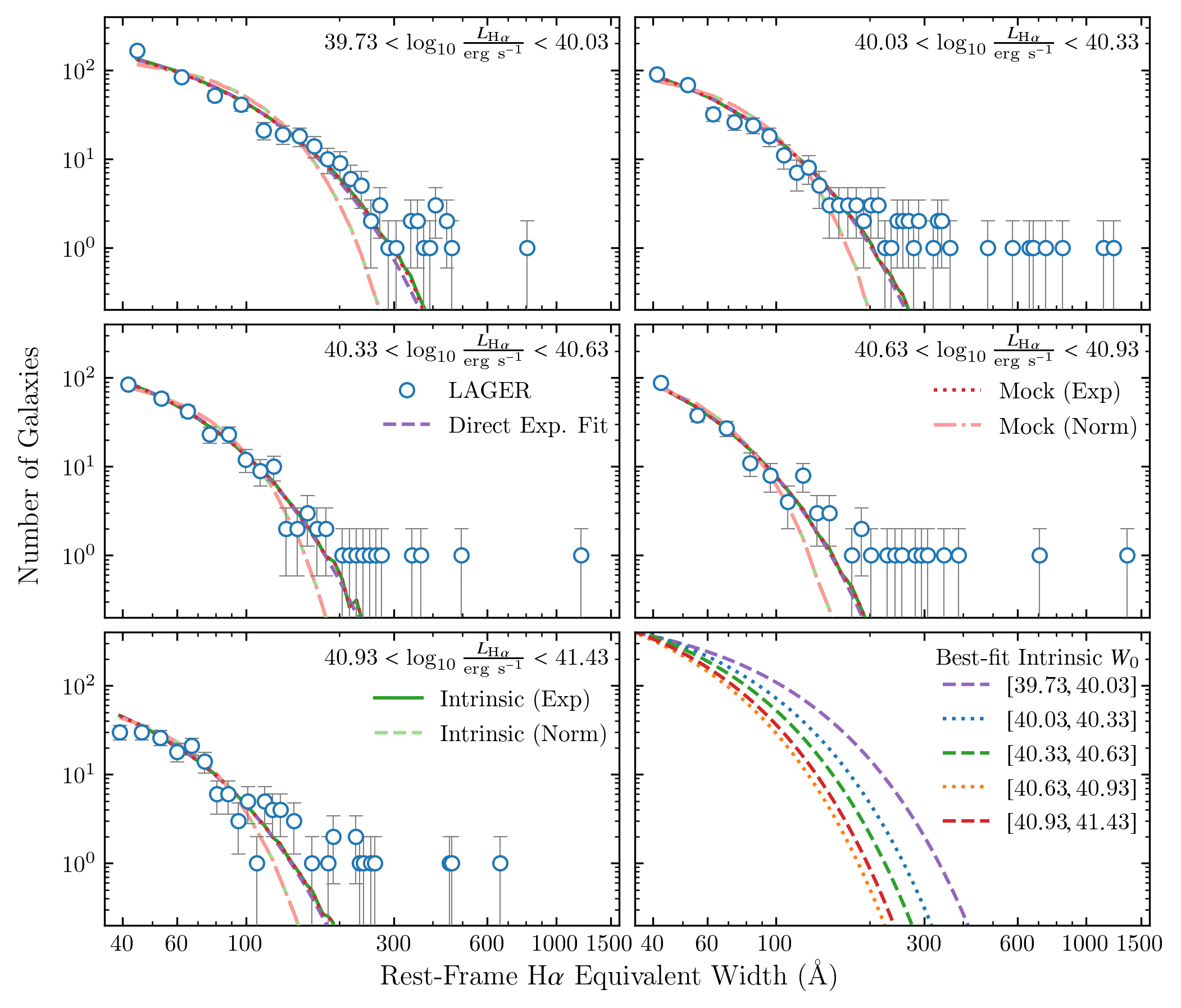}
	\caption{The EW distributions in subdivisions of \ha~luminosity. The distribution of LAGER \ha~emitters is shown as {\it open blue} circles. The direct exponential fit, shown as a {\it dashed purple} line, corresponds to ignoring selection bias corrections. The intrinsic EW distributions are shown assuming an exponential distribution and normal distribution as {\it solid green} and {\it dashed light green} lines, respectively. Each intrinsic distribution corresponds to the best-fit \wo~and \so~, respectively, assumed in Approach 1 (see \S\ref{sec:approach1}) that produces a mock sample that best represents the observations and are shown as {\it dotted red} and {\it dash-dotted light red} lines, respectively. Selection effects do not seem to affect the underlying measurements, although this is because the selection limits are uniform per given \ha~luminosity (see Figure \ref{fig:EW_Line}). The normal distribution is found to fail at high EWs where an exponential distribution is favored by observations. The {\it bottom right} panel shows the intrinsic EW probability distribution, defined by Equation \ref{eqn:exp_model} and the normalization rescaled to $\sim 10^4$, where faint emitters are 30 times more likely to have EW$_0 > 200$\AA~compared to the brightest emitters.}
	\label{fig:EWdistrib_line}
\end{figure*}

We first explore how EW correlates with \ha~luminosity using the approach highlighted in \S\ref{sec:approach1}. Figure \ref{fig:EW_Line} shows the distribution of EW and \ha~luminosity with their rest-frame $R$-band luminosity color-coded. No clear trend is seen between the two physical properties. For L$_\textrm{\ha} < 10^{41}$ erg s$^{-1}$, we find 7 sources have rest-frame EW $>10^3$\AA, while the brighter population does not show any such sources. Figure \ref{fig:EW_Line} also shows the selection limits of the LAGER survey which shows a missing population of EW$ < 35$\AA~at all \ha~luminosities observed. 

We test for a correlation between EW and \ha~luminosity by subdividing both the LAGER and mock samples (based on the approach outlined in \S\ref{sec:approach1}) in \ha~luminosity bins of 0.3 dex in width and measuring the characteristic EW that best represents the sample. The brightest \ha~luminosity bin is set to a width 0.5 dex in order to increase the number statistics given that the bright-end of any galaxy sample will be the least populated.

The EW distributions are shown in Figure \ref{fig:EWdistrib_line} where we show the mock simulations for the case of an exponential and a normal distribution as a {\it dotted red} and {\it dashed-dotted light red} line, respectively. The intrinsic exponential and normal EW distributions used to generate the mocks samples are shown as a {\it solid green} and {\it dashed light green} line, respectively. We also show a direct exponential fit using Equation \ref{eqn:exp_model} as a {\it dashed purple} line. This represents a measurement where we directly fit the observations and ignore any selection corrects. We refer to this also as the `selection-biased' case throughout the paper.

\begin{figure}
	\centering
	\includegraphics[width=\columnwidth]{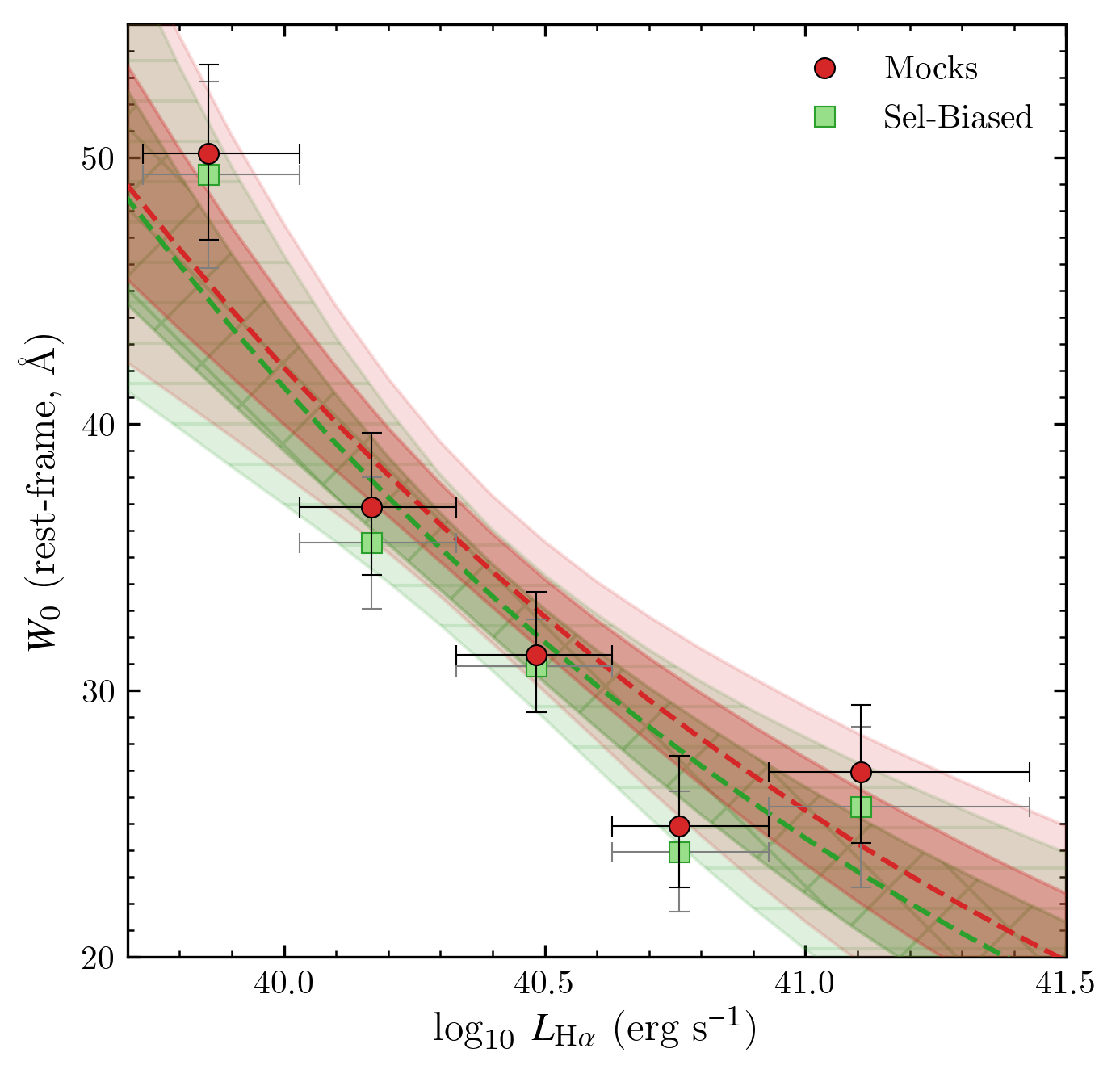}
	\caption{The \wo~-- \ha~luminosity correlation. The selection-biased \wo~are shown in {\it green squares} and are based on the direct exponential fits from Figure \ref{fig:EWdistrib_line}. Our mock simulation \wo~measurements are shown as {\it red circles}. Dark and light shaded regions correspond to the $1$ and $2\sigma$ confidence regions of our power-law fit, respectively. No discernible difference is seen between the two cases where power law fits show a slope of $-0.22^{+0.04}_{-0.05}$ and $-0.23\pm0.05$ based on the mock simulation and selection biased \wo, respectively. This is somewhat expected given that the selection cuts are uniform within the EW -- \ha~luminosity plane as shown in Figure \ref{fig:EW_Line}.}
	\label{fig:w0_line}
\end{figure}

Figure \ref{fig:EWdistrib_line} shows no statistically significant difference between directly fitting the data and using our modeled EW approach such that selection effects does not seem to affect the observed EW distributions. This is not surprising given how the selection limits are folded within the EW -- $L_\textrm{\ha}$ plane shown in Figure \ref{fig:EW_Line}. The main source of incompleteness in each subsample arises from the common $35$\AA~EW cut such that all subsamples are equally affected by a uniform selection criteria. 

The fits in Figure \ref{fig:EWdistrib_line} show that an exponential distribution best represents the observations. Our samples assuming a normal distribution matches the observations up to $\sim 150$\AA~for the faintest $L_\textrm{\ha}$ bin and $\sim 100$\AA~at the brightest bin, while assuming an exponential distribution pushes the limit to $\sim 300$\AA~and $\sim 150$\AA, respectively. Table \ref{table:w0_props} shows the reduced $\chi^2$ for both the exponential and normal EW distribution cases. For the case of assuming an intrinsic normal distribution, we find $\chi^2_{red} = 3.31$ and $1.38$ at $L_{\textrm{\ha}} = 10^{39.73 - 40.03}$ erg s$^{-1}$ and $10^{40.93 - 41.43}$ erg s$^{-1}$, respectively, while assuming an intrinsic exponential distribution results in  $\chi^2_{red} = 1.38$ and $1.20$, respectively. Each of our $L_{\textrm{\ha}}$ subsamples show a higher $\chi^2_{red}$ when using a normal distribution. We therefore rule out the normal distribution for the rest of our analysis and assume an exponential EW distribution for all subsequent measurements. The higher EW sources that are missed by an exponential profile are further explored in \S\ref{sec:high_EW} where we find they constitute a small fraction of the total population of \ha~emitters. These are also sources that have faint continuum and, therefore, have larger uncertainties.

Figure \ref{fig:w0_line} and Table \ref{table:w0_props} shows \wo~for each of the \ha~luminosity bins with the selection-biased case, shown as {\it green squares}, and the case where we use our mock simulations assuming an intrinsic exponential EW distribution, shown as {\it red circles}. The selection-biased \wo~are based on the direct exponential fits shown in Figure \ref{fig:EWdistrib_line} and ignore any selection effect corrections. We find no statistically significant difference between the two cases as was also seen in the individual EW distributions. This is due to how selection limits are uniform within the EW -- $L_\textrm{\ha}$ plane, such that each \ha~luminosity bin is affected equally by selection.

A clear correlation between \wo~and \ha~luminosity is observed in Figure \ref{fig:w0_line} where faint \ha~emitters are seen to exhibit an EW distribution skewed towards higher EWs. The intrinsic \wo~is found to increase from $26.93^{+2.51}_{-2.68}$\AA~at $L_\textrm{\ha} = 10^{40.93 - 41.43}$ erg s$^{-1}$ to $50.15^{+3.32}_{-3.24}$\AA~at $L_\textrm{\ha} = 10^{39.73 - 40.03}$ erg s$^{-1}$, which shows that the EW distribution widths increase by almost a factor of two by our faintest \ha~luminosity bin. The increase corresponds to $\textrm{\wo} \sim L_{\textrm{\ha}}^{-0.22\pm0.05}$ and $L_{\textrm{\ha}}^{-0.23\pm0.05}$ for the case of using the intrinsic \wo~and the selection-biased measurements, respectively, as shown in Table \ref{table:w0_model}. The {\it bottom right} panel of Figure \ref{fig:EWdistrib_line} shows the intrinsic EW distribution where we find that faint \ha~emitters are $\sim 30$ times more likely to have rest-frame EW $> 200$\AA~compared to bright \ha~emitters. Interpreting the EW as a proxy for sSFR would suggest that faint \ha~emitters cover a diverse population of star-forming galaxies in regards to their star-formation histories.

\subsection{Equivalent Width and Continuum Luminosity}
\label{sec:ew_cont}

\begin{figure}
	\includegraphics[width=\columnwidth]{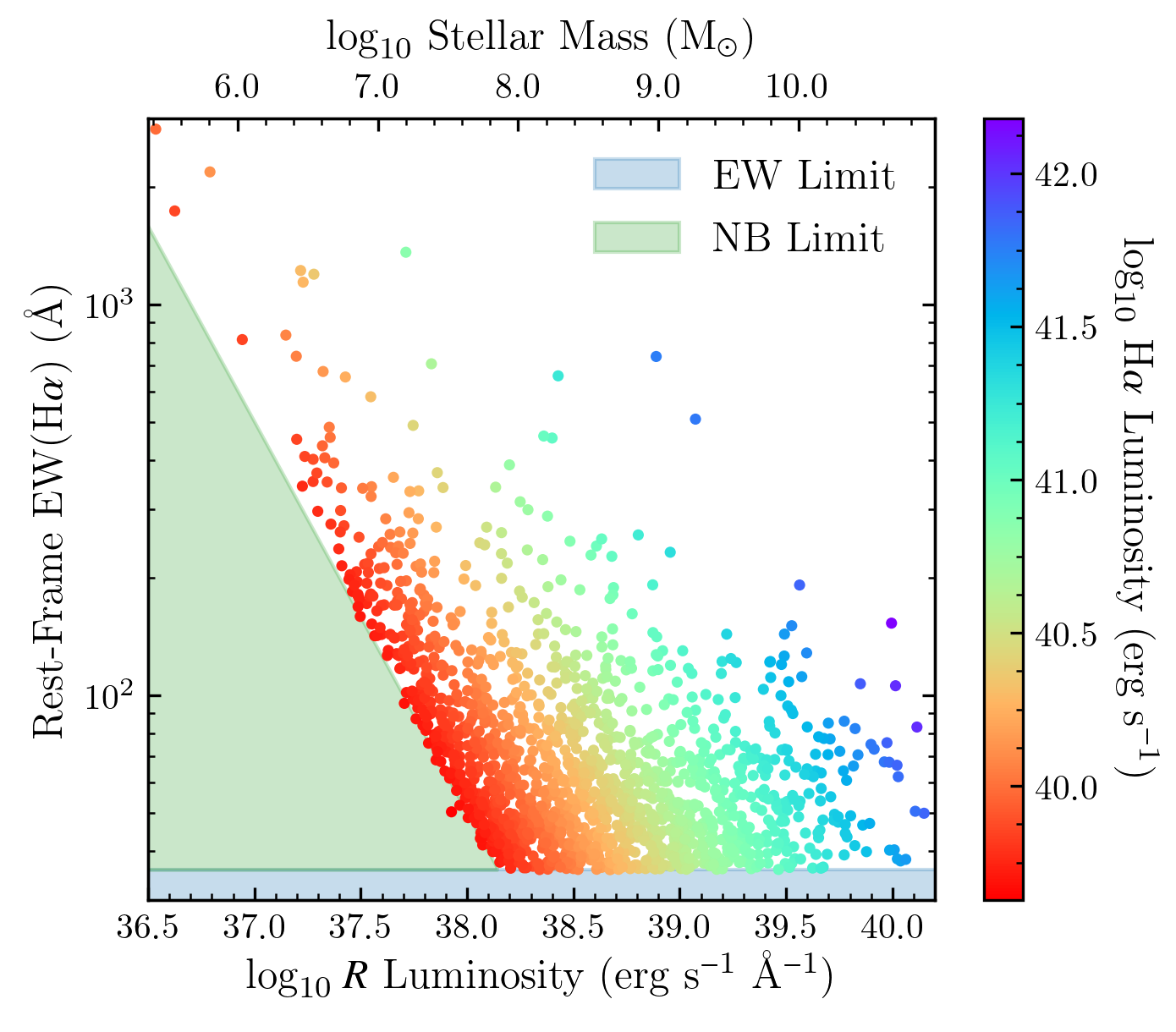}
	\caption{The comparison between rest-frame EW and rest-frame $R$-band luminosity with the \ha~luminosity of each source color coded. The selection limits are highlighted as well. The EW limit used in LAGER forms a uniform cut per given continuum luminosity, while the NB limit causes a non-uniform EW cut that becomes significant at luminosities $\lesssim 10^{38.1 - 38.2}$ erg s$^{-1}$ \AA$^{-1}$. In turn, it raises the concern of how selection effects can bias \wo~measurements at low continuum luminosities. The LAGER \ha~emitters show a wide range of EW ranging from $35$\AA~to$\sim 3000$\AA. Faint continuum sources tend to have higher EWs compared to bright continuum \ha~emitters such that no \ha~emitters are detected with EW$>200$\AA~for $L_R > 10^{39.2}$ erg s$^{-1}$ \AA$^{-1}$. This would suggest some EW -- stellar mass correlation, but it becomes crucial that selection effects are taken into account prior to making such measurements.}
	\label{fig:EW_cont}
\end{figure}

\begin{figure*}
	\centering
	\includegraphics[width=\textwidth]{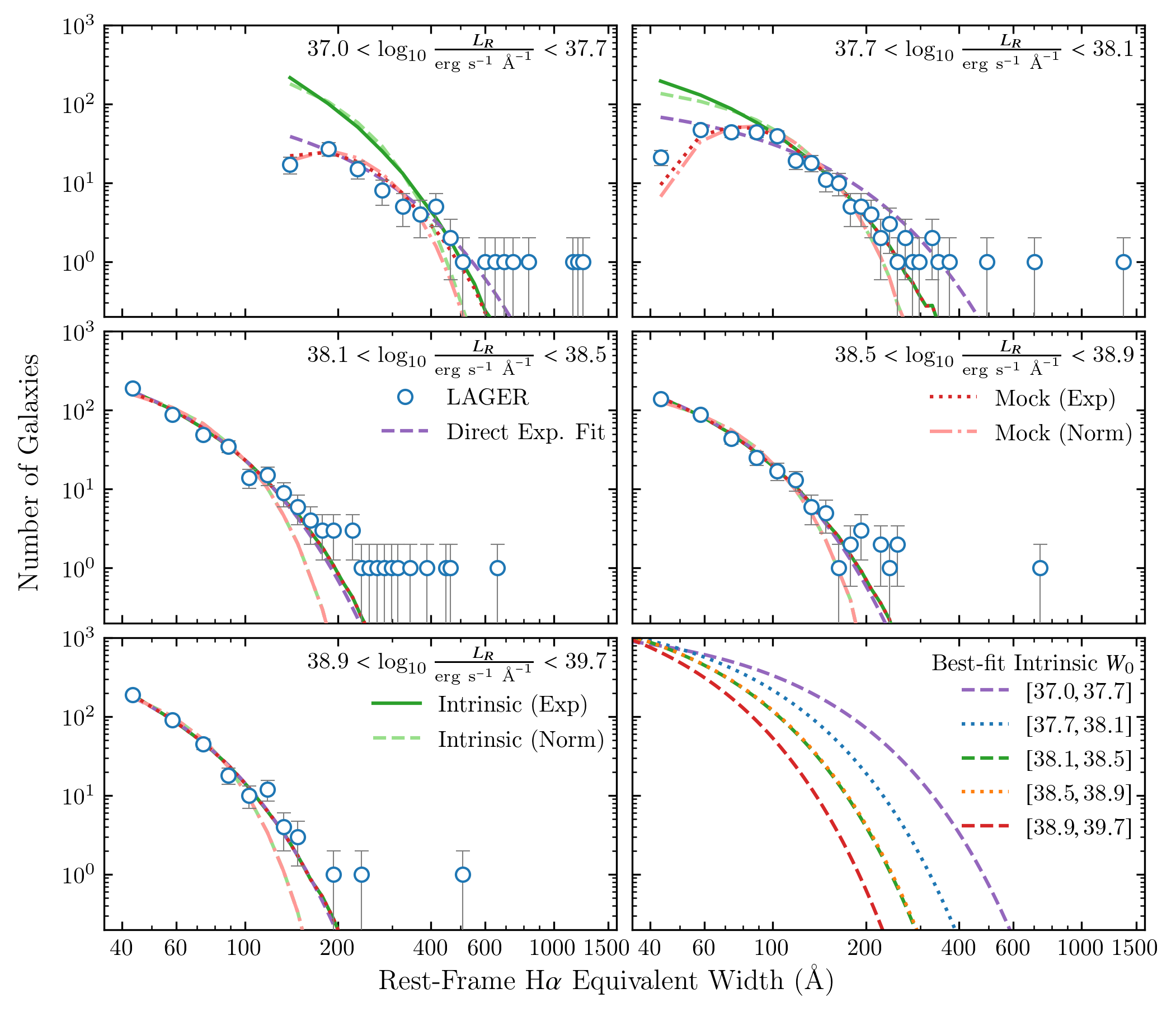}
	\caption{The EW distributions subdivided in rest-frame $R$-band luminosity bins. The color and axis labeling is the same as defined in Figure \ref{fig:EWdistrib_line}. The top two panels are the faintest continuum bins and are also the two subsamples that show evidence for incompleteness at lower EWs when comparing the intrinsic distribution to the observations. This is due to the nonuniform EW cut arising from the narrowband selection limit at $L_R < 10^{38.10}$ erg s$^{-1}$ \AA$^{-1}$. The direct exponential fit overestimates the width of the distribution as it tries to capture the incompleteness at lower EWs, while the mock simulations trace the incompleteness, especially the turnover in the the faintest bin ({\it top left panel}) at EW$< 200$\AA. The comparison between the mock simulation and its associated intrinsic distribution highlights the affect selection can have on measuring the shape of EW distributions. The {\it bottom right panel} shows the intrinsic EW probability distribution, with the normalization rescaled to 10$^5$, where faint continuum, low-mass emitters are $\sim 320$ times more likely to have EW$_0 > 200$\AA~compared to bright continuum, high-mass emitters.}
	\label{fig:EWdistrib_cont}
\end{figure*}

\begin{figure}
	\centering
	\includegraphics[width=\columnwidth]{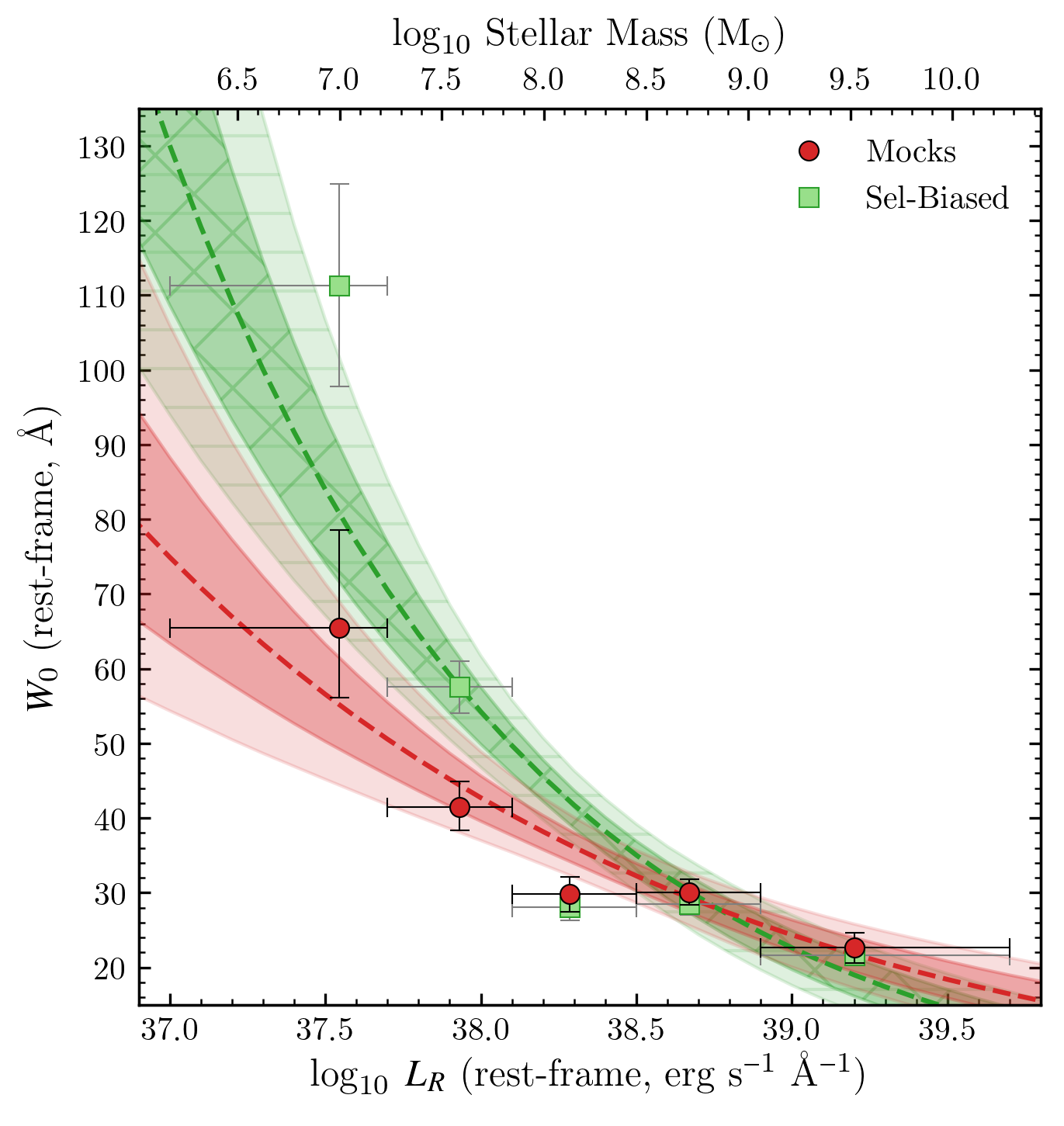}
	\caption{The \wo~-- stellar mass correlation. The selection-biased correlation is shown in {\it green} and is based on directly fitting an exponential distribution to the EW distributions. The mocks are shown in {\it red} and take into account selection effects such that the measurements represent the intrinsic \wo~and how it scales with stellar mass. The two methods show similar \wo~at $L_R > 10^{38.5}$, although this is expected since the selection limits within this continuum range is the common rest-frame EW limit of 35\AA. The disagreement between the two methods at fainter continuum are due to selection effects caused by the NB magnitude limit causing an increasing EW cut with decreasing continuum luminosity.}
	\label{fig:w0_cont}
\end{figure}

Past studies have observed a correlation between EW and stellar mass for \ha~(e.g., \citealt{Fumagalli2012,Sobral2014,Faisst2016,Marmol2016,Rasappu2016,Reddy2018}), \oiii~and \oii~(e.g., \citealt{Khostovan2016,Reddy2018}), and \lya~emitters (e.g., \citealt{Oyarzun2016,Oyarzun2017,Santos2020}). Given that the EW is also an observation proxy for the sSFR (e.g., \citealt{Fumagalli2012}), with the latter dependent directly on stellar mass, it is not surprising that such a correlation could exist between EW and stellar mass (or other continuum-related properties). The question that arises is how much does selection biases contribute to shaping this correlation?

Figure \ref{fig:EW_cont} shows the distribution of our $z = 0.47$ \ha~emitters in the EW -- $L_R$ plane with each source color-coded by their \ha~luminosity. The top axis shows the corresponding stellar mass when using Equation \ref{eqn:ML_ratio}, although we note that this does not take into account the scatter of the $M/L_R$ ratio and should be treated as a rough estimate of the stellar mass. A correlation between EW and continuum luminosity is present although we can see how it is shaped at the faint end by the $5\sigma$ NB magnitude limit (\ha~flux limit). On the other hand, we also see high EW emitters are more present at fainter continuum and lower stellar mass. Only 8/92 \ha~emitters with continuum luminosity $>10^{39.5}$ \cgs~($>10^{10}$ \msol) have an EW$> 100$\AA~ with a maximum EW $\sim 200$\AA~compared to 115/981 with $L_R \sim 10^{38.2 - 39.5}$ \cgs~($10^{7.9 - 10}$ \msol) and a maximum EW $\sim 730$\AA.  This shift to higher EW becomes even more evident at $L_R < 10^{38.1 - 38.2}$ \cgs~where 7 \ha~emitters have an EW in excess of 1000\AA. However, the $5\sigma$ NB limit causes an increasing EW limit with decreasing $L_R$ and stellar mass. 

Previous EW studies have focused on measuring the typical EW for a given stellar mass bin (e.g., \citealt{Fumagalli2012,Sobral2014,Khostovan2016,Reddy2018}), however each measurement would, in principle, be affected by selection limits. For example, a narrowband survey may measure mean/median EWs that are elevated due to the uniform EW selection limit. Surveys with constant/varying line flux limits would also have elevated typical EWs at lower stellar masses due to missing low EW systems in increasing numbers with decreasing continuum luminosity/stellar mass. We can see evidence of this in Figure \ref{fig:EW_cont}. For example, the median EW at $L_R \sim 10^{37.0 - 37.5}$ \cgs~is $339.5\pm186.3$\AA~compared to $98.5\pm69.8$\AA~at $L_R \sim 10^{37.5 - 38.1}$ \cgs, although the faint continuum bin is skewed to higher EW given that \ha~emitters with EW$< 200 - 500$\AA~are missing due to the line flux ($5\sigma$ NB magnitude) selection limit. On the bright continuum end, the median EW asymptotes given the lack of high EW sources with increasing $L_R$ and stellar mass and is shaped by the uniform 35\AA~EW selection limit. It is important then to address to what extent the EW -- stellar mass correlation is shaped by selection biases. Furthermore, it also raises the question of how representative high EW emitters are of the typical population of \ha~emitters at low stellar mass/faint continuum luminosities. 

To address this issue, we subdivide our sample in bins of rest-frame $R$-band luminosity as highlighted in Table \ref{table:w0_props} where we split the sample based on the intersection of the EW and NB limits at $\sim 10^{38.1}$ erg s$^{-1}$ \AA$^{-1}$. The EW distributions of each subsample is shown in Figure \ref{fig:EWdistrib_cont}, where we show the results of our simulations in the case of an intrinsic exponential distribution ({\it solid green} line) and an intrinsic normal distribution ({\it dashed light green} line) with the best-fit selection biased mock sample distribution shown as a {\it dotted red} and {\it dashed dotted light red} line, respectively. A direct exponential fit is also shown as a {\it dashed purple} line where we fit Equation \ref{eqn:exp_model} directly to the observations. 

We find that both the normal and exponential distributions strongly agree with the observations up to $\sim 300$\AA~for our $L_R \sim 10^{37.0 - 37.7}$ \cgs~bin and $\sim 100$\AA~for our $L_R \sim 10^{38.9 - 39.7}$ \cgs~bin. However, the normal distribution fails to represent the observations at higher EWs, while the exponential profile does well up to $\sim 600$\AA~and $\sim 200$\AA~for the same bins, respectively. Table \ref{table:w0_props} shows the reduced $\chi^2$ for both the exponential and the normal distributions. For every $L_R$ subsample, we find that the exponential distribution shows a lower $\chi^2$ in comparison to the normal distribution. We conclude that an exponential distribution best represents the LAGER observations and we, therefore, resort the rest of our analysis to the mock samples created using an assumed intrinsic exponential EW distribution.

The top panel of Figure \ref{fig:EWdistrib_cont} corresponds to our two faintest continuum luminosity bins and shows the incompleteness introduced by the NB magnitude limit. The LAGER EW distribution shows a complete turnover around $200$\AA~at $L_R \sim 10^{37.0 - 37.7}$ \cgs~and at $< 100$\AA~for our $10^{37.7 - 38.1}$ \cgs~sample. Directly fitting an exponential model to the observations without placing any constraints on the range of EWs used in the fit results in a distribution that is skewed towards higher EWs while trying to capture the flatter distribution towards low EW. On the other hand, our mock simulations nicely show the turnover in the faintest $L_R$ bin although it underestimates the number of \ha~emitters for the lowest EW bin in our $L_R \sim 10^{37.7 - 38.1}$ \cgs~sample. The intrinsic distributions that are overlaid in Figure \ref{fig:EWdistrib_cont} show how the NB limit causes the lack of low EW sources with increasing degree towards lower EWs. In our $L_R \sim 10^{37.0 - 37.7}$ \cgs~sample, we find that the observations are complete for selection down to $\sim 400$\AA~and as we go towards lower EWs the separation between the intrinsic and mock distributions increases due to the LAGER selection limit. The significant difference between the direct exponential fit and intrinsic EW distribution highlights the importance of taking selection into account when investigating EW properties of star-forming galaxies.

The two brightest continuum luminosity samples are shown in the bottom panel of Figure \ref{fig:EWdistrib_cont}. We find no difference between the direct exponential fit and mock simulations for the two bins, although this is expected as the selection limit affecting the $> 10^{38.1}$ \cgs~\ha~emitters is the common EW cut. This is similar to what we saw for every one of the \ha~luminosity subsamples in \S\ref{sec:ew_line}.

Figure \ref{fig:w0_cont} shows the best-fit \wo~for each of our $L_R$ samples with the stellar mass shown corresponding to our $M/L_R$ defined in Equation \ref{eqn:ML_ratio}. We find that the selection-biased and mock simulations have consistent \wo~for our $L_R > 10^{38.1}$ \cgs ($>10^{7.8}$ \msol) samples, which is expected given that the EW limit is the dominant selection effect and is uniform within the EW -- $L_R$ plane. At $L_R < 10^{38.1}$ erg s$^{-1}$ \AA$^{-1}$, we find a significant difference between the two approaches where the selection biased \wo~for our faintest bin is $111.30\pm13.56$\AA~in comparison to the mock simulations \wo~of $65.44^{+13.16}_{-9.28}$\AA. As discussed above, the source of this discrepancy is the $5\sigma$ NB magnitude selection limit which causes the lack of low EW with decreasing $L_R$ and stellar mass as seen by the turnover in the {\it top left} panel of Figure \ref{fig:EWdistrib_cont}. The direct exponential fit (selection-biased \wo) compensates for this turnover by increasing the best-fit \wo~which causes for a wider EW distribution (higher \wo). Our mock simulations take selection biases into account and favors an intrinsic \wo~lower than what would be measured if one simply fits an exponential profile to the observations.

\begin{table*}
	\centering
	\caption{Measurements of the equivalent width distributions assuming an exponential and a normal distribution in subsamples of \ha~luminosity and $R$-band luminosity. For the cases of subdivision in \ha~luminosity and $R$-band luminosity, we use Approach 1 and 2 as described in \S \ref{sec:modeling}, respectively, with $N_{gal}$ being the total number of observed \ha~sources per subsample. $W^{\textrm{direct}}_0$ is based on directly fitting an exponential profile to the LAGER EW distributions and ignoring selection effect corrections. The best-fit intrinsic \wo~and \so~are shown assuming an exponential and normal EW distribution constrained by the LAGER EW distributions. Included is the reduced $\chi^2$ for both the exponential and normal distributions, which shows that observations favor an exponential EW distribution model.}
	\def\arraystretch{1.3}%
	\begin{tabular*}{\textwidth}{@{\extracolsep{\fill}} l c c c c c c c}
		\hline
		Sample & Bin & $N_{gal}$ & $W^{\textrm{direct}}_0$ & $W_0$ & $\sigma_0$ & $\chi^2_{red}$ & $\chi^2_{red}$ \\
		&	&	& (\AA)	& (\AA)	& (\AA) & (exp) & (norm) \\
		\hline
		$\log_{10}$ \ha~Luminosity & $39.86^{+0.17}_{-0.13}$ & 466 & $49.34^{+3.49}_{-3.49}$ & $50.15^{+3.32}_{-3.24}$ & $71.34^{+6.17}_{-3.93}$ & 1.38 & 3.31\\
		& $40.17^{+0.16}_{-0.14}$ & 335 & $35.53^{+2.47}_{-2.47}$ & $36.86^{+2.81}_{-2.54}$ & $55.07^{+3.03}_{-2.63}$ & 1.11 & 1.84\\
		& $40.48^{+0.15}_{-0.15}$ & 285 & $30.91^{+1.74}_{-1.74}$ & $31.32^{+2.36}_{-2.15}$ & $50.03^{+2.39}_{-2.40}$ & 0.67 & 1.02\\
		& $40.76^{+0.17}_{-0.13}$ & 204 & $23.94^{+2.26}_{-2.26}$ & $24.91^{+2.62}_{-2.30}$ & $40.28^{+2.73}_{-2.61}$ & 1.07 & 1.56\\
		& $41.11^{+0.32}_{-0.18}$ & 189 & $25.62^{+3.02}_{-3.02}$ & $26.93^{+2.51}_{-2.68}$ & $42.17^{+2.67}_{-2.52}$ & 1.20 & 1.38\\
		\hline
		$\log_{10}$ $R$-band Luminosity & $37.54^{+0.16}_{-0.54}$ & 92 & $111.30^{+13.56}_{-13.56}$ & $65.44^{+13.16}_{-9.28}$ & $126.73^{+16.36}_{-15.55}$ & 0.79 & 1.29\\
		& $37.93^{+0.17}_{-0.23}$ & 284 & $57.52^{+3.51}_{-3.51}$ & $41.41^{+3.44}_{-3.03}$ & $71.01^{+5.05}_{-4.64}$ & 0.69 & 1.48\\
		& $38.28^{+0.22}_{-0.18}$ & 430 & $28.10^{+1.81}_{-1.81}$ & $29.77^{+2.31}_{-2.32}$ & $47.80^{+2.54}_{-2.73}$ & 1.15 & 2.20\\
		& $38.67^{+0.23}_{-0.17}$ & 350 & $28.51^{+1.52}_{-1.52}$ & $30.06^{+1.69}_{-1.66}$ & $49.83^{+2.32}_{-2.54}$ & 0.79 & 1.44\\
		& $39.20^{+0.50}_{-0.30}$ & 373 & $21.59^{+1.28}_{-1.28}$ & $22.64^{+2.01}_{-2.06}$ & $38.99^{+1.74}_{-1.74}$ & 0.87 & 1.81\\
		\hline
	\end{tabular*}
	\label{table:w0_props}
\end{table*}

\begin{table}
	\centering
	\caption{Best-fit power law properties relating \wo~to \ha~luminosity, $R$-band luminosity, and stellar mass using the measurements shown in Table \ref{table:w0_props} and Figures \ref{fig:w0_cont} and \ref{fig:w0_line}. The $R$-band luminosity and stellar mass assessments are from Approach 2 in \S \ref{sec:modeling} with the latter incorporating our mass-to-light ratio model highlighted in Equation \ref{eqn:ML_ratio}.}
	\def\arraystretch{1.3}%
	\begin{tabular*}{\columnwidth}{@{\extracolsep{\fill}} l c c c}
		\hline
		Property & Type & Slope & $\log_{10}$ Const. \\
		\hline
		\ha~Luminosity & Intrinsic & $-0.218^{+0.043}_{-0.050}$ & $10.34^{+2.03}_{-1.73}$\\
		& Sel. Bias & $-0.228^{+0.047}_{-0.051}$ & $10.75^{+2.05}_{-1.90}$\\
		$R$-band Luminosity & Intrinsic & $-0.244^{+0.046}_{-0.049}$ & $10.90^{+1.89}_{-1.78}$\\
		& Sel. Bias & $-0.380^{+0.054}_{-0.062}$ & $16.17^{+2.37}_{-2.09}$\\
		Stellar Mass & Intrinsic & $-0.160^{+0.030}_{-0.033}$ & $2.86^{+0.27}_{-0.26}$\\
		& Sel. Bias & $-0.249^{+0.036}_{-0.041}$ & $3.65^{+0.33}_{-0.29}$\\
		\hline
	\end{tabular*}
	\label{table:w0_model}
\end{table}

Figure \ref{fig:w0_cont} shows a strong correlation between \wo~and $L_R$ (stellar mass) where \wo~increases with decreasing continuum luminosity. The best-fit power law measurements are highlighted in Table \ref{table:w0_model} and shown in Figure \ref{fig:w0_cont} where we find that $\textrm{\wo} \sim L_R^{-0.24^{+0.05}_{-0.05}}$ and $\sim L_R^{-0.38^{+0.05}_{-0.06}}$ for the case where selection effect corrections are taken into account and when they are ignored, respectively. The steeper slope for the case where we ignore selection effect corrections is a result of the increasing EW limit at $L_R < 10^{38.1}$ \cgs~causing an overestimation of the underlying \wo. Convolving the $M/L_R$ ratio of Equation \ref{eqn:ML_ratio}, we find a \wo~-- stellar mass correlation with $\textrm{\wo} \sim M^{-0.16\pm0.03}$ and $\sim M^{-0.25\pm0.04}$ when correcting for selection effects and ignoring such corrections, respectively. The two cases are statistically different and highlights how selection biases are enhancing the EW -- stellar mass correlation. Taking selection corrections into account with our approach results in a weaker correlation but $\sim 5\sigma$ significance from a null correlation.

Our selection-biased slope of $-0.25\pm0.04$ is in perfect agreement to the EW$\sim M^{-0.25}$ correlation that  \citet{Sobral2014} found for the HiZELS \ha~narrowband samples at $z = 0.4$, $0.84$, $1.47$, and $2.23$ (the four NB slices; each $\Delta z \sim 0.01 - 0.02$). Given that HiZELS is also a narrowband survey, it also has similar selection limits such that a non-uniform EW limit is present towards lower stellar masses. Correcting for selection effects with the HiZELS sample following our approach may also result in a similar intrinsic EW -- stellar mass correlation that we find here with the LAGER sample. \citet{Reddy2018} used the spectroscopic MOSDEF sample and found slopes of $-0.378\pm0.004$ and $-0.286\pm0.003$ at $z \sim 1.5$ and $z\sim 2.3$, respectively, similar to HiZELS and our selection-biased measurement (assuming no redshift evolution in the \wo -- stellar mass slope). Although the selection function is quite different from our narrowband LAGER sample, \citet{Reddy2018} shows that their samples are line luminosity complete down to their mass-complete threshold of $10^{9.0}$ \msol~by doubling their nominal emission line flux threshold and re-assessing their EW -- stellar mass correlation where they find a $\lesssim 0.1$ dex change in the normalization. This may suggest that the slope could have a redshift evolution from $z = 0.47$ (LAGER) to $z \sim 1.5$ (MOSDEF).  Lastly, \citet{Fumagalli2012} used 3D$-${\it HST} grism data and found an EW -- stellar mass trend consistent with a slope of $\sim -0.50$ and $\sim -0.35$ at $z \sim 0.9$ and $1.2$, respectively. Using archival VVDS data, \citet{Fumagalli2012} also found a slope of $\sim -0.47$ at $z \sim 0.3$, steeper than the slope we find for our selection-biased case. They note that their 3D$-${\it HST} and VVDS samples are mass-complete down to $10^{10.0}$ \msol~and $10^{9.5}$ \msol, respectively, however do not mention if they are line luminosity complete which can impact their EW completeness. It is then evident that if we are to compare the EW -- stellar mass correlation between samples and also at various redshifts to investigate possible redshift evolutions, then we must carefully take selection limits into account and apply the necessary corrections. 

The intrinsic EW distributions shown in the {\it bottom right} panel of Figure \ref{fig:EWdistrib_cont} also signify how faint-continuum, low-mass \ha~emitters tend to have more high EW emitters compared to high-mass \ha~emitters, where we find \ha~emitters are $\sim 320$ more probable to have rest-frame EW$ > 200$\AA~compared to high-mass \ha~emitters. This also is seen in Figure \ref{fig:EW_cont} where we find a factor of $\sim 3$ change in \wo~from low- to high-mass. In comparison, faint \ha~emitters are $\sim 30$ times as likely to have rest-frame EW$ > 200$\AA~compared to bright \ha~emitters with a factor of $\sim 2$ change in \wo~from faint to bright \ha~luminosity. This highlights how EW strongly depends on strong stellar mass compared to \ha~luminosity.

\begin{figure}
	\centering
	\includegraphics[width=\columnwidth]{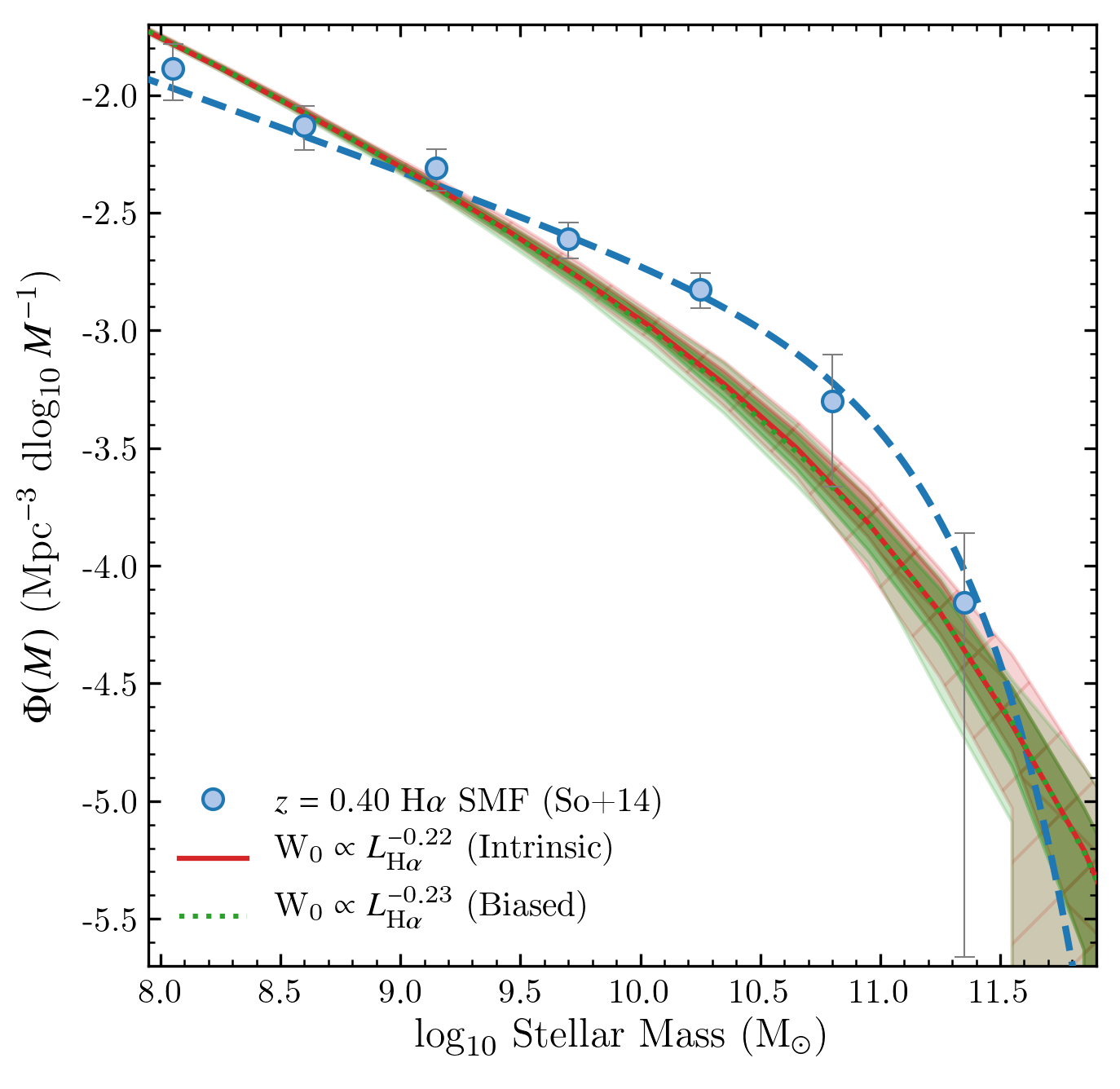}
	\caption{The stellar mass function of the mock samples in comparison to the $z = 0.4$ \ha~measurement of \citet{Sobral2014}. The mock samples assume the power law paramaterization used to fit the \wo~-- $L_\textrm{\ha}$ correlation with the selection-biased case shown in {\it green} and the intrinsic case shown in {\it red} and are generated using the approach define in \S\ref{sec:approach1}. Given that the selection limits are uniform in the EW -- $L_{\textrm{\ha}}$ plane, we find no significant difference between the stellar mass functions of both cases. In comparison to the \citet{Sobral2014} SMF, we find both cases underestimate the number densities at $10^{9.0} < M < 10^{11.5}$ \msol~and are within $1\sigma$ agreement at lower stellar masses. This would suggest that the \wo~-- \ha~luminosity correlation does not reproduce all three main statistical distributions and, therefore, can be the byproduct of the \wo~-- stellar mass correlation.}
	\label{fig:rep_SMF}
\end{figure}

\begin{figure}
	\centering
	\includegraphics[width=\columnwidth]{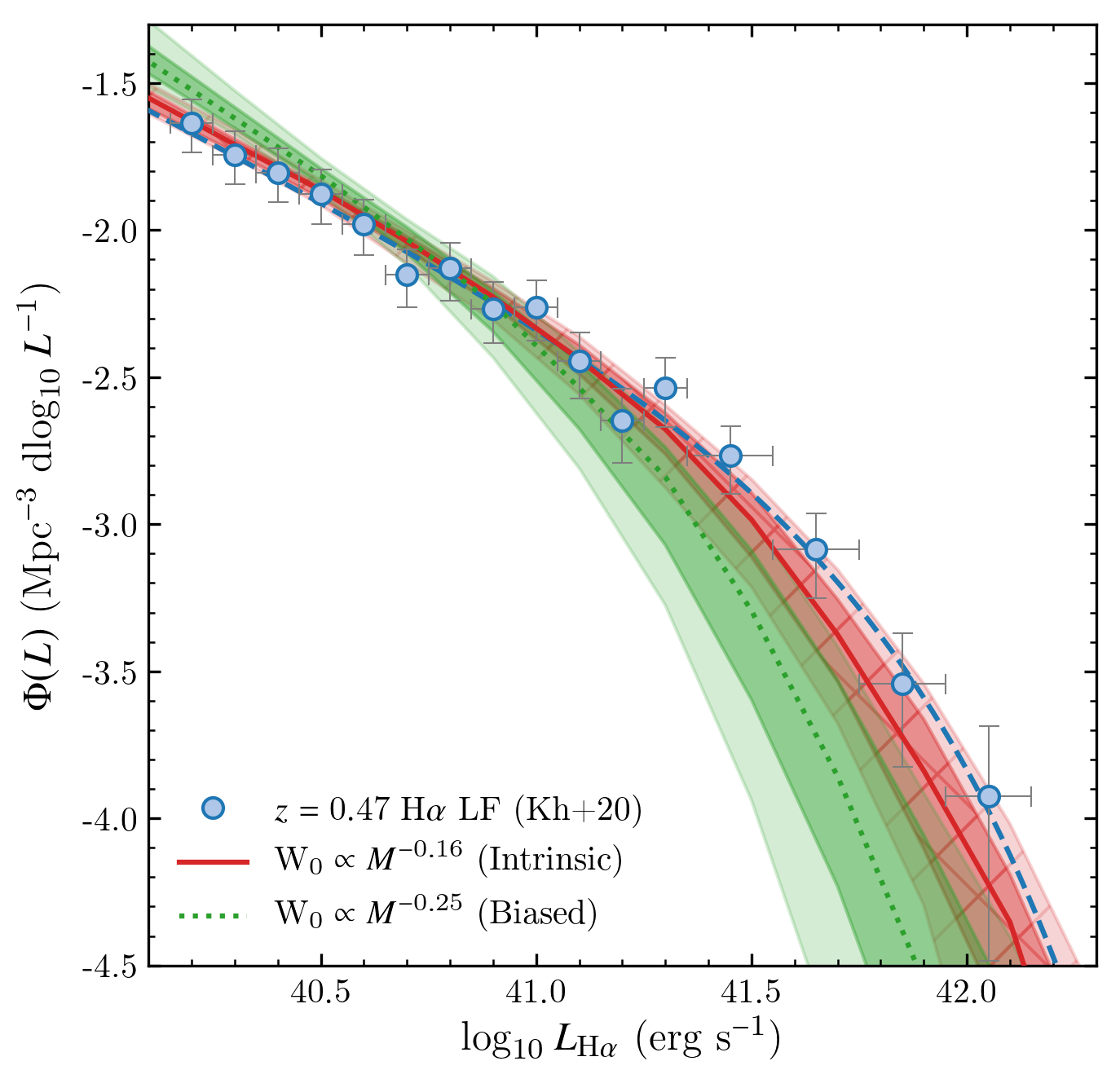}
	\caption{The comparison between the luminosity functions of our mock samples and \citet{Khostovan2020}. The mock samples were generated using our best-fit \wo~-- stellar mass model following the approach define in \S\ref{sec:approach2}. The intrinsic, selection-effect corrected correlation ({\it red}) between \wo~and stellar mass shows a nearly perfect agreement with the observed \ha~luminosity function of \citet{Khostovan2020} while the selection-biased correlation ({\it green}) increasingly underestimates the bright-end with increasing $L_\textrm{\ha}$. This highlights the importance of correcting for selection effects in EW distribution measurements and also provides strong evidence that the \wo~and stellar mass correlation is physical as it can reproduce all three main statistical distributions of galaxies.}
	\label{fig:rep_LF}
\end{figure}

\subsection{Which is it? Stellar Mass or \ha~luminosity}

In \S\ref{sec:ew_line} and \ref{sec:ew_cont}, we investigated how EW correlates with \ha~and continuum luminosity (stellar mass), respectively. In the case of ignoring selection corrections, we found EW scales as $L_{\textrm{\ha}}^{-0.23\pm0.05}$ and $L_R^{-0.38^{+0.05}_{-0.06}}$ ($M^{-0.25\pm0.04}$). For the case where we model the intrinsic distributions and fit to the observations by correcting for selection effects, we find EW scales as $L_{\textrm{\ha}}^{-0.22^{+0.04}_{-0.05}}$ and $L_R^{-0.24\pm0.05}$ ($M^{-0.16\pm0.03}$). The results of \S\ref{sec:ew_line} and \ref{sec:ew_cont} show that \ha~luminosity and stellar mass are correlated with EW. However, \ha~luminosity (instantaneous SFR) and stellar mass are also observed to be correlated as the SFR -- stellar mass correlation that has been extensively covered in the literature (e.g., \citealt{Whitaker2012,Speagle2014}). The question that arises is does EW depend more on \ha~luminosity or stellar mass? Is the correlation with stellar mass or $L_\textrm{\ha}$ influenced/shaped by the other?

To address this issue, we argue that for a correlation to be considered as the dominant/independent correlation would require that it can reproduce the \ha~luminosity function, stellar mass function, and EW distribution. The color-coding of Figure \ref{fig:EW_Line} shows there lies a range of continuum luminosities (stellar mass) at a given \ha~luminosity. This suggests that at each point within the EW -- \ha~luminosity correlation also lies the correlation between EW -- stellar mass. If the EW -- \ha~luminosity correlation can then reproduce the \ha~LF, SMF, and EW distribution while the EW -- stellar mass correlation can not, then that would be evidence the latter is a shaped by the former, which is then the dominant/independent correlation.

In each approach described in \S\ref{sec:approach1} and \ref{sec:approach2}, we assume one of the following statistical distributions (LF or SMF) and use observations to constrain the intrinsic EW distribution, as shown in \S\ref{sec:ew_line} and \ref{sec:ew_cont}. We now test the correlations found in the respective sections and see if we can reproduce the third statistical distribution. In the case of Approach 1 (LF + EW scaling with \ha~luminosity), we seek to reproduce the SMF while in Approach 2 (SMF + EW scaling with stellar mass), we seek to reproduce the \ha~LF. If only one approach reproduces the third property, then it would suggest that the EW is primarily dependent on a specific property (\ha~luminosity or stellar mass), while the secondary correlation is shaped by the primary correlation via the \ha~luminosity (SFR) -- stellar mass correlation. It would also provide evidence that such an EW scaling is physical (driven by physical processes and is the primary trend) and is representative of a complete population of star-forming galaxies.

Figure \ref{fig:rep_SMF} shows the SMF using the mock sample generated from Approach 1 (\S\ref{sec:approach1}) where we populate the sample using the $z = 0.47$ \ha~LF \citep{Khostovan2020} and the \wo -- \ha~luminosity correlations shown in Table \ref{table:w0_model} and Figures \ref{fig:w0_line} to assign EWs. Using the combined information of EW and \ha~luminosity, we determine the continuum luminosity and stellar mass for each source where the latter is determined using Equation \ref{eqn:ML_ratio}. We find no significant difference between the predicted SMFs when using the intrinsic and selection-biased \wo~-- $L_\textrm{\ha}$ correlations, which is expected given that the two are similar to one another.

We compare our predicted SMFs to the $z = 0.40$ \citet{Sobral2014} \ha~SMF and find that we are within $1\sigma$ agreement at lower stellar masses up to $\sim 10^{9.0}$ \msol. At higher stellar masses, we find our predicted SMFs underpredict the number densities until $\sim 10^{11.5}$ \msol. Given that the characteristic stellar mass of the \citet{Sobral2014} SMF is $M^\star = 10^{11.07\pm0.54}$ \msol, we see the number densities drop significantly while our predicted SMFs show a shallower decrease and are more consistent with $M^\star$ of $\sim 10^{11.8}$ \msol. We also find our predicted SMF to have a steeper faint-end slope of $\alpha \sim -1.76$ in comparison to the $-1.37\pm0.02$ measured by \citet{Sobral2014}. Relying on the EW -- $L_\textrm{\ha}$ correlation seems then to overpredict the number of $>10^{11.5}$ \msol~emitters, while underpredicting the number of $10^{9.0 - 11.5}$ \msol~emitters. Although the correlation reproduces the \ha~luminosity function and EW distributions, we find that it can not reproduce the \ha~stellar mass function.

We next test if we can reproduce the \ha~LFs by using our mock samples from Approach 2, where we randomly sample from an intrinsic stellar mass function (assuming the $z = 0.4$ \citet{Sobral2014} SMF) and an EW distribution scaled with stellar mass based on our results in \S\ref{sec:ew_cont}. Figure \ref{fig:rep_LF} shows our predicted LFs with the intrinsic EW -- stellar mass $\textrm{\wo} \sim M^{-0.16}$ shown in {\it red} and the selection-biased $\textrm{\wo} \sim M^{-0.25}$ shown in {\it green}. We compare our predicted LFs to the $z = 0.47$ \citet{Khostovan2020} LAGER \ha~LF shown in {\it blue} where the characteristic number density and \ha~luminosity are $\phi^\star = 10^{-3.16\pm0.09}$ Mpc$^{-3}$, $L^\star = 10^{41.72\pm0.09}$ erg s$^{-1}$, respectively, with a fixed faint-end slope $\alpha = -1.75$. 

The selection-biased correlation fails to reproduce the bright-end where number densities become increasingly underestimated with increasing \ha~luminosity where the predicted LF is consistent with an $L^\star \sim 41.35$, which is $0.37$ dex lower than the observed \ha~LF. There is some agreement at $L_\textrm{\ha} < 10^{41.0}$ erg s$^{-1}$, however, the number densities become slightly overpredicted at $L_\textrm{\ha} < 10^{40.5}$ erg s$^{-1}$ signifying a steeper faint-end slope ($\alpha \sim -1.88$) in comparison to the observed LF.

We find the predicted LF using the intrinsic $\textrm{\wo} \sim M^{-0.16}$ correlation provides a near perfect match to the observed LF. The bright-end is slightly below the observations and is consistent with an $L^\star \sim 10^{41.62}$ erg s$^{-1}$ ($\sim 0.10$ dex lower than the LAGER LF) and a faint-end slope of $\alpha \sim -1.75$ consistent with the observed LF. However, the predicted LF is within $1\sigma$ agreement for all \ha~luminosities probed. This raises three key points. First, the intrinsic correlation is able to reproduce the EW distribution (SF history), stellar mass function (integrated SF history), and luminosity function (instantaneous SF). Since all three trace physical processes associated with star-formation activity, it suggests that the $\textrm{\wo} \sim M^{-0.16}$ correlation is also shaped by the same physical processes as well. Secondly, given that the selection-biased correlation failed to reproduce the \ha~LF, it also shows the importance of taking selection biases into account when investigating the EW properties of star-forming galaxies. Lastly, this also shows that EW seems to primarily depend on stellar mass as it can reproduce all three statistical distributions, while the correlation with \ha~luminosity could be a result of the EW -- stellar mass trend given that \ha~luminosity (SFR) and stellar mass are also correlated (e.g., SFR -- stellar mass correlation, `main sequence').

\subsection{Are low-mass galaxies more bursty?}
\label{sec:high_EW}

\begin{figure}
	\centering
	\includegraphics[width=\columnwidth]{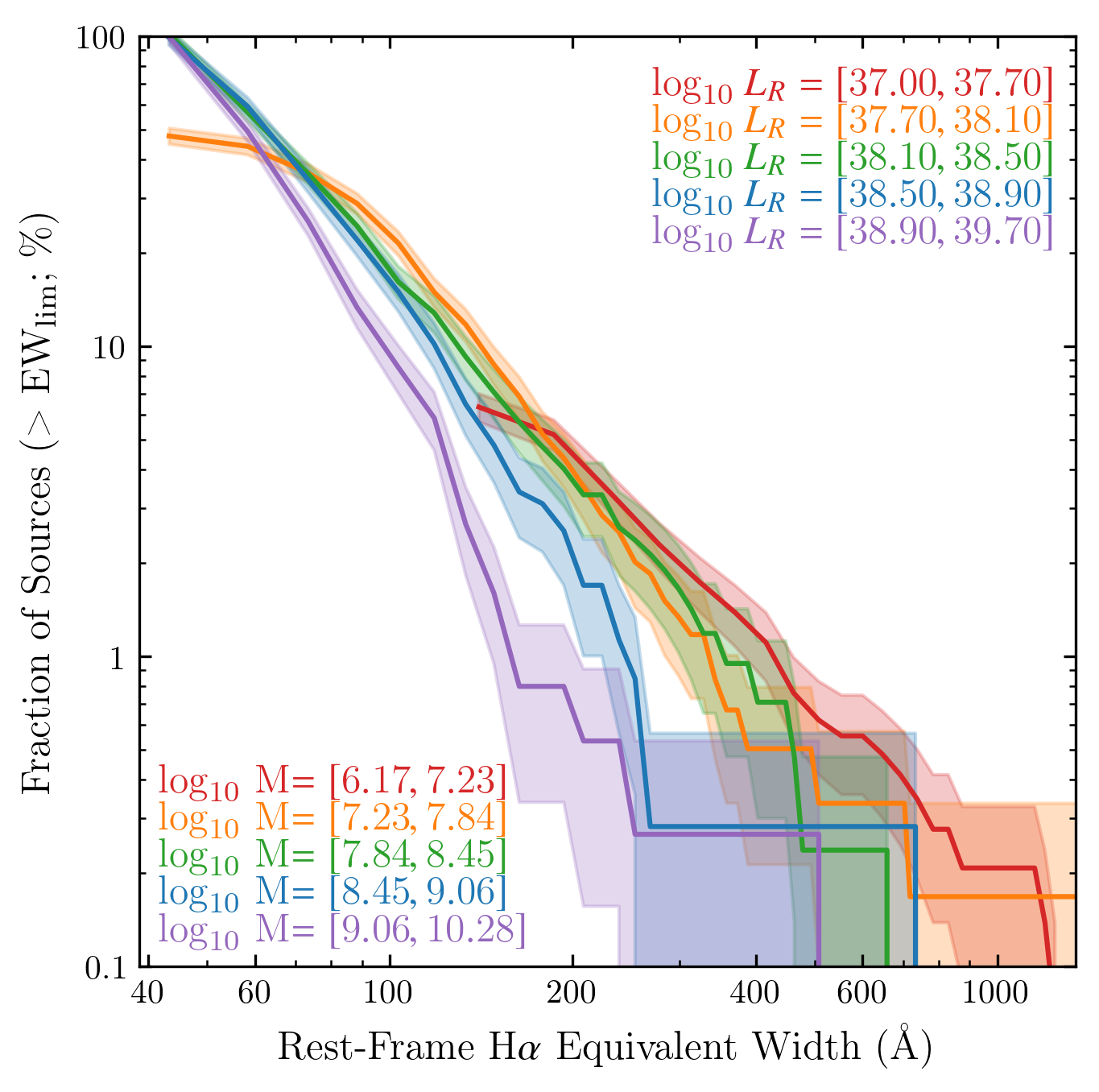}
	\caption{The fraction of \ha~emitters in the LAGER sample above an EW threshold subdivided in $R$-band luminosity and corrected for selection bias using the intrinsic EW distirbutions. We find an increasing fraction of high EW outliers with decreasing continuum luminosity with our faintest $L_R$ bin having the highest fractions. These are also low mass \ha~emitters ($< 10^8$ M$_\odot$) and may be undergoing a phase of bursty star-formation activity given their high equivalent widths.}
	\label{fig:fractions}
\end{figure}

Studies of low-mass galaxies suggest they tend to be systems undergoing periods of bursty star formation activity in comparison to high-mass galaxies (e.g., \citealt{Glazebrook1999,Iglesias2004,Lee2011,Weisz2012,Dominguez2015,Guo2016,Broussard2019,Emami2019}). The typical observatonal proxy for burstiness used in such studies is the $L_\textrm{\ha}$/$L_{\textrm{UV}}$ ratio, where \ha~traces the instantaneous star formation activity (timescales of 5 -- 10 Myr; $O-$ and $B-$type stars) and the UV continuum traces a longer timescale of activity ($\sim 100$ Myr; $O$, $B$, and $A-$type stars; e.g., \citealt{Kennicutt1998}). One major caveat to $L_\textrm{\ha}$/$L_{\textrm{UV}}$~measurements is the assumptions made for dust corrections, which greatly affects the UV continuum (e.g., \citealt{Kewley2002,Lee2009,Faisst2019}).

Here we use \ha~EW as an observational proxy for burstiness, which has a few advantages in comparison to using the $L_\textrm{\ha}$/$L_{\textrm{UV}}$ ratio. One advantage is that no dust corrections are required in the case that $E(B-V)_{nebular} \sim E(B-V)_{stellar}$ (e.g., \citealt{Erb2006,Reddy2015,Puglisi2016}). Also, the \ha~EW uses rest-frame $L_R$, which is redwards of the $4000$\AA~break. This makes the continuum measurement used in the \ha~EWs even more sensitive to larger timescales of star-formation as rest-frame $R$ also traces older stellar populations. This would make the \ha~EW even more sensitive to periods of bursty star formation activity in respect to traditional measurements.

Figures \ref{fig:EWdistrib_line} and \ref{fig:EWdistrib_cont} show that we can strongly constrain the EW distributions using exponential models up to an upper EW limit, which is between $200 - 600$\AA~depending on the subsample used. Beyond this limit, we find high EW outliers that can be potentially systems undergoing a period of burstiness. Figure \ref{fig:EWdistrib_cont} shows our brightest continuum sample ($\log_{10} L_R = 10^{38.9 - 39.7}$ erg s$^{-1}$ \AA$^{-1}$ corresponding to stellar masses of $10^{9.0 - 10.25}$ \msol) having only two outliers above 200\AA, while the faintest sample ($\log_{10} L_R = 10^{37.0 - 37.7}$ erg s$^{-1}$ \AA$^{-1}$ corresponding to stellar masses of $10^{6.1 - 7.2}$ \msol) has 8 outliers with EW $>600$\AA, with 3 sources having EW $> 1000$\AA. This would suggest that there are more high EW sources at lower stellar masses, although selection biases behave differently in each continuum bin, as discussed in \S\ref{sec:ew_cont}, such that it could potentially drive the results. Therefore, are low-mass galaxies really exhibiting evidence for bursty star formation activity in comparison to high-mass systems? Given that we have constrained the intrinsic EW distributions using our mock simulations, we can estimate the total number of \ha~emitters we expect for the whole sample, which can then be used to calculate the intrinsic fraction of \ha~emitters at a given limiting EW threshold. 

Figure \ref{fig:fractions} shows the fraction of \ha~emitters above a limiting EW threshold for each continuum luminosity subsample. We calculate the fraction as being the number of observed \ha~emitters above a given EW threshold divided by the total number of \ha~emitters defined by the intrinsic EW distribution above $35$\AA~(LAGER selection cut; \citealt{Khostovan2020}). The level of incompleteness in the observed samples can be seen for the faintest two $L_R$ samples. Our faintest bin, shown in {\it red}, reaches about $6$ percent by $\sim 150$\AA~and the second faintest bin, shown in {\it orange}, reaches about $50$ percent by $\sim 45$\AA. These are the two $L_R$ bins that suffer from the $5\sigma$ NB magnitude limit causing a non-uniform EW cut as shown in Figure \ref{fig:EW_cont}.

We find the selection-corrected distributions are shifted towards higher EW with decreasing continuum/stellar mass. Our brightest continuum sample has $1$ percent of the sample at rest-frame EW$\gtrsim 175$\AA~while our faintest continuum sample has the same fraction of sources with rest-frame EW$\gtrsim 500$\AA. The continuum samples in between show a progression towards higher EWs for the same $1$ fraction with decreasing continuum. Interpreting these results in the scope of stellar mass, we see that $> 10^{8.5 - 9}$ \msol~\ha~emitters have fewer high EW systems compared to $<10^{8.5}$\msol~emitters. In the scope of burstiness traced by EW, this shows evidence that low-mass, faint continuum \ha~emitters have higher outliers on sSFR indicative of systems undergoing episodic star-formation activity in comparison to high mass, bright continuum \ha~emitters.

Our results are in agreement with local measurements of burstiness done by \citet{Emami2019}, which investigated the star formation histories of dwarf galaxies using both the \ha/UV ratio (timescale) and $\Delta\log_{10} L_\textrm{\ha}$ (amplitude), with the latter being \ha~luminosity (SFR) offset from the main sequence per associated stellar mass. They concluded that $< 10^8$ \msol~galaxies experience intense levels of burstiness that rapidly occurs on timescales $< 30$ Myr, while $>10^8$ \msol~galaxies experience slower ($>300$ Myr) and shallower burst amplitudes. Indeed, we find increasing fractions of outliers starting with our $10^{38.10 - 38.50}$ \cgs~sample, which corresponds to stellar masses of $\sim 10^{8.5}$ \msol. At $<10^{38.1}$ \cgs~($<10^8$ \msol), we find an increasing fraction of high EW \ha~outliers consistent with where \citet{Emami2019} is also finding evidence for intense, episodic star formation activity. We note that the possibility of IMF variations in some sources could also contribute to high \ha~EW outliers (e.g., \citealt{Hoversten2008,Meurer2009,Nanayakkara2017}). Furthermore, comparing \ha~to continuum fluxes (e.g., \ha/UV~ratio, \ha~EW) as a tool for probing stochastic SFHs can be dependent on the presence of binary stellar populations although studies still find the scatter in \ha/UV ratios is attributed to bursty SF activity (e.g., \citealt{Eldridge2012,Sparre2017}). Overall, our results suggest that $z = 0.47$ low-mass \ha~emitters have higher fractions of high EW outliers and may be indicative of being intrinsically more bursty than high-mass emitters.

\section{Discussion}
\label{sec:discussion}
\subsection{Ando Effect -- Lack of massive, high EW galaxies}

\begin{figure}
	\centering
	\includegraphics[width=\columnwidth]{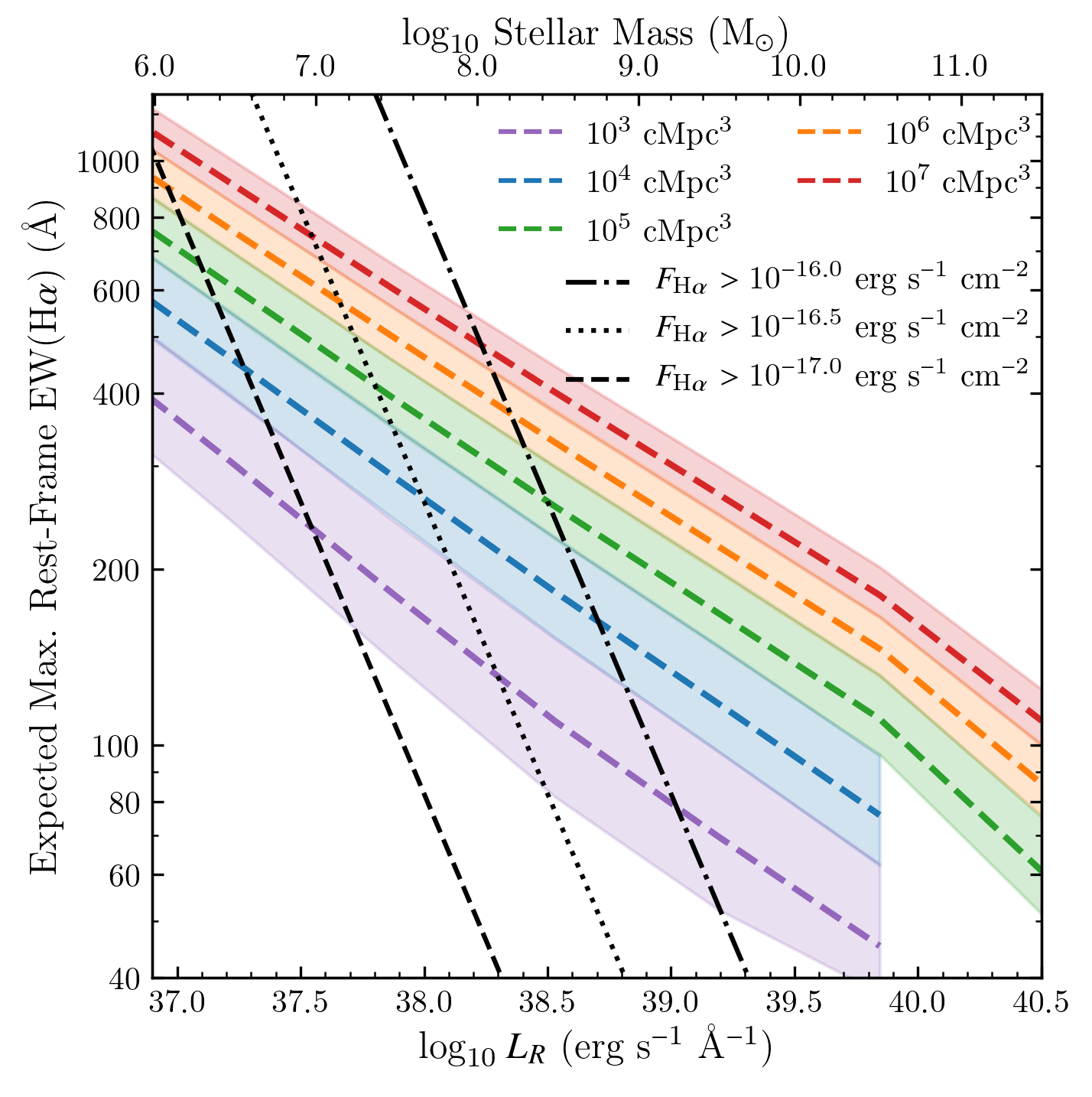}
	\caption{The expected maximum rest-frame \ha~EW at $z = 0.47$ for which a single galaxy is expected within a given comoving volume assuming the $z = 0.40$ \ha~\citet{Sobral2014} SMF, the intrinsic $\textrm{\wo}\sim M^{-0.16}$ correlation from this study, and a minimum EW cutoff of $35$\AA. A total of 10,000 mock surveys was done per comoving volume to taken into account low number statistics at the massive-end, especially for the smaller surveys. The shaded regions corresponds to the $1\sigma$ scatter around the median maximum EW per stellar mass and continuum luminosity. Increasing the survey volume still results in a lack of massive, high EW emitters consistent with the Ando effect. Overlaid are \ha~line flux survey limits where a $10^{-16.0}$ erg s$^{-1}$ cm$^{-2}$ limit and survey volume of $10^{5.0}$ cMpc$^3$ would limit the survey to a maximum of $\sim 250$\AA~corresponding to a stellar mass of $10^{8.5}$ \msol.}
	\label{fig:ando}
\end{figure}

Early \lya~and LBG studies reported a lack of bright UV continuum, high EW emitters and an EW -- UV continuum correlation (e.g., \citealt{Ouchi2003,Shapley2003, Shimasaku2006}), commonly referred to as the `Ando' effect where \citet{Ando2006} found a correlation between \lya~EW and rest-frame 1400\AA~continuum for a spectroscopic sample of $z \sim 5 - 6$ LBGs and reported a deficiency of high \lya~EW, bright UV systems. Subsequent work on \lya~emitters and LBG samples at $0.2 < z < 7$ also find a lack of bright continuum, high EW \lya~emitters (e.g., \citealt{Stanway2007,Deharveng2008,Ouchi2008,Vanzella2009,Stark2010,Kashikawa2011,Zheng2014,Furusawa2016,Ota2017,Hashimoto2017,Santos2020}). \citet{Ando2006} suggests the deficiency is due to star-formation activity occurring at earlier times in massive LBGs resulting in low \lya~EW/sSFR (e.g., older stellar populations) consistent with clustering studies where massive, bright continuum \lya~emitters and LBGs tend to reside in $> 10^{12}$ \msol~dark matter halos (e.g., \citealt{Kashikawa2006,Hildebrandt2009,Harikane2016,Harikane2018,Khostovan2019}). Similar high EW, bright continuum deficiencies are also reported for \ha, \oiii, and \oii~emission line galaxies (e.g., \citealt{Fumagalli2012,Sobral2014,Khostovan2016,Reddy2018}) with such systems found to reside in massive dark matter halos (e.g., \citealt{Sobral2010, Cochrane2018,Khostovan2018}).

However, \citet{Nilsson2009} showed how the `Ando' effect could arise from \lya~flux limits (selection) and small survey volumes (sample variance) by simulating \lya~EW distributions constrained by observations of $z \sim 3$ \lya~emitters and LBGs. \citet{Zheng2014} simulated $z \sim 4.5$ \lya~emitters and found that the \lya~EW -- UV continuum anti-correlation could be artificially generated by \lya~selection in narrowband surveys. \citet{Hashimoto2017} also finds that \lya~flux limits shape the EW -- UV continuum correlation while the upper bound in EW at bright UV is due to the rarity of sources.

Our results as shown in Figures \ref{fig:EWdistrib_cont} and \ref{fig:w0_cont} have taken into account selection effects, consist of a large sample of 1572 \ha~emitters, and a wide $1.1\times10^5$ Mpc$^3$ survey that mitigates sample/cosmic variance effects. We confirm an intrinsic EW -- continuum/stellar mass correlation where we find a deficiency of high EW, bright continuum emitters. Figure \ref{fig:w0_cont} shows a $\sim5\sigma$ difference and a factor of $\sim 3$ change in \wo~between the lowest and highest stellar mass bin. We also find in Figure \ref{fig:fractions} that low-mass galaxies to have higher number of EW/sSFR outliers compared to high-mass galaxies, even when corrected for selection effects. Furthermore, our measured intrinsic, selection bias-corrected correlation of \wo~$\sim M^{-0.16\pm0.03}$ is at $\sim 5\sigma$ significance from a null correlation suggesting that that an EW -- continuum correlation is a physical property of \ha~emitters and points towards a physical origin of the `Ando' effect rather than a selection bias/sample variance origin. We therefore ask the question, how wide of a survey does one require in order to attain a population of massive, high EW \ha~emitters?

We address this question by predicting the maximum \ha~EW that is observable for a mock survey with a given comoving volume, a lower rest-frame EW threshold of $35$\AA, and an EW distribution that scales as $\textrm{\wo} \propto M^{-0.16}$. We assume the $z = 0.4$ \citet{Sobral2014} SMF as the main galaxy distribution for our mock survey with comoving volumes varying between $10^{3 - 7}$ Mpc$^3$. For each survey, we calculate the total number of expected \ha~emitters within the given comoving volume for a stellar mass range between $10^{6 - 12}$ \msol. We then determine the corresponding EW per each source by randomly drawing from an exponential EW distribution with $\textrm{\wo} \propto M^{-0.16}$. The maximum EW at a given stellar mass is then measured and is defined as the highest EW for which one \ha~emitter is observable given these assumptions. For each mock survey, we repeat these measurements 10,000 times to take into account the spread introduced by low number of galaxies, specifically in the smaller surveys ($< 10^5$ Mpc$^3$) and at the bright continuum, massive-end.

Figure \ref{fig:ando} shows our maximum EW predictions as a function of continuum luminosity (stellar mass) and survey volumes. We note these predictions are only based on an exponential model that does not take into account high EW outliers that can have enhanced line emission in comparison to their continuum. Our predictions show that increasing the comoving volume of a given survey increases the range of EWs observed at all continuum luminosities and stellar masses. The maximum EW expected increases per increasing magnitude of comoving volume by $\sim 100 - 200$\AA~at the faintest $L_R$ and by $\sim 20 - 30$\AA~at the brightest $L_R$. We find that the number densities at $M > 10^{10.5}$ \msol~for volumes $10^4$ Mpc$^3$ are too small to simulate \ha~emitters, highlighting the importance of large survey volumes. However, we find that increasing the comoving volume only marginally helps in observing higher EW sources at brighter $L_R$.

Our predictions suggest, given the assumption of an exponential EW distribution, a LAGER-like narrowband survey with a comoving volume of $10^6$ Mpc$^3$ should be able to observe at least one $>10^{10}$\msol~\ha~emitter with a rest-frame EW$ > 200$\AA, while a $10^7$ Mpc$^3$ survey would observe around 4 \ha~emitters. The complete 24 deg$^2$ LAGER is planned to cover $10^6$ Mpc$^3$ which would allow for the investigation of massive, high sSFR \ha~emitters. Future space-missions, such as the {\it Nancy Grace Roman Space Telescope}, would also be capable of observing such rare, massive star-forming galaxies. Such sources may also be massive, dusty starbursts that reside in cluster-like environments (e.g., \citealt{Koyama2013,Dannerbauer2014,Overzier2016,Sobral2016,Shimakawa2018}).

\subsection{Implications on Main Sequence Measurements}

\begin{figure}
	\centering
	\includegraphics[width=\columnwidth]{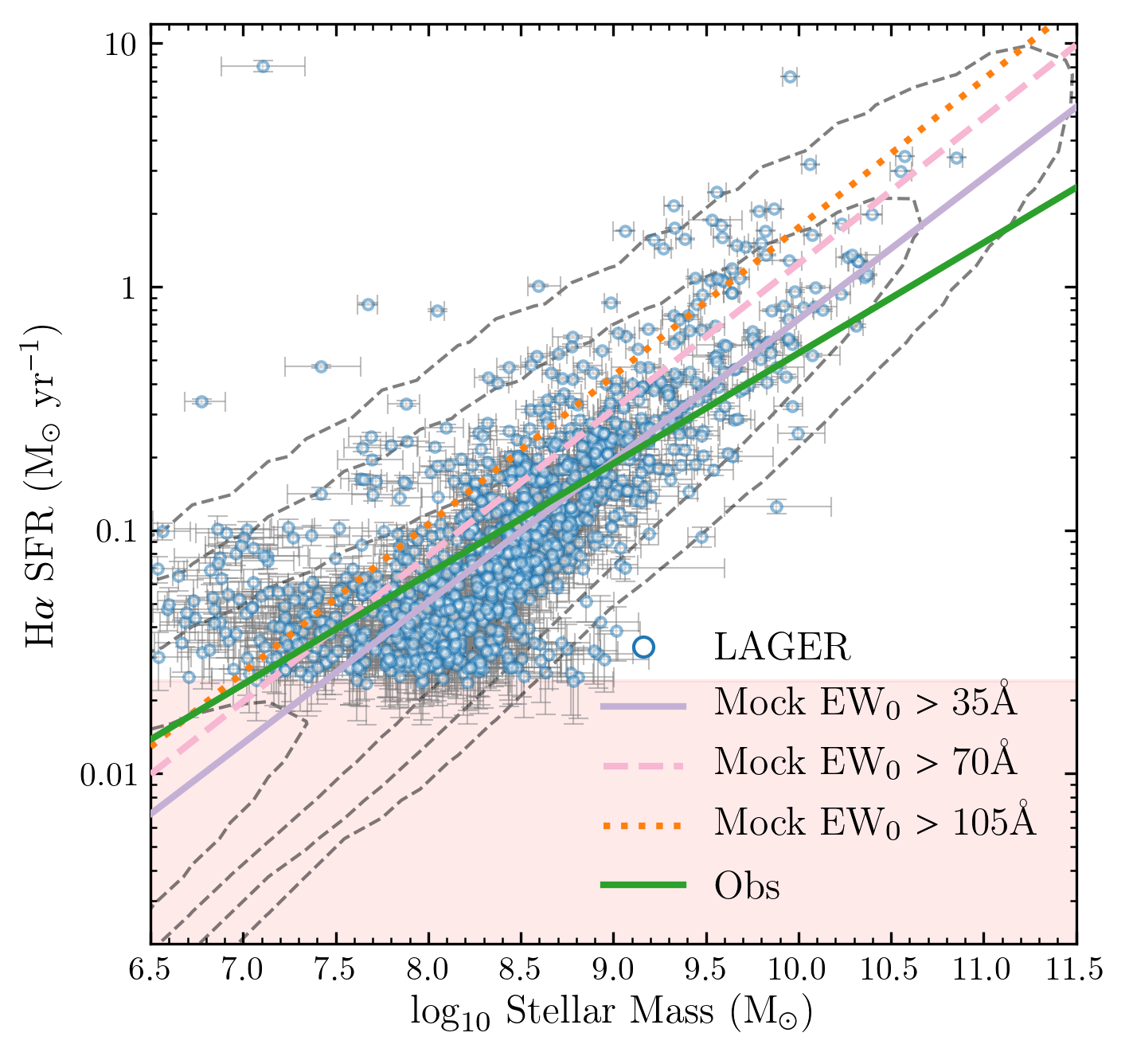}
	\caption{The correlation between observed (uncorrected for dust) \ha~star formation rate and stellar mass. The \ha~SFRs assume the \citet{Kennicutt1998} calibration corrected for a \citet{Chabrier2003} IMF to match the stellar mass IMF assumption of the COSMOS2015 \citep{Laigle2016} catalog. The observed \ha~emitters are shown as {\it blue circles} and show a linear correlation down to $10^{8.5}$ \msol, where the narrowband limit causes a sharp cutoff at $\sim 0.025$ \msol~yr$^{-1}$. The distribution of the mock sample assuming an EW$>35$\AA~cut and \wo$\propto M^{-0.16}$ are shown as contours and highlight the continuing linear correlation down to lower stellar masses. Linear fits show a shallower slope for the LAGER sample which is directly caused by the narrowband selection limit. Increasing the EW cut of the mock sample from 35\AA~to 70\AA~and 105\AA~results in an increase of the normalization. This illustrates how selection effects can drive main sequence measurements and highlights the importance of correction for said biases.}
	\label{fig:MS}
\end{figure}

The empirical correlation between star formation activity and stellar mass, commonly referred to as the `main sequence' (MS), has been extensively studied with various samples over the past decade at low ($z \lesssim 1$; e.g., \citealt{Brinchmann2004,Elbaz2007,Salim2007}), intermediate ($z \sim 1 - 3$; e.g., \citealt{Daddi2007,Noeske2007,Reddy2012}), and high redshifts ($z \gtrsim 3$; e.g., \citealt{Lee2011K,Bouwens2012,Steinhardt2014}) and follows a simple power law up to $z \sim 5$ (e.g., \citealt{Speagle2014}). However, recent work find a steeper (shallower) slope at low (high) stellar masses (e.g., \citealt{Whitaker2014,Schreiber2015,Tomczak2016}), however uncertainties in dust corrections and sample variance at the high-mass end and selection biases towards the low-mass end could affect the underlying correlations. Using our intrinsic EW -- stellar mass correlation, we investigate the implications of selection effects on the shape of the SFR -- stellar mass correlation.

Figure \ref{fig:MS} shows the main sequence where \ha~star formation rates are observed (uncorrected for dust) and are measured directly from the combined narrowband and broadband photometry along with the \citet{Kennicutt1998} \ha~calibration corrected for a \citet{Chabrier2003} IMF. Stellar mass is taken directly from the COSMOS2015 \citep{Laigle2016} catalog and also assumes a \citet{Chabrier2003} IMF. The contours show the distribution of our mock sources that best represents the \ha~LAGER sample ($W_0 \propto M^{-0.16}$) with a rest-frame EW$>35$\AA. We note that not applying a dust correction would result in a lower normalization and shallower slope compared to measurements in the literature, especially since high-mass galaxies are generally dustier in comparison to low-mass galaxies (e.g., \citealt{Garn2010,Sobral2012,Dominguez2013,Kashino2013}). However, we emphasize that the main objective is to investigate the influence of selection biases, especially at low mass, on measurements of the main sequence and not make a definitive main sequence measurement.

The LAGER \ha~emitters show a linear correlation between SFR and stellar mass from high stellar masses down to $\sim 10^{8.5}$ \msol. At lower stellar masses, the observed \ha~emitters are limited by the narrowband (\ha~flux) limit which causes a sharp horizontal cut at $\sim 0.025$ \msol~yr$^{-1}$ (e.g., Malmquist Bias) and is highlighted in Figure \ref{fig:MS} as a {\it light red} shaded region. For a given stellar mass below $<10^{8}$ \msol, the SFR ranges between the selection cut and $\sim 0.1$ \msol~yr$^{-1}$ with a few outliers at higher SFRs. If the linear correlation that is observed at higher stellar masses does continue with decreasing stellar mass, then the presence of \ha~emitters with $< 10^8$ \msol~above the selection limit suggests that the spread of the main sequence could be increasing. Other studies have found tentative evidence for an increase in the main sequence scatter with decreasing stellar mass \citep{Salmon2015,Santini2017,Boogaard2018}. This may be due to low mass galaxies being more susceptible to bursty star formation activity in comparison to high mass galaxies as we found in \S\ref{sec:high_EW}. 

The mock simulations show a linear correlation that extends to all stellar masses shown in Figure \ref{fig:MS}. More importantly, it shows the extent of the main sequence at stellar masses $< 10^{8.5}$ \msol~and SFR below the selection limit of $0.025$ \msol~yr$^{-1}$. The simulations highlight the incompleteness of typical narrowband surveys where an increasing fraction of sources are missing towards lower stellar masses, while still recovering the high SFR end of the distribution at these stellar masses. This has implications when fitting the main sequence.

Figure \ref{fig:MS} shows the best-fit power law using the LAGER \ha~emitters and the mock simulation with EW$>35$\AA~shown as a {\it green} and {\it purple} line, respectively. The LAGER main sequence fit is consistent with a slope of $\sim 0.45$, which is shallower to the $\sim 0.58$ that we find using our mock simulation. The shallower slope in the observations is directly due to the influence of the \ha~luminosity cut. Since the incompleteness of observed \ha~emitters increases with decreasing stellar mass, fitting a simple power law to the observations results in a shallower slope as it tries to capture the low mass, high SFR end of the intrinsic distribution at a given stellar mass. The mock simulation on the other hand shows a slope that is consistent with the LAGER sources above $10^{8.5}$ \msol. One could argue then that a simple stellar mass cut can be placed at a level where the narrowband selection causes incompleteness, but such a cut can be subjective and also limits the scope of a main sequence studies to the massive end and discards potential science that can be done with low-mass, dwarf-like systems.

Another selection bias that can affect main sequence studies, at least in the scope of narrowband surveys, is the EW cut which can also be thought of as a sSFR cut. Increasing the EW/sSFR limits would essentially increase the normalization. We see this behavior in Figure \ref{fig:MS} where we show the MS given EW limits of 35\AA, 70\AA, and 105\AA~shown as {\it solid purple}, {\it pink dashed}, and {\it orange dotted} lines, respectively. As expected, the normalizations increase with increasing EW limits while the slope is marginally affected. 

Figure \ref{fig:MS} shows the importance selection effects can have on both the normalization and slope. These implications are not limited to only narrowband surveys, which have simple selection functions that can be easily modeled. Photometric and spectroscopic surveys can have even more complicated selection functions that can bias main sequence measurements, however recent progress has been made to assess selection biases in such data sets and recalibrate measurements to uniform assumptions (e.g., \citealt{Speagle2014}). Our results have the important implication that to understand the main sequence, especially at lower stellar masses, requires careful assessments on selection biases and corrections.

\subsection{Implications for Future Surveys}

The next decade will see the introduction of several next-generation surveys, such as those with the Rubin Observatory (formerly LSST; \citealt{Ivezic2019}),  {\it JWST} \citep{Gardner2006}, {\it Euclid} \citep{Laureijs2011}, and {\it Nancy Grace Roman Space Telescope} (\textit{NGRST}; formerly {\it WFIRST}; \citealt{Spergel2015}), all of which will present us with large samples of star-forming galaxies at various cosmic epochs. Given the importance of such surveys, careful planning is necessary for survey design and an estimation of the scientific output. For example, gauging the number of sources expected in a blind survey would require information of the expected luminosity (stellar mass) function down to a given flux (stellar mass) limit, along with the comoving volume based on survey design. 

Our results provide additional input on survey predictions/expectations, specifically for slitless spectroscopic surveys. Such surveys are limited by their resolving power, $R = \lambda/\Delta\lambda$, where $\Delta\lambda$ is the limiting wavelength width required to resolve two spectral features at a given wavelength, $\lambda$. Since the EW is defined as the amount of continuum wavelength coverage needed to match with the corresponding emission line flux, the spectroscopic resolution is linked to an effective EW threshold.

We showed in Figure \ref{fig:w0_cont} that \wo~decreases with increasing continuum luminosity, which could affect how slitless spectroscopic surveys can capture the bright (massive) end of a galaxy population. If we consider an {\it HST}/ACS G800L grism survey with a limiting resolution of 80\AA~for point-like sources (covers $5500 - 11000$\AA), then such a survey will be sensitive to $z = 0.47$ \ha~emitters ($9640$\AA) with rest-frame EW $> 54$\AA~(assuming EW $\sim \Delta\lambda$; e.g., \citealt{Xu2007,Pirzkal2018}). In the scope of Figure \ref{fig:w0_cont}, such a survey would detect 50 percent and 30 percent of \ha~emitters with $L_R \sim 10^{38.5}$ erg s$^{-1}$ \AA$^{-1}$ and $10^{39.5}$ erg s$^{-1}$ \AA$^{-1}$, respectively, in comparison to a survey with limiting EW of 35\AA~(e.g., LAGER). This would suggest that information regarding the EW distribution is also crucial on top of luminosity/stellar mass functions when it comes to predicting expected source counts for future grism surveys, such as those planned with {\it JWST}, {\it Euclid}, and {\it NGRST}. 

Furthermore, our maximum EW predictions in Figure \ref{fig:ando} also highlight the importance of limiting line fluxes where we show the limitations in EW and continuum luminosity introduced by \ha~line flux limits. We stress the point that these maximum EW limits are based on the assumption that galaxies follow an exponential EW distribution. However, we saw in Figures \ref{fig:EWdistrib_line} and \ref{fig:EWdistrib_cont} a number of high EW outliers that depart from an exponential distribution. Therefore, these predictions should be considered as the expected maximum EWs given the model parameters where we expect to have a single \ha~emitter in the underlying sample. We find that a survey with a line flux limit of $10^{-16.0}$ erg s$^{-1}$ cm$^{-2}$ and a volume coverage of $10^5$ Mpc$^3$ will be able to observe up to EW $\sim 250$\AA, while decreasing the line flux limit to $10^{-16.5}$ erg s$^{-1}$ cm$^{-2}$ for the same volume coverage almost doubles the expected maximum EW limit and also pushes towards lower stellar masses by a dex. However, this does not rule out high EW outliers which can populate samples, especially at volumes larger than the LAGER survey ($\sim 10^5$ Mpc$^3$). Therefore, our predictions are solely based on galaxies following an exponential EW distribution and suggest to what EW limit can we expect an \ha~emitter depending on survey parameters.

Overall, this has implications for future survey design where we find an intricate balance between survey volume, line flux limits, and stellar mass limits can set the range of expected EWs that could be observed based on an exponential EW distribution. Given that the \ha~EW is also a tracer of star formation activity, the wider range of expected EWs would in turn mean a population of star-forming galaxies with a diverse variety of star formation histories that could be explored to understand the underlying physics. Our results then implies that careful planning that takes into account information regarding EW distributions can also be beneficial in designing future surveys of star-forming galaxies.

\section{Conclusions}
\label{sec:conclusions}
We have presented our new methodology of measuring the shape of EW distributions by modeling the intrinsic properties of emission-line galaxy samples, propagating selection effects, and then comparing to observations. The $z = 0.47$ \ha~EW distributions are measured using our technique and are observationally constrained using the large sample of 1572 \ha-selected emitters from the $\sim 3$ deg$^2$ LAGER survey coverage of the COSMOS field. Here we highlight the main points of this study:

\begin{enumerate}[leftmargin=*]
	
	\item Selection limits within the EW -- \ha~luminosity plane is uniform such that for any given \ha~luminosity, the range of EWs probed is the same. In the EW -- $L_R$ (stellar mass) plane, a common EW cut is seen down to $L_R \sim 10^{38.1 - 38.2}$ erg s$^{-1}$ \AA$^{-1}$ while the narrowband magnitude (line flux) limit dominates at fainter continuum. This causes a non-uniform EW cut that increases with decreasing continuum luminosity.
	
	\item In terms of \ha~luminosity, the EW distributions are best represented by an exponential distribution up to $\sim 300$\AA~and $\sim 150$\AA~at $L_\textrm{\ha} \sim 10^{39.73 - 40.03}$ erg s$^{-1}$ and $10^{40.93 - 41.43}$ erg s$^{-1}$. respectively. We find that an exponential distribution also best represents the EW distributions in bins of continuum luminosity (stellar mass) up to $\sim 600$\AA~and $200$\AA~at $L_R \sim 10^{37.0 - 37.7}$ erg s$^{-1}$ \AA$^{-1}$ and $10^{38.9 - 39.7}$ erg s$^{-1}$ \AA$^{-1}$, respectively.
	
	\item We find selection limits to affect the shape of EW distributions when the samples are subdivided in continuum luminosity (stellar mass). At $L_R > 10^{38.1}$ erg s$^{-1}$ \AA$^{-1}$, the modeled intrinsic EW distribution and the direct exponential fit show no clear difference which is due to the uniform EW cut of $\sim 35$\AA. Incompleteness from selection limits start to affect the EW distributions at $\lesssim 100$\AA~and $\lesssim400$\AA~at $L_R \sim 10^{37.7 - 38.1}$ erg s$^{-1}$ \AA$^{-1}$ and $10^{37.0 - 37.7}$ erg s$^{-1}$ \AA$^{-1}$, respectively. 
	
	\item We find an EW -- \ha~luminosity correlation where $\textrm{\wo} \sim L_\textrm{\ha}^{-0.23\pm0.05}$ and $\textrm{\wo} \sim L_\textrm{\ha}^{-0.22^{+0.04}_{-0.05}}$ in the case of ignoring selection effect corrections and using our mock sample, respectively. The agreement between the two suggests that selection does not affect the EW -- \ha~correlation, although this is primarily due to the uniform selection cuts within the EW -- $L_\textrm{\ha}$ plane.

	\item An intrinsic EW correlation of $\textrm{\wo} \sim L_R^{-0.24\pm0.05}$ ($M^{-0.16\pm0.03}$) is found. Not accounting for selection effects results in a steeper slope of $\textrm{\wo} \sim L_R^{-0.38^{+0.05}_{-0.06}}$ ($M^{-0.25\pm0.04}$). This highlights the importance and need of taking selection corrections into account when investigating correlations between EW and galaxy properties, such as stellar mass.
	
	\item The predicted stellar mass function assuming an EW distribution scaled as $\textrm{\wo} \sim L_\textrm{\ha}^{-0.22^{+0.04}_{-0.05}}$ does not agree with the $z = 0.4$ \ha~stellar mass function of \citet{Sobral2014}. Given that \ha~luminosity and stellar mass are also correlated with one another, this could suggest that the EW -- $L_\textrm{\ha}$ correlation could be shaped by the trend with stellar mass.
	
	\item The predicted \ha~LF when assuming $\textrm{\wo} \sim M^{-0.16\pm0.03}$ strongly agrees with the observed $z = 0.47$ \ha~LAGER LF \citep{Khostovan2020}. Using the best-fit correlation in the case where selection corrections are ignored, we find the predicted \ha~luminosity function is steeper and underestimates number densities for $L_\textrm{\ha} > 10^{40.6}$ erg s$^{-1}$ ($L > 0.08L^\star$). The strong agreement between the predicted \ha~LF, assuming $\textrm{\wo} \sim M^{-0.16\pm0.03}$, and observation suggests that EW is primarily dependent on stellar mass. This also suggests the EW -- stellar mass correlation is most likely driven by physical processes that also shape the \ha~LF and SMF.

	\item Correcting for incompleteness by assuming an intrinsic EW distribution scaled as $M^{-0.16\pm0.03}$, we find a higher fraction of high EW outliers at faint continuum luminosities. At the brightest continuum bin, $1$ percent of sources have EW$\gtrsim 175$\AA~compared to $\gtrsim 500$\AA~for the faintest continuum bin. This suggests that faint continuum, low-mass \ha~emitters will tend to be high sSFR outliers indicative of systems undergoing bursty star formation activity in respect to bright continuum, high-mass emitters.
	
	\item The shape of the SFR -- stellar mass correlation is also found to be dependent on selection where directly fitting the LAGER sources results in a shallower slope compared to using our intrinsic sample assuming $\textrm{\wo} \sim M^{-0.16\pm0.03}$. We find varying the EW limit, which serves as a proxy for the specific SFR, causes an increase/decrease of the normalization. This highlights the importance of selection corrections to resolve contentions in the shape of the SFR `main sequence'.

\end{enumerate}
	
Our results show that the correlation between EW, \ha~luminosity, and stellar mass are not selection-effect driven, although the correlation with stellar mass seems to best represent $z = 0.47$ \ha~emitters as it can reproduce all three major statistical properties of star-forming galaxies: LF, SMF, and EW distribution. Future investigation of what processes contribute to the shape of these correlations would be of great importance in understanding the underlying physics via observations and simulations. The results shown also are useful for survey planning of future EW-limited grism surveys such as those planned with {\it NGRST}.

\section*{Acknowledgments}
 AAK is supported by an appointment to the NASA Postdoctoral Program at the Goddard Space Flight Center, administered by the Universities Space Research Association (USRA) through a contract with NASA. J.X.W. is supported by National Science Foundation of China (grants No. 11421303 \& 11890693) and CAS Frontier Science Key Research Program (QYZDJ-SSW-SLH006). Z.Y.Z. thanks the National Science Foundation of China (11773051, 12022303) and the CAS Pioneer Hundred Talents Program.
 
 This project used data obtained with the Dark Energy Camera (DECam), which was constructed by the Dark Energy Survey (DES) collaboration. Funding for the DES Projects has been provided by the U.S. Department of Energy, the U.S. National Science Foundation, the Ministry of Science and Education of Spain, the Science and Technology Facilities Council of the United Kingdom, the Higher Education Funding Council for England, the National Center for Supercomputing Applications at the University of Illinois at Urbana-Champaign, the Kavli Institute of Cosmological Physics at the University of Chicago, Center for Cosmology and Astro-Particle Physics at the Ohio State University, the Mitchell Institute for Fundamental Physics and Astronomy at Texas A\&M University, Financiadora de Estudos e Projetos, Funda\c{c}\~{a}o Carlos Chagas Filho de Amparo, Financiadora de Estudos e Projetos, Funda\c{c}\~{a}o Carlos Chagas Filho de Amparo \`{a} Pesquisa do Estado do Rio de Janeiro, Conselho Nacional de Desenvolvimento Cient\'{i}fico e Tecnol\'{o}gico and the Minist\'{e}rio da Ci\^{e}ncia, Tecnologia e Inova\c{c}\~{a}o, the Deutsche Forschungsgemeinschaft and the Collaborating Institutions in the Dark Energy Survey. 
 
 The Collaborating Institutions are Argonne National Laboratory, the University of California at Santa Cruz, the University of Cambridge, Centro de Investigaciones En\'{e}rgeticas, Medioambientales y Tecnol\'{o}gicas-Madrid, the University of Chicago, University College London, the DES-Brazil Consortium, the University of Edinburgh, the Eidgen\"{o}ssische Technische Hochschule (ETH) Z\"{u}rich, Fermi National Accelerator Laboratory, the University of Illinois at Urbana-Champaign, the Institut de Ci\`{e}ncies de l'Espai (IEEC/CSIC), the Institut de F\'{i}sica d'Altes Energies, Lawrence Berkeley National Laboratory, the Ludwig-Maximilians Universit\"{a}t M\"{u}nchen and the associated Excellence Cluster Universe, the University of Michigan, the National Optical Astronomy Observatory, the University of Nottingham, the Ohio State University, the OzDES Membership Consortium, the University of Pennsylvania, the University of Portsmouth, SLAC National Accelerator Laboratory, Stanford University, the University of Sussex, and Texas A\&M University. 
 
 Based on observations at Cerro Tololo Inter-American Observatory, National Optical Astronomy Observatory (NOAO 2017A-0366, 2017B-0330, 2018A-0371, 2018B-0327; PI: S.~Malhotra; NOAO 2019B-1008, PI: L.~F.~Barrientos), which is operated by the Association of Universities for Research in Astronomy (AURA) under a cooperative agreement with the National Science Foundation.

\section*{Data Availability}
Data used in this study comes from CTIO/DECam NB964 and $z$ imaging as part of the LAGER survey. Raw images are publicly available in the NOAO Science Archives. 

\bibliography{EW_modeling}

\begin{thebibliography}{}
\makeatletter
\relax
\def\mn@urlcharsother{\let\do\@makeother \do\$\do\&\do\#\do\^\do\_\do\%\do\~}
\def\mn@doi{\begingroup\mn@urlcharsother \@ifnextchar [ {\mn@doi@}
  {\mn@doi@[]}}
\def\mn@doi@[#1]#2{\def\@tempa{#1}\ifx\@tempa\@empty \href
  {http://dx.doi.org/#2} {doi:#2}\else \href {http://dx.doi.org/#2} {#1}\fi
  \endgroup}
\def\mn@eprint#1#2{\mn@eprint@#1:#2::\@nil}
\def\mn@eprint@arXiv#1{\href {http://arxiv.org/abs/#1} {{\tt arXiv:#1}}}
\def\mn@eprint@dblp#1{\href {http://dblp.uni-trier.de/rec/bibtex/#1.xml}
  {dblp:#1}}
\def\mn@eprint@#1:#2:#3:#4\@nil{\def\@tempa {#1}\def\@tempb {#2}\def\@tempc
  {#3}\ifx \@tempc \@empty \let \@tempc \@tempb \let \@tempb \@tempa \fi \ifx
  \@tempb \@empty \def\@tempb {arXiv}\fi \@ifundefined
  {mn@eprint@\@tempb}{\@tempb:\@tempc}{\expandafter \expandafter \csname
  mn@eprint@\@tempb\endcsname \expandafter{\@tempc}}}

\bibitem[\protect\citeauthoryear{{Abbott} et~al.,}{{Abbott}
  et~al.}{2018}]{Abbott2018}
{Abbott} T.~M.~C.,  et~al., 2018, \mn@doi [\apjs] {10.3847/1538-4365/aae9f0},
  \href {https://ui.adsabs.harvard.edu/abs/2018ApJS..239...18A} {239, 18}

\bibitem[\protect\citeauthoryear{{Ando}, {Ohta}, {Iwata}, {Akiyama}, {Aoki}  \&
  {Tamura}}{{Ando} et~al.}{2006}]{Ando2006}
{Ando} M.,  {Ohta} K.,  {Iwata} I.,  {Akiyama} M.,  {Aoki} K.,   {Tamura} N.,
  2006, \mn@doi [\apjl] {10.1086/505652}, \href
  {https://ui.adsabs.harvard.edu/abs/2006ApJ...645L...9A} {645, L9}

\bibitem[\protect\citeauthoryear{{Atek} et~al.,}{{Atek}
  et~al.}{2011}]{Atek2011}
{Atek} H.,  et~al., 2011, \mn@doi [\apj] {10.1088/0004-637X/743/2/121}, \href
  {https://ui.adsabs.harvard.edu/abs/2011ApJ...743..121A} {743, 121}

\bibitem[\protect\citeauthoryear{{Atek} et~al.,}{{Atek}
  et~al.}{2014}]{Atek2014}
{Atek} H.,  et~al., 2014, \mn@doi [\apj] {10.1088/0004-637X/789/2/96}, \href
  {https://ui.adsabs.harvard.edu/abs/2014ApJ...789...96A} {789, 96}

\bibitem[\protect\citeauthoryear{{Balogh} et~al.,}{{Balogh}
  et~al.}{2014}]{Balogh2014}
{Balogh} M.~L.,  et~al., 2014, \mn@doi [\mnras] {10.1093/mnras/stu1332}, \href
  {https://ui.adsabs.harvard.edu/abs/2014MNRAS.443.2679B} {443, 2679}

\bibitem[\protect\citeauthoryear{{Boogaard} et~al.,}{{Boogaard}
  et~al.}{2018}]{Boogaard2018}
{Boogaard} L.~A.,  et~al., 2018, \mn@doi [\aap] {10.1051/0004-6361/201833136},
  \href {https://ui.adsabs.harvard.edu/abs/2018A&A...619A..27B} {619, A27}

\bibitem[\protect\citeauthoryear{{Bouwens} et~al.,}{{Bouwens}
  et~al.}{2012}]{Bouwens2012}
{Bouwens} R.~J.,  et~al., 2012, \mn@doi [\apj] {10.1088/0004-637X/754/2/83},
  \href {https://ui.adsabs.harvard.edu/abs/2012ApJ...754...83B} {754, 83}

\bibitem[\protect\citeauthoryear{{Brammer} et~al.,}{{Brammer}
  et~al.}{2012}]{Brammer2012}
{Brammer} G.~B.,  et~al., 2012, \mn@doi [\apjs] {10.1088/0067-0049/200/2/13},
  \href {https://ui.adsabs.harvard.edu/abs/2012ApJS..200...13B} {200, 13}

\bibitem[\protect\citeauthoryear{{Brinchmann}, {Charlot}, {White}, {Tremonti},
  {Kauffmann}, {Heckman}  \& {Brinkmann}}{{Brinchmann}
  et~al.}{2004}]{Brinchmann2004}
{Brinchmann} J.,  {Charlot} S.,  {White} S.~D.~M.,  {Tremonti} C.,  {Kauffmann}
  G.,  {Heckman} T.,   {Brinkmann} J.,  2004, \mn@doi [\mnras]
  {10.1111/j.1365-2966.2004.07881.x}, \href
  {https://ui.adsabs.harvard.edu/abs/2004MNRAS.351.1151B} {351, 1151}

\bibitem[\protect\citeauthoryear{{Broussard} et~al.,}{{Broussard}
  et~al.}{2019}]{Broussard2019}
{Broussard} A.,  et~al., 2019, \mn@doi [\apj] {10.3847/1538-4357/ab04ad}, \href
  {https://ui.adsabs.harvard.edu/abs/2019ApJ...873...74B} {873, 74}

\bibitem[\protect\citeauthoryear{{Bunker}, {Warren}, {Hewett}  \&
  {Clements}}{{Bunker} et~al.}{1995}]{Bunker1995}
{Bunker} A.~J.,  {Warren} S.~J.,  {Hewett} P.~C.,   {Clements} D.~L.,  1995,
  \mn@doi [\mnras] {10.1093/mnras/273.2.513}, \href
  {https://ui.adsabs.harvard.edu/abs/1995MNRAS.273..513B} {273, 513}

\bibitem[\protect\citeauthoryear{{Calabr{\`o}} et~al.,}{{Calabr{\`o}}
  et~al.}{2017}]{Calabro2017}
{Calabr{\`o}} A.,  et~al., 2017, \mn@doi [\aap] {10.1051/0004-6361/201629762},
  \href {https://ui.adsabs.harvard.edu/abs/2017A&A...601A..95C} {601, A95}

\bibitem[\protect\citeauthoryear{{Caputi} et~al.,}{{Caputi}
  et~al.}{2017}]{Caputi2017}
{Caputi} K.~I.,  et~al., 2017, \mn@doi [\apj] {10.3847/1538-4357/aa901e}, \href
  {https://ui.adsabs.harvard.edu/abs/2017ApJ...849...45C} {849, 45}

\bibitem[\protect\citeauthoryear{{Chabrier}}{{Chabrier}}{2003}]{Chabrier2003}
{Chabrier} G.,  2003, \mn@doi [\pasp] {10.1086/376392}, \href
  {https://ui.adsabs.harvard.edu/abs/2003PASP..115..763C} {115, 763}

\bibitem[\protect\citeauthoryear{{Ciardullo} et~al.,}{{Ciardullo}
  et~al.}{2012}]{Ciardullo2012}
{Ciardullo} R.,  et~al., 2012, \mn@doi [\apj] {10.1088/0004-637X/744/2/110},
  \href {https://ui.adsabs.harvard.edu/abs/2012ApJ...744..110C} {744, 110}

\bibitem[\protect\citeauthoryear{{Cochrane}, {Best}, {Sobral}, {Smail},
  {Geach}, {Stott}  \& {Wake}}{{Cochrane} et~al.}{2018}]{Cochrane2018}
{Cochrane} R.~K.,  {Best} P.~N.,  {Sobral} D.,  {Smail} I.,  {Geach} J.~E.,
  {Stott} J.~P.,   {Wake} D.~A.,  2018, \mn@doi [\mnras]
  {10.1093/mnras/stx3345}, \href
  {https://ui.adsabs.harvard.edu/abs/2018MNRAS.475.3730C} {475, 3730}

\bibitem[\protect\citeauthoryear{{Comparat} et~al.,}{{Comparat}
  et~al.}{2015}]{Comparat2015}
{Comparat} J.,  et~al., 2015, \mn@doi [\aap] {10.1051/0004-6361/201424767},
  \href {https://ui.adsabs.harvard.edu/abs/2015A&A...575A..40C} {575, A40}

\bibitem[\protect\citeauthoryear{{Cool} et~al.,}{{Cool}
  et~al.}{2013}]{Cool2013}
{Cool} R.~J.,  et~al., 2013, \mn@doi [\apj] {10.1088/0004-637X/767/2/118},
  \href {https://ui.adsabs.harvard.edu/abs/2013ApJ...767..118C} {767, 118}

\bibitem[\protect\citeauthoryear{{Cowie}, {Barger}  \& {Hu}}{{Cowie}
  et~al.}{2010}]{Cowie2010}
{Cowie} L.~L.,  {Barger} A.~J.,   {Hu} E.~M.,  2010, \mn@doi [\apj]
  {10.1088/0004-637X/711/2/928}, \href
  {https://ui.adsabs.harvard.edu/abs/2010ApJ...711..928C} {711, 928}

\bibitem[\protect\citeauthoryear{{Daddi} et~al.,}{{Daddi}
  et~al.}{2007}]{Daddi2007}
{Daddi} E.,  et~al., 2007, \mn@doi [\apj] {10.1086/521818}, \href
  {https://ui.adsabs.harvard.edu/abs/2007ApJ...670..156D} {670, 156}

\bibitem[\protect\citeauthoryear{{Damen}, {Labb{\'e}}, {Franx}, {van Dokkum},
  {Taylor}  \& {Gawiser}}{{Damen} et~al.}{2009}]{Damen2009}
{Damen} M.,  {Labb{\'e}} I.,  {Franx} M.,  {van Dokkum} P.~G.,  {Taylor} E.~N.,
    {Gawiser} E.~J.,  2009, \mn@doi [\apj] {10.1088/0004-637X/690/1/937}, \href
  {https://ui.adsabs.harvard.edu/abs/2009ApJ...690..937D} {690, 937}

\bibitem[\protect\citeauthoryear{{Dannerbauer} et~al.,}{{Dannerbauer}
  et~al.}{2014}]{Dannerbauer2014}
{Dannerbauer} H.,  et~al., 2014, \mn@doi [\aap] {10.1051/0004-6361/201423771},
  \href {https://ui.adsabs.harvard.edu/abs/2014A&A...570A..55D} {570, A55}

\bibitem[\protect\citeauthoryear{{Davidzon} et~al.,}{{Davidzon}
  et~al.}{2017}]{Davidzon2017}
{Davidzon} I.,  et~al., 2017, \mn@doi [\aap] {10.1051/0004-6361/201730419},
  \href {https://ui.adsabs.harvard.edu/abs/2017A&A...605A..70D} {605, A70}

\bibitem[\protect\citeauthoryear{{Davies} et~al.,}{{Davies}
  et~al.}{2019}]{Davies2019}
{Davies} L.~J.~M.,  et~al., 2019, \mn@doi [\mnras] {10.1093/mnras/sty2957},
  \href {https://ui.adsabs.harvard.edu/abs/2019MNRAS.483.1881D} {483, 1881}

\bibitem[\protect\citeauthoryear{{Deharveng} et~al.,}{{Deharveng}
  et~al.}{2008}]{Deharveng2008}
{Deharveng} J.-M.,  et~al., 2008, \mn@doi [\apj] {10.1086/587953}, \href
  {https://ui.adsabs.harvard.edu/abs/2008ApJ...680.1072D} {680, 1072}

\bibitem[\protect\citeauthoryear{{Dekel} et~al.,}{{Dekel}
  et~al.}{2009}]{Dekel2009}
{Dekel} A.,  et~al., 2009, \mn@doi [\nat] {10.1038/nature07648}, \href
  {https://ui.adsabs.harvard.edu/abs/2009Natur.457..451D} {457, 451}

\bibitem[\protect\citeauthoryear{{Dom{\'\i}nguez} et~al.,}{{Dom{\'\i}nguez}
  et~al.}{2013}]{Dominguez2013}
{Dom{\'\i}nguez} A.,  et~al., 2013, \mn@doi [\apj]
  {10.1088/0004-637X/763/2/145}, \href
  {https://ui.adsabs.harvard.edu/abs/2013ApJ...763..145D} {763, 145}

\bibitem[\protect\citeauthoryear{{Dom{\'\i}nguez}, {Siana}, {Brooks},
  {Christensen}, {Bruzual}, {Stark}  \& {Alavi}}{{Dom{\'\i}nguez}
  et~al.}{2015}]{Dominguez2015}
{Dom{\'\i}nguez} A.,  {Siana} B.,  {Brooks} A.~M.,  {Christensen} C.~R.,
  {Bruzual} G.,  {Stark} D.~P.,   {Alavi} A.,  2015, \mn@doi [\mnras]
  {10.1093/mnras/stv1001}, \href
  {https://ui.adsabs.harvard.edu/abs/2015MNRAS.451..839D} {451, 839}

\bibitem[\protect\citeauthoryear{{Elbaz} et~al.,}{{Elbaz}
  et~al.}{2007}]{Elbaz2007}
{Elbaz} D.,  et~al., 2007, \mn@doi [\aap] {10.1051/0004-6361:20077525}, \href
  {https://ui.adsabs.harvard.edu/abs/2007A&A...468...33E} {468, 33}

\bibitem[\protect\citeauthoryear{{Eldridge}}{{Eldridge}}{2012}]{Eldridge2012}
{Eldridge} J.~J.,  2012, \mn@doi [\mnras] {10.1111/j.1365-2966.2012.20662.x},
  \href {https://ui.adsabs.harvard.edu/abs/2012MNRAS.422..794E} {422, 794}

\bibitem[\protect\citeauthoryear{{Emami}, {Siana}, {Weisz}, {Johnson}, {Ma}  \&
  {El-Badry}}{{Emami} et~al.}{2019}]{Emami2019}
{Emami} N.,  {Siana} B.,  {Weisz} D.~R.,  {Johnson} B.~D.,  {Ma} X.,
  {El-Badry} K.,  2019, \mn@doi [\apj] {10.3847/1538-4357/ab211a}, \href
  {https://ui.adsabs.harvard.edu/abs/2019ApJ...881...71E} {881, 71}

\bibitem[\protect\citeauthoryear{{Erb}, {Steidel}, {Shapley}, {Pettini},
  {Reddy}  \& {Adelberger}}{{Erb} et~al.}{2006}]{Erb2006}
{Erb} D.~K.,  {Steidel} C.~C.,  {Shapley} A.~E.,  {Pettini} M.,  {Reddy} N.~A.,
    {Adelberger} K.~L.,  2006, \mn@doi [\apj] {10.1086/505341}, \href
  {https://ui.adsabs.harvard.edu/abs/2006ApJ...647..128E} {647, 128}

\bibitem[\protect\citeauthoryear{{Faisst} et~al.,}{{Faisst}
  et~al.}{2016}]{Faisst2016}
{Faisst} A.~L.,  et~al., 2016, \mn@doi [\apj] {10.3847/0004-637X/821/2/122},
  \href {https://ui.adsabs.harvard.edu/abs/2016ApJ...821..122F} {821, 122}

\bibitem[\protect\citeauthoryear{{Faisst}, {Capak}, {Emami}, {Tacchella}  \&
  {Larson}}{{Faisst} et~al.}{2019}]{Faisst2019}
{Faisst} A.~L.,  {Capak} P.~L.,  {Emami} N.,  {Tacchella} S.,   {Larson} K.~L.,
   2019, \mn@doi [\apj] {10.3847/1538-4357/ab425b}, \href
  {https://ui.adsabs.harvard.edu/abs/2019ApJ...884..133F} {884, 133}

\bibitem[\protect\citeauthoryear{{Fumagalli} et~al.,}{{Fumagalli}
  et~al.}{2012}]{Fumagalli2012}
{Fumagalli} M.,  et~al., 2012, \mn@doi [\apjl] {10.1088/2041-8205/757/2/L22},
  \href {https://ui.adsabs.harvard.edu/abs/2012ApJ...757L..22F} {757, L22}

\bibitem[\protect\citeauthoryear{{Furusawa} et~al.,}{{Furusawa}
  et~al.}{2016}]{Furusawa2016}
{Furusawa} H.,  et~al., 2016, \mn@doi [\apj] {10.3847/0004-637X/822/1/46},
  \href {https://ui.adsabs.harvard.edu/abs/2016ApJ...822...46F} {822, 46}

\bibitem[\protect\citeauthoryear{{Gardner} et~al.,}{{Gardner}
  et~al.}{2006}]{Gardner2006}
{Gardner} J.~P.,  et~al., 2006, \mn@doi [\ssr] {10.1007/s11214-006-8315-7},
  \href {https://ui.adsabs.harvard.edu/abs/2006SSRv..123..485G} {123, 485}

\bibitem[\protect\citeauthoryear{{Garn} \& {Best}}{{Garn} \&
  {Best}}{2010}]{Garn2010}
{Garn} T.,  {Best} P.~N.,  2010, \mn@doi [\mnras]
  {10.1111/j.1365-2966.2010.17321.x}, \href
  {https://ui.adsabs.harvard.edu/abs/2010MNRAS.409..421G} {409, 421}

\bibitem[\protect\citeauthoryear{{Glazebrook}, {Blake}, {Economou}, {Lilly}  \&
  {Colless}}{{Glazebrook} et~al.}{1999}]{Glazebrook1999}
{Glazebrook} K.,  {Blake} C.,  {Economou} F.,  {Lilly} S.,   {Colless} M.,
  1999, \mn@doi [\mnras] {10.1046/j.1365-8711.1999.02576.x}, \href
  {https://ui.adsabs.harvard.edu/abs/1999MNRAS.306..843G} {306, 843}

\bibitem[\protect\citeauthoryear{{Gonz{\'a}lez}, {Labb{\'e}}, {Bouwens},
  {Illingworth}, {Franx}, {Kriek}  \& {Brammer}}{{Gonz{\'a}lez}
  et~al.}{2010}]{Gonzalez2010}
{Gonz{\'a}lez} V.,  {Labb{\'e}} I.,  {Bouwens} R.~J.,  {Illingworth} G.,
  {Franx} M.,  {Kriek} M.,   {Brammer} G.~B.,  2010, \mn@doi [\apj]
  {10.1088/0004-637X/713/1/115}, \href
  {https://ui.adsabs.harvard.edu/abs/2010ApJ...713..115G} {713, 115}

\bibitem[\protect\citeauthoryear{{Gonz{\'a}lez}, {Bouwens}, {Illingworth},
  {Labb{\'e}}, {Oesch}, {Franx}  \& {Magee}}{{Gonz{\'a}lez}
  et~al.}{2014}]{Gonzalez2014}
{Gonz{\'a}lez} V.,  {Bouwens} R.,  {Illingworth} G.,  {Labb{\'e}} I.,  {Oesch}
  P.,  {Franx} M.,   {Magee} D.,  2014, \mn@doi [\apj]
  {10.1088/0004-637X/781/1/34}, \href
  {https://ui.adsabs.harvard.edu/abs/2014ApJ...781...34G} {781, 34}

\bibitem[\protect\citeauthoryear{{Gronwall} et~al.,}{{Gronwall}
  et~al.}{2007}]{Gronwall2007}
{Gronwall} C.,  et~al., 2007, \mn@doi [\apj] {10.1086/520324}, \href
  {https://ui.adsabs.harvard.edu/abs/2007ApJ...667...79G} {667, 79}

\bibitem[\protect\citeauthoryear{{Guaita} et~al.,}{{Guaita}
  et~al.}{2010}]{Guaita2010}
{Guaita} L.,  et~al., 2010, \mn@doi [\apj] {10.1088/0004-637X/714/1/255}, \href
  {https://ui.adsabs.harvard.edu/abs/2010ApJ...714..255G} {714, 255}

\bibitem[\protect\citeauthoryear{{Guo} et~al.,}{{Guo} et~al.}{2016}]{Guo2016}
{Guo} Y.,  et~al., 2016, \mn@doi [\apj] {10.3847/1538-4357/833/1/37}, \href
  {https://ui.adsabs.harvard.edu/abs/2016ApJ...833...37G} {833, 37}

\bibitem[\protect\citeauthoryear{{Harikane} et~al.,}{{Harikane}
  et~al.}{2016}]{Harikane2016}
{Harikane} Y.,  et~al., 2016, \mn@doi [\apj] {10.3847/0004-637X/821/2/123},
  \href {https://ui.adsabs.harvard.edu/abs/2016ApJ...821..123H} {821, 123}

\bibitem[\protect\citeauthoryear{{Harikane} et~al.,}{{Harikane}
  et~al.}{2018}]{Harikane2018}
{Harikane} Y.,  et~al., 2018, \mn@doi [\pasj] {10.1093/pasj/psx097}, \href
  {https://ui.adsabs.harvard.edu/abs/2018PASJ...70S..11H} {70, S11}

\bibitem[\protect\citeauthoryear{{Hashimoto} et~al.,}{{Hashimoto}
  et~al.}{2017}]{Hashimoto2017}
{Hashimoto} T.,  et~al., 2017, \mn@doi [\aap] {10.1051/0004-6361/201731579},
  \href {https://ui.adsabs.harvard.edu/abs/2017A&A...608A..10H} {608, A10}

\bibitem[\protect\citeauthoryear{{Hasinger} et~al.,}{{Hasinger}
  et~al.}{2018}]{Hasinger2018}
{Hasinger} G.,  et~al., 2018, \mn@doi [\apj] {10.3847/1538-4357/aabacf}, \href
  {http://adsabs.harvard.edu/abs/2018ApJ...858...77H} {858, 77}

\bibitem[\protect\citeauthoryear{{Heinis} et~al.,}{{Heinis}
  et~al.}{2014}]{Heinis2014}
{Heinis} S.,  et~al., 2014, \mn@doi [\mnras] {10.1093/mnras/stt1960}, \href
  {https://ui.adsabs.harvard.edu/abs/2014MNRAS.437.1268H} {437, 1268}

\bibitem[\protect\citeauthoryear{{Hildebrandt}, {Pielorz}, {Erben}, {van
  Waerbeke}, {Simon}  \& {Capak}}{{Hildebrandt} et~al.}{2009}]{Hildebrandt2009}
{Hildebrandt} H.,  {Pielorz} J.,  {Erben} T.,  {van Waerbeke} L.,  {Simon} P.,
   {Capak} P.,  2009, \mn@doi [\aap] {10.1051/0004-6361/200811042}, \href
  {https://ui.adsabs.harvard.edu/abs/2009A&A...498..725H} {498, 725}

\bibitem[\protect\citeauthoryear{{Hopkins}, {Kere{\v{s}}}, {O{\~n}orbe},
  {Faucher-Gigu{\`e}re}, {Quataert}, {Murray}  \& {Bullock}}{{Hopkins}
  et~al.}{2014}]{Hopkins2014}
{Hopkins} P.~F.,  {Kere{\v{s}}} D.,  {O{\~n}orbe} J.,  {Faucher-Gigu{\`e}re}
  C.-A.,  {Quataert} E.,  {Murray} N.,   {Bullock} J.~S.,  2014, \mn@doi
  [\mnras] {10.1093/mnras/stu1738}, \href
  {https://ui.adsabs.harvard.edu/abs/2014MNRAS.445..581H} {445, 581}

\bibitem[\protect\citeauthoryear{{Hoversten} \& {Glazebrook}}{{Hoversten} \&
  {Glazebrook}}{2008}]{Hoversten2008}
{Hoversten} E.~A.,  {Glazebrook} K.,  2008, \mn@doi [\apj] {10.1086/524095},
  \href {https://ui.adsabs.harvard.edu/abs/2008ApJ...675..163H} {675, 163}

\bibitem[\protect\citeauthoryear{{Hu} et~al.,}{{Hu} et~al.}{2019}]{Hu2019}
{Hu} W.,  et~al., 2019, \mn@doi [\apj] {10.3847/1538-4357/ab4cf4}, \href
  {https://ui.adsabs.harvard.edu/abs/2019ApJ...886...90H} {886, 90}

\bibitem[\protect\citeauthoryear{{Iglesias-P{\'a}ramo}, {Boselli}, {Gavazzi}
  \& {Zaccardo}}{{Iglesias-P{\'a}ramo} et~al.}{2004}]{Iglesias2004}
{Iglesias-P{\'a}ramo} J.,  {Boselli} A.,  {Gavazzi} G.,   {Zaccardo} A.,  2004,
  \mn@doi [\aap] {10.1051/0004-6361:20034572}, \href
  {https://ui.adsabs.harvard.edu/abs/2004A&A...421..887I} {421, 887}

\bibitem[\protect\citeauthoryear{{Ilbert} et~al.,}{{Ilbert}
  et~al.}{2013}]{Ilbert2013}
{Ilbert} O.,  et~al., 2013, \mn@doi [\aap] {10.1051/0004-6361/201321100}, \href
  {https://ui.adsabs.harvard.edu/abs/2013A&A...556A..55I} {556, A55}

\bibitem[\protect\citeauthoryear{{Ivezi{\'c}} et~al.,}{{Ivezi{\'c}}
  et~al.}{2019}]{Ivezic2019}
{Ivezi{\'c}} {\v{Z}}.,  et~al., 2019, \mn@doi [\apj]
  {10.3847/1538-4357/ab042c}, \href
  {https://ui.adsabs.harvard.edu/abs/2019ApJ...873..111I} {873, 111}

\bibitem[\protect\citeauthoryear{{Jiang} et~al.,}{{Jiang}
  et~al.}{2016}]{Jiang2016}
{Jiang} L.,  et~al., 2016, \mn@doi [\apj] {10.3847/0004-637X/816/1/16}, \href
  {https://ui.adsabs.harvard.edu/abs/2016ApJ...816...16J} {816, 16}

\bibitem[\protect\citeauthoryear{{Juneau} et~al.,}{{Juneau}
  et~al.}{2005}]{Juneau2005}
{Juneau} S.,  et~al., 2005, \mn@doi [\apjl] {10.1086/427937}, \href
  {https://ui.adsabs.harvard.edu/abs/2005ApJ...619L.135J} {619, L135}

\bibitem[\protect\citeauthoryear{{Karim} et~al.,}{{Karim}
  et~al.}{2011}]{Karim2011}
{Karim} A.,  et~al., 2011, \mn@doi [\apj] {10.1088/0004-637X/730/2/61}, \href
  {https://ui.adsabs.harvard.edu/abs/2011ApJ...730...61K} {730, 61}

\bibitem[\protect\citeauthoryear{{Kashikawa} et~al.,}{{Kashikawa}
  et~al.}{2006}]{Kashikawa2006}
{Kashikawa} N.,  et~al., 2006, \mn@doi [\apj] {10.1086/498403}, \href
  {https://ui.adsabs.harvard.edu/abs/2006ApJ...637..631K} {637, 631}

\bibitem[\protect\citeauthoryear{{Kashikawa} et~al.,}{{Kashikawa}
  et~al.}{2011}]{Kashikawa2011}
{Kashikawa} N.,  et~al., 2011, \mn@doi [\apj] {10.1088/0004-637X/734/2/119},
  \href {https://ui.adsabs.harvard.edu/abs/2011ApJ...734..119K} {734, 119}

\bibitem[\protect\citeauthoryear{{Kashino} et~al.,}{{Kashino}
  et~al.}{2013}]{Kashino2013}
{Kashino} D.,  et~al., 2013, \mn@doi [\apjl] {10.1088/2041-8205/777/1/L8},
  \href {https://ui.adsabs.harvard.edu/abs/2013ApJ...777L...8K} {777, L8}

\bibitem[\protect\citeauthoryear{{Katsianis} et~al.,}{{Katsianis}
  et~al.}{2019}]{Katsianis2019}
{Katsianis} A.,  et~al., 2019, \mn@doi [\apj] {10.3847/1538-4357/ab1f8d}, \href
  {https://ui.adsabs.harvard.edu/abs/2019ApJ...879...11K} {879, 11}

\bibitem[\protect\citeauthoryear{{Kennicutt}}{{Kennicutt}}{1998}]{Kennicutt1998}
{Kennicutt} Robert~C. J.,  1998, \mn@doi [\araa]
  {10.1146/annurev.astro.36.1.189}, \href
  {https://ui.adsabs.harvard.edu/abs/1998ARA&A..36..189K} {36, 189}

\bibitem[\protect\citeauthoryear{{Kewley}, {Geller}, {Jansen}  \&
  {Dopita}}{{Kewley} et~al.}{2002}]{Kewley2002}
{Kewley} L.~J.,  {Geller} M.~J.,  {Jansen} R.~A.,   {Dopita} M.~A.,  2002,
  \mn@doi [\aj] {10.1086/344487}, \href
  {https://ui.adsabs.harvard.edu/abs/2002AJ....124.3135K} {124, 3135}

\bibitem[\protect\citeauthoryear{{Khochfar} \& {Silk}}{{Khochfar} \&
  {Silk}}{2011}]{Khochfar2011}
{Khochfar} S.,  {Silk} J.,  2011, \mn@doi [\mnras]
  {10.1111/j.1745-3933.2010.00976.x}, \href
  {https://ui.adsabs.harvard.edu/abs/2011MNRAS.410L..42K} {410, L42}

\bibitem[\protect\citeauthoryear{{Khostovan}, {Sobral}, {Mobasher}, {Smail},
  {Darvish}, {Nayyeri}, {Hemmati}  \& {Stott}}{{Khostovan}
  et~al.}{2016}]{Khostovan2016}
{Khostovan} A.~A.,  {Sobral} D.,  {Mobasher} B.,  {Smail} I.,  {Darvish} B.,
  {Nayyeri} H.,  {Hemmati} S.,   {Stott} J.~P.,  2016, \mn@doi [\mnras]
  {10.1093/mnras/stw2174}, \href
  {https://ui.adsabs.harvard.edu/abs/2016MNRAS.463.2363K} {463, 2363}

\bibitem[\protect\citeauthoryear{{Khostovan} et~al.,}{{Khostovan}
  et~al.}{2018}]{Khostovan2018}
{Khostovan} A.~A.,  et~al., 2018, \mn@doi [\mnras] {10.1093/mnras/sty925},
  \href {https://ui.adsabs.harvard.edu/abs/2018MNRAS.478.2999K} {478, 2999}

\bibitem[\protect\citeauthoryear{{Khostovan} et~al.,}{{Khostovan}
  et~al.}{2019}]{Khostovan2019}
{Khostovan} A.~A.,  et~al., 2019, \mn@doi [\mnras] {10.1093/mnras/stz2149},
  \href {https://ui.adsabs.harvard.edu/abs/2019MNRAS.489..555K} {489, 555}

\bibitem[\protect\citeauthoryear{{Khostovan} et~al.,}{{Khostovan}
  et~al.}{2020}]{Khostovan2020}
{Khostovan} A.~A.,  et~al., 2020, arXiv e-prints, \href
  {https://ui.adsabs.harvard.edu/abs/2020arXiv200104989K} {p. arXiv:2001.04989}

\bibitem[\protect\citeauthoryear{{Koyama}, {Kodama}, {Tadaki}, {Hayashi},
  {Tanaka}, {Smail}, {Tanaka}  \& {Kurk}}{{Koyama} et~al.}{2013}]{Koyama2013}
{Koyama} Y.,  {Kodama} T.,  {Tadaki} K.-i.,  {Hayashi} M.,  {Tanaka} M.,
  {Smail} I.,  {Tanaka} I.,   {Kurk} J.,  2013, \mn@doi [\mnras]
  {10.1093/mnras/sts133}, \href
  {https://ui.adsabs.harvard.edu/abs/2013MNRAS.428.1551K} {428, 1551}

\bibitem[\protect\citeauthoryear{{Kriek} et~al.,}{{Kriek}
  et~al.}{2015}]{Kriek2015}
{Kriek} M.,  et~al., 2015, \mn@doi [\apjs] {10.1088/0067-0049/218/2/15}, \href
  {https://ui.adsabs.harvard.edu/abs/2015ApJS..218...15K} {218, 15}

\bibitem[\protect\citeauthoryear{{Labb{\'e}} et~al.,}{{Labb{\'e}}
  et~al.}{2013}]{Labbe2013}
{Labb{\'e}} I.,  et~al., 2013, \mn@doi [\apjl] {10.1088/2041-8205/777/2/L19},
  \href {https://ui.adsabs.harvard.edu/abs/2013ApJ...777L..19L} {777, L19}

\bibitem[\protect\citeauthoryear{{Laigle} et~al.,}{{Laigle}
  et~al.}{2016}]{Laigle2016}
{Laigle} C.,  et~al., 2016, \mn@doi [\apjs] {10.3847/0067-0049/224/2/24}, \href
  {https://ui.adsabs.harvard.edu/abs/2016ApJS..224...24L} {224, 24}

\bibitem[\protect\citeauthoryear{{Laureijs} et~al.,}{{Laureijs}
  et~al.}{2011}]{Laureijs2011}
{Laureijs} R.,  et~al., 2011, arXiv e-prints, \href
  {https://ui.adsabs.harvard.edu/abs/2011arXiv1110.3193L} {p. arXiv:1110.3193}

\bibitem[\protect\citeauthoryear{{Lee} et~al.,}{{Lee} et~al.}{2009}]{Lee2009}
{Lee} J.~C.,  et~al., 2009, \mn@doi [\apj] {10.1088/0004-637X/706/1/599}, \href
  {https://ui.adsabs.harvard.edu/abs/2009ApJ...706..599L} {706, 599}

\bibitem[\protect\citeauthoryear{{Lee} et~al.,}{{Lee} et~al.}{2011a}]{Lee2011}
{Lee} J.~C.,  et~al., 2011a, \mn@doi [\apjs] {10.1088/0067-0049/192/1/6}, \href
  {https://ui.adsabs.harvard.edu/abs/2011ApJS..192....6L} {192, 6}

\bibitem[\protect\citeauthoryear{{Lee} et~al.,}{{Lee} et~al.}{2011b}]{Lee2011K}
{Lee} K.-S.,  et~al., 2011b, \mn@doi [\apj] {10.1088/0004-637X/733/2/99}, \href
  {https://ui.adsabs.harvard.edu/abs/2011ApJ...733...99L} {733, 99}

\bibitem[\protect\citeauthoryear{{Lilly} et~al.,}{{Lilly}
  et~al.}{2009}]{Lilly2009}
{Lilly} S.~J.,  et~al., 2009, \mn@doi [\apjs] {10.1088/0067-0049/184/2/218},
  \href {http://adsabs.harvard.edu/abs/2009ApJS..184..218L} {184, 218}

\bibitem[\protect\citeauthoryear{{M{\'a}rmol-Queralt{\'o}}, {McLure}, {Cullen},
  {Dunlop}, {Fontana}  \& {McLeod}}{{M{\'a}rmol-Queralt{\'o}}
  et~al.}{2016}]{Marmol2016}
{M{\'a}rmol-Queralt{\'o}} E.,  {McLure} R.~J.,  {Cullen} F.,  {Dunlop} J.~S.,
  {Fontana} A.,   {McLeod} D.~J.,  2016, \mn@doi [\mnras]
  {10.1093/mnras/stw1212}, \href
  {https://ui.adsabs.harvard.edu/abs/2016MNRAS.460.3587M} {460, 3587}

\bibitem[\protect\citeauthoryear{{Maseda} et~al.,}{{Maseda}
  et~al.}{2014}]{Maseda2014}
{Maseda} M.~V.,  et~al., 2014, \mn@doi [\apj] {10.1088/0004-637X/791/1/17},
  \href {https://ui.adsabs.harvard.edu/abs/2014ApJ...791...17M} {791, 17}

\bibitem[\protect\citeauthoryear{{Masters}, {Stern}, {Cohen}, {Capak},
  {Rhodes}, {Castander}  \& {Paltani}}{{Masters} et~al.}{2017}]{Masters2017}
{Masters} D.~C.,  {Stern} D.~K.,  {Cohen} J.~G.,  {Capak} P.~L.,  {Rhodes}
  J.~D.,  {Castander} F.~J.,   {Paltani} S.,  2017, \mn@doi [\apj]
  {10.3847/1538-4357/aa6f08}, \href
  {http://adsabs.harvard.edu/abs/2017ApJ...841..111M} {841, 111}

\bibitem[\protect\citeauthoryear{{Matthee} \& {Schaye}}{{Matthee} \&
  {Schaye}}{2019}]{Matthee2019}
{Matthee} J.,  {Schaye} J.,  2019, \mn@doi [\mnras] {10.1093/mnras/stz030},
  \href {https://ui.adsabs.harvard.edu/abs/2019MNRAS.484..915M} {484, 915}

\bibitem[\protect\citeauthoryear{{McLinden} et~al.,}{{McLinden}
  et~al.}{2011}]{McLinden2011}
{McLinden} E.~M.,  et~al., 2011, \mn@doi [\apj] {10.1088/0004-637X/730/2/136},
  \href {https://ui.adsabs.harvard.edu/abs/2011ApJ...730..136M} {730, 136}

\bibitem[\protect\citeauthoryear{{Mehta} et~al.,}{{Mehta}
  et~al.}{2017}]{Mehta2017}
{Mehta} V.,  et~al., 2017, \mn@doi [\apj] {10.3847/1538-4357/aa6259}, \href
  {https://ui.adsabs.harvard.edu/abs/2017ApJ...838...29M} {838, 29}

\bibitem[\protect\citeauthoryear{{Meurer} et~al.,}{{Meurer}
  et~al.}{2009}]{Meurer2009}
{Meurer} G.~R.,  et~al., 2009, \mn@doi [\apj] {10.1088/0004-637X/695/1/765},
  \href {https://ui.adsabs.harvard.edu/abs/2009ApJ...695..765M} {695, 765}

\bibitem[\protect\citeauthoryear{{Momcheva} et~al.,}{{Momcheva}
  et~al.}{2016}]{Momcheva2016}
{Momcheva} I.~G.,  et~al., 2016, \mn@doi [\apjs] {10.3847/0067-0049/225/2/27},
  \href {https://ui.adsabs.harvard.edu/abs/2016ApJS..225...27M} {225, 27}

\bibitem[\protect\citeauthoryear{{Muzzin} et~al.,}{{Muzzin}
  et~al.}{2013}]{Muzzin2013}
{Muzzin} A.,  et~al., 2013, \mn@doi [\apj] {10.1088/0004-637X/777/1/18}, \href
  {https://ui.adsabs.harvard.edu/abs/2013ApJ...777...18M} {777, 18}

\bibitem[\protect\citeauthoryear{{Nanayakkara} et~al.,}{{Nanayakkara}
  et~al.}{2017}]{Nanayakkara2017}
{Nanayakkara} T.,  et~al., 2017, \mn@doi [\mnras] {10.1093/mnras/stx605}, \href
  {https://ui.adsabs.harvard.edu/abs/2017MNRAS.468.3071N} {468, 3071}

\bibitem[\protect\citeauthoryear{{Neistein} \& {Dekel}}{{Neistein} \&
  {Dekel}}{2008}]{Neistein2008}
{Neistein} E.,  {Dekel} A.,  2008, \mn@doi [\mnras]
  {10.1111/j.1365-2966.2007.12570.x}, \href
  {https://ui.adsabs.harvard.edu/abs/2008MNRAS.383..615N} {383, 615}

\bibitem[\protect\citeauthoryear{{Nilsson}, {M{\"o}ller-Nilsson}, {M{\o}ller},
  {Fynbo}  \& {Shapley}}{{Nilsson} et~al.}{2009}]{Nilsson2009}
{Nilsson} K.~K.,  {M{\"o}ller-Nilsson} O.,  {M{\o}ller} P.,  {Fynbo} J.~P.~U.,
   {Shapley} A.~E.,  2009, \mn@doi [\mnras] {10.1111/j.1365-2966.2009.15439.x},
  \href {https://ui.adsabs.harvard.edu/abs/2009MNRAS.400..232N} {400, 232}

\bibitem[\protect\citeauthoryear{{Noeske} et~al.,}{{Noeske}
  et~al.}{2007}]{Noeske2007}
{Noeske} K.~G.,  et~al., 2007, \mn@doi [\apjl] {10.1086/517926}, \href
  {https://ui.adsabs.harvard.edu/abs/2007ApJ...660L..43N} {660, L43}

\bibitem[\protect\citeauthoryear{{Ota} et~al.,}{{Ota} et~al.}{2017}]{Ota2017}
{Ota} K.,  et~al., 2017, \mn@doi [\apj] {10.3847/1538-4357/aa7a0a}, \href
  {https://ui.adsabs.harvard.edu/abs/2017ApJ...844...85O} {844, 85}

\bibitem[\protect\citeauthoryear{{Ouchi} et~al.,}{{Ouchi}
  et~al.}{2003}]{Ouchi2003}
{Ouchi} M.,  et~al., 2003, \mn@doi [\apj] {10.1086/344476}, \href
  {https://ui.adsabs.harvard.edu/abs/2003ApJ...582...60O} {582, 60}

\bibitem[\protect\citeauthoryear{{Ouchi} et~al.,}{{Ouchi}
  et~al.}{2008}]{Ouchi2008}
{Ouchi} M.,  et~al., 2008, \mn@doi [\apjs] {10.1086/527673}, \href
  {https://ui.adsabs.harvard.edu/abs/2008ApJS..176..301O} {176, 301}

\bibitem[\protect\citeauthoryear{{Overzier}}{{Overzier}}{2016}]{Overzier2016}
{Overzier} R.~A.,  2016, \mn@doi [\aapr] {10.1007/s00159-016-0100-3}, \href
  {https://ui.adsabs.harvard.edu/abs/2016A&ARv..24...14O} {24, 14}

\bibitem[\protect\citeauthoryear{{Oyarz{\'u}n} et~al.,}{{Oyarz{\'u}n}
  et~al.}{2016}]{Oyarzun2016}
{Oyarz{\'u}n} G.~A.,  et~al., 2016, \mn@doi [\apjl]
  {10.3847/2041-8205/821/1/L14}, \href
  {https://ui.adsabs.harvard.edu/abs/2016ApJ...821L..14O} {821, L14}

\bibitem[\protect\citeauthoryear{{Oyarz{\'u}n}, {Blanc}, {Gonz{\'a}lez},
  {Mateo}  \& {Bailey}}{{Oyarz{\'u}n} et~al.}{2017}]{Oyarzun2017}
{Oyarz{\'u}n} G.~A.,  {Blanc} G.~A.,  {Gonz{\'a}lez} V.,  {Mateo} M.,
  {Bailey} John~I. I.,  2017, \mn@doi [\apj] {10.3847/1538-4357/aa7552}, \href
  {https://ui.adsabs.harvard.edu/abs/2017ApJ...843..133O} {843, 133}

\bibitem[\protect\citeauthoryear{{Pirzkal} et~al.,}{{Pirzkal}
  et~al.}{2018}]{Pirzkal2018}
{Pirzkal} N.,  et~al., 2018, \mn@doi [\apj] {10.3847/1538-4357/aae585}, \href
  {https://ui.adsabs.harvard.edu/abs/2018ApJ...868...61P} {868, 61}

\bibitem[\protect\citeauthoryear{{Puglisi} et~al.,}{{Puglisi}
  et~al.}{2016}]{Puglisi2016}
{Puglisi} A.,  et~al., 2016, \mn@doi [\aap] {10.1051/0004-6361/201526782},
  \href {https://ui.adsabs.harvard.edu/abs/2016A&A...586A..83P} {586, A83}

\bibitem[\protect\citeauthoryear{{Rasappu}, {Smit}, {Labb{\'e}}, {Bouwens},
  {Stark}, {Ellis}  \& {Oesch}}{{Rasappu} et~al.}{2016}]{Rasappu2016}
{Rasappu} N.,  {Smit} R.,  {Labb{\'e}} I.,  {Bouwens} R.~J.,  {Stark} D.~P.,
  {Ellis} R.~S.,   {Oesch} P.~A.,  2016, \mn@doi [\mnras]
  {10.1093/mnras/stw1484}, \href
  {https://ui.adsabs.harvard.edu/abs/2016MNRAS.461.3886R} {461, 3886}

\bibitem[\protect\citeauthoryear{{Reddy}, {Pettini}, {Steidel}, {Shapley},
  {Erb}  \& {Law}}{{Reddy} et~al.}{2012}]{Reddy2012}
{Reddy} N.~A.,  {Pettini} M.,  {Steidel} C.~C.,  {Shapley} A.~E.,  {Erb} D.~K.,
    {Law} D.~R.,  2012, \mn@doi [\apj] {10.1088/0004-637X/754/1/25}, \href
  {https://ui.adsabs.harvard.edu/abs/2012ApJ...754...25R} {754, 25}

\bibitem[\protect\citeauthoryear{{Reddy} et~al.,}{{Reddy}
  et~al.}{2015}]{Reddy2015}
{Reddy} N.~A.,  et~al., 2015, \mn@doi [\apj] {10.1088/0004-637X/806/2/259},
  \href {https://ui.adsabs.harvard.edu/abs/2015ApJ...806..259R} {806, 259}

\bibitem[\protect\citeauthoryear{{Reddy} et~al.,}{{Reddy}
  et~al.}{2018}]{Reddy2018}
{Reddy} N.~A.,  et~al., 2018, \mn@doi [\apj] {10.3847/1538-4357/aaed1e}, \href
  {https://ui.adsabs.harvard.edu/abs/2018ApJ...869...92R} {869, 92}

\bibitem[\protect\citeauthoryear{{Salim} et~al.,}{{Salim}
  et~al.}{2007}]{Salim2007}
{Salim} S.,  et~al., 2007, \mn@doi [\apjs] {10.1086/519218}, \href
  {https://ui.adsabs.harvard.edu/abs/2007ApJS..173..267S} {173, 267}

\bibitem[\protect\citeauthoryear{{Salmon} et~al.,}{{Salmon}
  et~al.}{2015}]{Salmon2015}
{Salmon} B.,  et~al., 2015, \mn@doi [\apj] {10.1088/0004-637X/799/2/183}, \href
  {https://ui.adsabs.harvard.edu/abs/2015ApJ...799..183S} {799, 183}

\bibitem[\protect\citeauthoryear{{Santini} et~al.,}{{Santini}
  et~al.}{2017}]{Santini2017}
{Santini} P.,  et~al., 2017, \mn@doi [\apj] {10.3847/1538-4357/aa8874}, \href
  {https://ui.adsabs.harvard.edu/abs/2017ApJ...847...76S} {847, 76}

\bibitem[\protect\citeauthoryear{{Santos} et~al.,}{{Santos}
  et~al.}{2020}]{Santos2020}
{Santos} S.,  et~al., 2020, \mn@doi [\mnras] {10.1093/mnras/staa093}, \href
  {https://ui.adsabs.harvard.edu/abs/2020MNRAS.493..141S} {493, 141}

\bibitem[\protect\citeauthoryear{{Schenker}, {Ellis}, {Konidaris}  \&
  {Stark}}{{Schenker} et~al.}{2013}]{Schenker2013}
{Schenker} M.~A.,  {Ellis} R.~S.,  {Konidaris} N.~P.,   {Stark} D.~P.,  2013,
  \mn@doi [\apj] {10.1088/0004-637X/777/1/67}, \href
  {https://ui.adsabs.harvard.edu/abs/2013ApJ...777...67S} {777, 67}

\bibitem[\protect\citeauthoryear{{Schreiber} et~al.,}{{Schreiber}
  et~al.}{2015}]{Schreiber2015}
{Schreiber} C.,  et~al., 2015, \mn@doi [\aap] {10.1051/0004-6361/201425017},
  \href {https://ui.adsabs.harvard.edu/abs/2015A&A...575A..74S} {575, A74}

\bibitem[\protect\citeauthoryear{{Shapley}, {Steidel}, {Pettini}  \&
  {Adelberger}}{{Shapley} et~al.}{2003}]{Shapley2003}
{Shapley} A.~E.,  {Steidel} C.~C.,  {Pettini} M.,   {Adelberger} K.~L.,  2003,
  \mn@doi [\apj] {10.1086/373922}, \href
  {https://ui.adsabs.harvard.edu/abs/2003ApJ...588...65S} {588, 65}

\bibitem[\protect\citeauthoryear{{Shim}, {Chary}, {Dickinson}, {Lin},
  {Spinrad}, {Stern}  \& {Yan}}{{Shim} et~al.}{2011}]{Shim2011}
{Shim} H.,  {Chary} R.-R.,  {Dickinson} M.,  {Lin} L.,  {Spinrad} H.,  {Stern}
  D.,   {Yan} C.-H.,  2011, \mn@doi [\apj] {10.1088/0004-637X/738/1/69}, \href
  {https://ui.adsabs.harvard.edu/abs/2011ApJ...738...69S} {738, 69}

\bibitem[\protect\citeauthoryear{{Shimakawa} et~al.,}{{Shimakawa}
  et~al.}{2018}]{Shimakawa2018}
{Shimakawa} R.,  et~al., 2018, \mn@doi [\mnras] {10.1093/mnras/sty2618}, \href
  {https://ui.adsabs.harvard.edu/abs/2018MNRAS.481.5630S} {481, 5630}

\bibitem[\protect\citeauthoryear{{Shimasaku} et~al.,}{{Shimasaku}
  et~al.}{2006}]{Shimasaku2006}
{Shimasaku} K.,  et~al., 2006, \mn@doi [\pasj] {10.1093/pasj/58.2.313}, \href
  {https://ui.adsabs.harvard.edu/abs/2006PASJ...58..313S} {58, 313}

\bibitem[\protect\citeauthoryear{{Shivaei} et~al.,}{{Shivaei}
  et~al.}{2015}]{Shivaei2015}
{Shivaei} I.,  et~al., 2015, \mn@doi [\apj] {10.1088/0004-637X/815/2/98}, \href
  {https://ui.adsabs.harvard.edu/abs/2015ApJ...815...98S} {815, 98}

\bibitem[\protect\citeauthoryear{{Silverman} et~al.,}{{Silverman}
  et~al.}{2015}]{Silverman2015}
{Silverman} J.~D.,  et~al., 2015, \mn@doi [\apjs] {10.1088/0067-0049/220/1/12},
  \href {https://ui.adsabs.harvard.edu/abs/2015ApJS..220...12S} {220, 12}

\bibitem[\protect\citeauthoryear{{Smit} et~al.,}{{Smit}
  et~al.}{2014}]{Smit2014}
{Smit} R.,  et~al., 2014, \mn@doi [\apj] {10.1088/0004-637X/784/1/58}, \href
  {https://ui.adsabs.harvard.edu/abs/2014ApJ...784...58S} {784, 58}

\bibitem[\protect\citeauthoryear{{Smit}, {Bouwens}, {Labb{\'e}}, {Franx},
  {Wilkins}  \& {Oesch}}{{Smit} et~al.}{2016}]{Smit2016}
{Smit} R.,  {Bouwens} R.~J.,  {Labb{\'e}} I.,  {Franx} M.,  {Wilkins} S.~M.,
  {Oesch} P.~A.,  2016, \mn@doi [\apj] {10.3847/1538-4357/833/2/254}, \href
  {https://ui.adsabs.harvard.edu/abs/2016ApJ...833..254S} {833, 254}

\bibitem[\protect\citeauthoryear{{Sobral}, {Best}, {Geach}, {Smail},
  {Cirasuolo}, {Garn}, {Dalton}  \& {Kurk}}{{Sobral} et~al.}{2010}]{Sobral2010}
{Sobral} D.,  {Best} P.~N.,  {Geach} J.~E.,  {Smail} I.,  {Cirasuolo} M.,
  {Garn} T.,  {Dalton} G.~B.,   {Kurk} J.,  2010, \mn@doi [\mnras]
  {10.1111/j.1365-2966.2010.16364.x}, \href
  {https://ui.adsabs.harvard.edu/abs/2010MNRAS.404.1551S} {404, 1551}

\bibitem[\protect\citeauthoryear{{Sobral}, {Best}, {Matsuda}, {Smail}, {Geach}
  \& {Cirasuolo}}{{Sobral} et~al.}{2012}]{Sobral2012}
{Sobral} D.,  {Best} P.~N.,  {Matsuda} Y.,  {Smail} I.,  {Geach} J.~E.,
  {Cirasuolo} M.,  2012, \mn@doi [\mnras] {10.1111/j.1365-2966.2011.19977.x},
  \href {https://ui.adsabs.harvard.edu/abs/2012MNRAS.420.1926S} {420, 1926}

\bibitem[\protect\citeauthoryear{{Sobral}, {Smail}, {Best}, {Geach}, {Matsuda},
  {Stott}, {Cirasuolo}  \& {Kurk}}{{Sobral} et~al.}{2013}]{Sobral2013}
{Sobral} D.,  {Smail} I.,  {Best} P.~N.,  {Geach} J.~E.,  {Matsuda} Y.,
  {Stott} J.~P.,  {Cirasuolo} M.,   {Kurk} J.,  2013, \mn@doi [\mnras]
  {10.1093/mnras/sts096}, \href
  {https://ui.adsabs.harvard.edu/abs/2013MNRAS.428.1128S} {428, 1128}

\bibitem[\protect\citeauthoryear{{Sobral}, {Best}, {Smail}, {Mobasher}, {Stott}
   \& {Nisbet}}{{Sobral} et~al.}{2014}]{Sobral2014}
{Sobral} D.,  {Best} P.~N.,  {Smail} I.,  {Mobasher} B.,  {Stott} J.,
  {Nisbet} D.,  2014, \mn@doi [\mnras] {10.1093/mnras/stt2159}, \href
  {https://ui.adsabs.harvard.edu/abs/2014MNRAS.437.3516S} {437, 3516}

\bibitem[\protect\citeauthoryear{{Sobral}, {Stroe}, {Koyama}, {Darvish},
  {Calhau}, {Afonso}, {Kodama}  \& {Nakata}}{{Sobral}
  et~al.}{2016}]{Sobral2016}
{Sobral} D.,  {Stroe} A.,  {Koyama} Y.,  {Darvish} B.,  {Calhau} J.,  {Afonso}
  A.,  {Kodama} T.,   {Nakata} F.,  2016, \mn@doi [\mnras]
  {10.1093/mnras/stw534}, \href
  {https://ui.adsabs.harvard.edu/abs/2016MNRAS.458.3443S} {458, 3443}

\bibitem[\protect\citeauthoryear{{Sparre}, {Hayward}, {Feldmann},
  {Faucher-Gigu{\`e}re}, {Muratov}, {Kere{\v{s}}}  \& {Hopkins}}{{Sparre}
  et~al.}{2017}]{Sparre2017}
{Sparre} M.,  {Hayward} C.~C.,  {Feldmann} R.,  {Faucher-Gigu{\`e}re} C.-A.,
  {Muratov} A.~L.,  {Kere{\v{s}}} D.,   {Hopkins} P.~F.,  2017, \mn@doi
  [\mnras] {10.1093/mnras/stw3011}, \href
  {https://ui.adsabs.harvard.edu/abs/2017MNRAS.466...88S} {466, 88}

\bibitem[\protect\citeauthoryear{{Speagle}, {Steinhardt}, {Capak}  \&
  {Silverman}}{{Speagle} et~al.}{2014}]{Speagle2014}
{Speagle} J.~S.,  {Steinhardt} C.~L.,  {Capak} P.~L.,   {Silverman} J.~D.,
  2014, \mn@doi [\apjs] {10.1088/0067-0049/214/2/15}, \href
  {https://ui.adsabs.harvard.edu/abs/2014ApJS..214...15S} {214, 15}

\bibitem[\protect\citeauthoryear{{Spergel} et~al.,}{{Spergel}
  et~al.}{2015}]{Spergel2015}
{Spergel} D.,  et~al., 2015, arXiv e-prints, \href
  {https://ui.adsabs.harvard.edu/abs/2015arXiv150303757S} {p. arXiv:1503.03757}

\bibitem[\protect\citeauthoryear{{Stanway} et~al.,}{{Stanway}
  et~al.}{2007}]{Stanway2007}
{Stanway} E.~R.,  et~al., 2007, \mn@doi [\mnras]
  {10.1111/j.1365-2966.2007.11469.x}, \href
  {https://ui.adsabs.harvard.edu/abs/2007MNRAS.376..727S} {376, 727}

\bibitem[\protect\citeauthoryear{{Stark}, {Ellis}, {Bunker}, {Bundy},
  {Targett}, {Benson}  \& {Lacy}}{{Stark} et~al.}{2009}]{Stark2009}
{Stark} D.~P.,  {Ellis} R.~S.,  {Bunker} A.,  {Bundy} K.,  {Targett} T.,
  {Benson} A.,   {Lacy} M.,  2009, \mn@doi [\apj]
  {10.1088/0004-637X/697/2/1493}, \href
  {https://ui.adsabs.harvard.edu/abs/2009ApJ...697.1493S} {697, 1493}

\bibitem[\protect\citeauthoryear{{Stark}, {Ellis}, {Chiu}, {Ouchi}  \&
  {Bunker}}{{Stark} et~al.}{2010}]{Stark2010}
{Stark} D.~P.,  {Ellis} R.~S.,  {Chiu} K.,  {Ouchi} M.,   {Bunker} A.,  2010,
  \mn@doi [\mnras] {10.1111/j.1365-2966.2010.17227.x}, \href
  {https://ui.adsabs.harvard.edu/abs/2010MNRAS.408.1628S} {408, 1628}

\bibitem[\protect\citeauthoryear{{Stark}, {Ellis}  \& {Ouchi}}{{Stark}
  et~al.}{2011}]{Stark2011}
{Stark} D.~P.,  {Ellis} R.~S.,   {Ouchi} M.,  2011, \mn@doi [\apjl]
  {10.1088/2041-8205/728/1/L2}, \href
  {https://ui.adsabs.harvard.edu/abs/2011ApJ...728L...2S} {728, L2}

\bibitem[\protect\citeauthoryear{{Stark}, {Schenker}, {Ellis}, {Robertson},
  {McLure}  \& {Dunlop}}{{Stark} et~al.}{2013}]{Stark2013}
{Stark} D.~P.,  {Schenker} M.~A.,  {Ellis} R.,  {Robertson} B.,  {McLure} R.,
  {Dunlop} J.,  2013, \mn@doi [\apj] {10.1088/0004-637X/763/2/129}, \href
  {https://ui.adsabs.harvard.edu/abs/2013ApJ...763..129S} {763, 129}

\bibitem[\protect\citeauthoryear{{Steinhardt} et~al.,}{{Steinhardt}
  et~al.}{2014}]{Steinhardt2014}
{Steinhardt} C.~L.,  et~al., 2014, \mn@doi [\apjl]
  {10.1088/2041-8205/791/2/L25}, \href
  {https://ui.adsabs.harvard.edu/abs/2014ApJ...791L..25S} {791, L25}

\bibitem[\protect\citeauthoryear{{Straatman} et~al.,}{{Straatman}
  et~al.}{2018}]{Straatman2018}
{Straatman} C. M.~S.,  et~al., 2018, \mn@doi [\apjs]
  {10.3847/1538-4365/aae37a}, \href
  {https://ui.adsabs.harvard.edu/abs/2018ApJS..239...27S} {239, 27}

\bibitem[\protect\citeauthoryear{{Tasca} et~al.,}{{Tasca}
  et~al.}{2015}]{Tasca2015}
{Tasca} L.~A.~M.,  et~al., 2015, \mn@doi [\aap] {10.1051/0004-6361/201425379},
  \href {https://ui.adsabs.harvard.edu/abs/2015A&A...581A..54T} {581, A54}

\bibitem[\protect\citeauthoryear{{Tomczak} et~al.,}{{Tomczak}
  et~al.}{2014}]{Tomczak2014}
{Tomczak} A.~R.,  et~al., 2014, \mn@doi [\apj] {10.1088/0004-637X/783/2/85},
  \href {https://ui.adsabs.harvard.edu/abs/2014ApJ...783...85T} {783, 85}

\bibitem[\protect\citeauthoryear{{Tomczak} et~al.,}{{Tomczak}
  et~al.}{2016}]{Tomczak2016}
{Tomczak} A.~R.,  et~al., 2016, \mn@doi [\apj] {10.3847/0004-637X/817/2/118},
  \href {https://ui.adsabs.harvard.edu/abs/2016ApJ...817..118T} {817, 118}

\bibitem[\protect\citeauthoryear{{Vanzella} et~al.,}{{Vanzella}
  et~al.}{2009}]{Vanzella2009}
{Vanzella} E.,  et~al., 2009, \mn@doi [\apj] {10.1088/0004-637X/695/2/1163},
  \href {https://ui.adsabs.harvard.edu/abs/2009ApJ...695.1163V} {695, 1163}

\bibitem[\protect\citeauthoryear{{Weinmann}, {Neistein}  \& {Dekel}}{{Weinmann}
  et~al.}{2011}]{Weinmann2011}
{Weinmann} S.~M.,  {Neistein} E.,   {Dekel} A.,  2011, \mn@doi [\mnras]
  {10.1111/j.1365-2966.2011.19440.x}, \href
  {https://ui.adsabs.harvard.edu/abs/2011MNRAS.417.2737W} {417, 2737}

\bibitem[\protect\citeauthoryear{{Weisz} et~al.,}{{Weisz}
  et~al.}{2012}]{Weisz2012}
{Weisz} D.~R.,  et~al., 2012, \mn@doi [\apj] {10.1088/0004-637X/744/1/44},
  \href {https://ui.adsabs.harvard.edu/abs/2012ApJ...744...44W} {744, 44}

\bibitem[\protect\citeauthoryear{{Whitaker}, {van Dokkum}, {Brammer}  \&
  {Franx}}{{Whitaker} et~al.}{2012}]{Whitaker2012}
{Whitaker} K.~E.,  {van Dokkum} P.~G.,  {Brammer} G.,   {Franx} M.,  2012,
  \mn@doi [\apjl] {10.1088/2041-8205/754/2/L29}, \href
  {https://ui.adsabs.harvard.edu/abs/2012ApJ...754L..29W} {754, L29}

\bibitem[\protect\citeauthoryear{{Whitaker} et~al.,}{{Whitaker}
  et~al.}{2014}]{Whitaker2014}
{Whitaker} K.~E.,  et~al., 2014, \mn@doi [\apj] {10.1088/0004-637X/795/2/104},
  \href {https://ui.adsabs.harvard.edu/abs/2014ApJ...795..104W} {795, 104}

\bibitem[\protect\citeauthoryear{{Willett} et~al.,}{{Willett}
  et~al.}{2015}]{Willett2015}
{Willett} K.~W.,  et~al., 2015, \mn@doi [\mnras] {10.1093/mnras/stv307}, \href
  {https://ui.adsabs.harvard.edu/abs/2015MNRAS.449..820W} {449, 820}

\bibitem[\protect\citeauthoryear{{Wold}, {Barger}  \& {Cowie}}{{Wold}
  et~al.}{2014}]{Wold2014}
{Wold} I. G.~B.,  {Barger} A.~J.,   {Cowie} L.~L.,  2014, \mn@doi [\apj]
  {10.1088/0004-637X/783/2/119}, \href
  {https://ui.adsabs.harvard.edu/abs/2014ApJ...783..119W} {783, 119}

\bibitem[\protect\citeauthoryear{{Wold}, {Finkelstein}, {Barger}, {Cowie}  \&
  {Rosenwasser}}{{Wold} et~al.}{2017}]{Wold2017}
{Wold} I. G.~B.,  {Finkelstein} S.~L.,  {Barger} A.~J.,  {Cowie} L.~L.,
  {Rosenwasser} B.,  2017, \mn@doi [\apj] {10.3847/1538-4357/aa8d6b}, \href
  {https://ui.adsabs.harvard.edu/abs/2017ApJ...848..108W} {848, 108}

\bibitem[\protect\citeauthoryear{{Xu} et~al.,}{{Xu} et~al.}{2007}]{Xu2007}
{Xu} C.,  et~al., 2007, \mn@doi [\aj] {10.1086/513512}, \href
  {https://ui.adsabs.harvard.edu/abs/2007AJ....134..169X} {134, 169}

\bibitem[\protect\citeauthoryear{{Zheng}, {Bell}, {Papovich}, {Wolf},
  {Meisenheimer}, {Rix}, {Rieke}  \& {Somerville}}{{Zheng}
  et~al.}{2007}]{Zheng2007}
{Zheng} X.~Z.,  {Bell} E.~F.,  {Papovich} C.,  {Wolf} C.,  {Meisenheimer} K.,
  {Rix} H.-W.,  {Rieke} G.~H.,   {Somerville} R.,  2007, \mn@doi [\apjl]
  {10.1086/518690}, \href
  {https://ui.adsabs.harvard.edu/abs/2007ApJ...661L..41Z} {661, L41}

\bibitem[\protect\citeauthoryear{{Zheng}, {Wang}, {Malhotra}, {Rhoads},
  {Finkelstein}  \& {Finkelstein}}{{Zheng} et~al.}{2014}]{Zheng2014}
{Zheng} Z.-Y.,  {Wang} J.-X.,  {Malhotra} S.,  {Rhoads} J.~E.,  {Finkelstein}
  S.~L.,   {Finkelstein} K.,  2014, \mn@doi [\mnras] {10.1093/mnras/stu054},
  \href {https://ui.adsabs.harvard.edu/abs/2014MNRAS.439.1101Z} {439, 1101}

\bibitem[\protect\citeauthoryear{{Zheng} et~al.,}{{Zheng}
  et~al.}{2019}]{Zheng2019}
{Zheng} Z.-Y.,  et~al., 2019, \mn@doi [\pasp] {10.1088/1538-3873/ab1c32}, \href
  {https://ui.adsabs.harvard.edu/abs/2019PASP..131g4502Z} {131, 074502}

\bibitem[\protect\citeauthoryear{{de Barros}, {Schaerer}  \& {Stark}}{{de
  Barros} et~al.}{2014}]{deBarros2014}
{de Barros} S.,  {Schaerer} D.,   {Stark} D.~P.,  2014, \mn@doi [\aap]
  {10.1051/0004-6361/201220026}, \href
  {https://ui.adsabs.harvard.edu/abs/2014A&A...563A..81D} {563, A81}

\makeatother
\end{thebibliography}

\bsp	
\label{lastpage}
\end{document}